\documentclass[10pt,preprint]{aastex}
\def\lesssim{\mathrel{\hbox{\rlap{\hbox{\lower4pt\hbox{$\sim$}}}\hbox{$<$}}}}
\def\grtsim{\mathrel{\hbox{\rlap{\hbox{\lower4pt\hbox{$\sim$}}}\hbox{$>$}}}}

\def\ninh{\mathrel{(N_{\rm i}/N_{\rm H})}}

\begin{document}

\title{10,000 Standard Solar Models: a Monte Carlo Simulation}

\author{ John N.   Bahcall\footnote{ The results presented in  this work are a
corolary of the  continuous effort John Bahcall has put for  about 40 years in
studying and understanding  the Sun. This paper took shape  and was written by
John 
during April and May, before most of the calculations were done (see
\S~\ref{subsec:howwritten}).  Most results were  already incorporated in it by
July.  John N. 
Bahcall passed  away on August  17, 2005.  With  great pain, A.M.S.   and S.B.
finished the preparation of the paper, 
particularly \S~\ref{sec:soundsppeddensityprofiles}, during October.  
 John will be deeply missed.} 
and  Aldo   M.  Serenelli}  \affil{Institute  for   Advanced  Study,  Einstein
Drive,Princeton,  NJ 08540}  \and \author{Sarbani  Basu}  \affil{Department of
Astronomy, Yale University, New Haven, CT 06520-8101}

\begin{abstract}
We  have  evolved 10,000  solar  models using  21  input  parameters that  are
randomly  drawn for  each model  from separate  probability  distributions for
every  parameter.   We  use the  results  of  these  models to  determine  the
theoretical  uncertainties  in the  predicted  surface  helium abundance,  the
profile of  the sound speed versus  radius, the profile of  the density versus
radius,  the depth of  the solar  convective zone,  the eight  principal solar
neutrino fluxes, and the fractions of  nuclear reactions that occur in the CNO
cycle  or in  the three  branches of  the p-p  chains. We  also  determine the
correlation coefficients of  the neutrino fluxes for use  in analysis of solar
neutrino oscillations.   Our calculations include the  most accurate available
input parameters, including radiative  opacity, equation of state, and nuclear
cross  sections. We  incorporate both  the recently  determined  heavy element
abundances recommended  by Asplund, Grevesse,  \& Sauval (2005) and  the older
(higher) heavy element abundances recommended by Grevesse \& Sauval (1998). We
present best-estimates of many characteristics of the standard solar model for
both sets of recommended heavy element compositions.
\end{abstract}


\maketitle

\section{INTRODUCTION}
\label{sec:introduction}

The primary purpose of this paper is to provide a quantitative
basis for deciding if a given prediction from solar models agrees
or disagrees with a measured value. We proceed by constructing
solar models in which, for every model separately, each of 21
input parameters is drawn randomly from a corresponding
probability distribution that describes our knowledge of the
parameter. We evolve models with many different sets of input
parameters and use the calculated distributions of different
theoretical quantities to describe the statistical significance of
comparisons between solar model predictions and helioseismological
or neutrino measurements. To give an explicit example, the
calculated probability distribution of the surface helium
abundance is determined by evolving many different solar models,
each with its own set of 21 randomly chosen input parameters, and
counting how many solar models yield helium abundances within each
specified bin or range of values.

The exquisite precision that has been obtained in helioseismology
over the past decade and the revolutionary advances in
understanding the properties of solar neutrinos make it
appropriate to develop the best-possible analysis techniques. New
and more powerful measurements of helioseismological parameters
and of solar neutrinos will be available in the next decade.  The
Monte Carlo simulations described in this paper will help to
position us to take full advantage of the new data.

To the best of our knowledge, the calculations described in this
paper are the first systematic attempt to use Monte Carlo
simulations to determine the uncertainties in solar model
predictions of parameters measured by helioseismology. The
helioseismological parameters we study are the depth of the
convective zone, the surface helium abundance, and the profiles of
the sound speed and density versus radius. \citet{bah88}
used a less extensive Monte Carlo
simulation, 1,000 solar models and 5 input parameters, to determine
the principal uncertainties in solar neutrino predictions. The
Monte Carlo simulations described in the present paper provide a
quantitative statistical basis for deciding if solar model
predictions agree, or disagree, with helioseismological
measurements. We do not know of any other statistical measure of
the agreement, or lack of it, between solar models and
helioseismology.

As astroseismology continues to develop, Monte Carlo simulations
of the kind described in this paper will be necessary to determine
the statistical measure of agreement between stellar models and
astroseismological measurements.

We provide in this paper the first full determination of the
correlation coefficients of the predicted solar model neutrino fluxes,
including correlations imposed by the evolution of the solar model as
well as correlations introduced by specific input parameters.
Previous discussions of correlations between neutrino fluxes have
mostly been based upon power-law approximations to the dependence of
individual neutrino fluxes upon specific input parameters 
\citep{fog95,fog02}. The correlation coefficients
determined here will make possible a simpler and somewhat more
powerful analysis of solar neutrino propagation.

 In subsequent papers, we will use the models
calculated for this paper to discuss the uncertainties in
quantities that require more extensive analysis to derive standard
deviations.  Examples of the quantities that will be studied later
include the shapes of the production probabilities versus radius
for solar neutrino fluxes, the shape of the electron distribution
versus radius, and the shape of the neutron distribution versus
radius.  These three quantities are all necessary for a precise
analysis of solar neutrino oscillations.

At present, the uncertainties in the heavy element abundances on
the surface of the Sun represent the dominant uncertainties in the
prediction with solar models of many quantities of interest. We
therefore carry out simulations using three different choices for
the heavy element abundances and their uncertainties (see
discussion below). We use the same uncertainties for
non-composition parameters for all three choices of the heavy
element abundances. Table~\ref{tab:10parameters} lists the
uncertainties in 10 important input parameters;
\S~\ref{sec:opacityeos}) discusses the uncertainties due to
radiative opacity and equation of state.

\subsection{The dilemma posed by the heavy element abundances}
\label{subsec:dilemmaheavyelements}

New and much  improved determinations for volatile elements  have led to lower
estimated photospheric abundances for the very important elements C, N, O, Ne,
and  Ar \citep[see][]{ags05,lod03,asp00b,all01,all02,asp04}.  Reductions range
between 0.13 and 0.24 dex.  The photospheric
abundance  of Si  has also  been reported  \citep{asp00a} to  be  smaller than
previous determinations by 0.05dex. 
Si  is  usually used  as  reference element  to link  the
photospheric and  meteoritic abundance scales  \citep{lod03}.  As a  result, a
lower  value  of the  photospheric  Si  leads to  an  equal  reduction of  the
meteoritic abundances of other important elements (e.g. Mg, S, Ca, Fe, Ni).

These  new determinations use  3-D calculations  (not 1-D  as in  the previous
calculations)  which  solve  the  MHD equations  consistently  with  radiative
transfer and which  correctly predict observed line widths.  Moreover, the new
calculations frequently include non-LTE effects; observational effects such as
blends are treated carefully. The net  result is that for the new calculations
the  abundances inferred  from molecular  and  atomic lines  are generally  in
agreement, whereas this was often not the case in previous abundance studies.

Surprisingly,  these new (lower)  heavy element  abundances, when  included in
solar   model   calculations,    lead   to   best-estimate   predictions   for
helioseismologically  measured quantities  like  the depth  of the  convective
zone,  the surface  helium abundance,  and the  radial distributions  of sound
speeds   and   densities   that   are   in  strong   disagreement   with   the
helioseismological measurements \citep{bp04,bah05a,bs05,basu04}.  So far there
has  not  been  a  successful  resolution  of  this  problem  \citep[see,  for
example,][]{bp04,  bah04b, bs05,  basu04, ant05,  tur04, guz04,  guz05, sea04,
bad05, mon04}.

Given what  we know about  the input parameters  of the solar models,  are the
disagreements  between solar models  that incorporate  the new  abundances (as
summarized    in   \citealt{ags05},    hereafter    AGS05   abundances)    and
helioseismological 
measurements  statistically  significant? And,  if  so,  at what  significance
level? The  Monte Carlo calculations described  in this paper  are required to
answer these questions.

Quite remarkably,  the older (higher) heavy element  abundances (as summarized
in \citealp{gs98}, hereafter GS98 abundances) lead to good
agreement with helioseismological  measurements when incorporated into precise
solar  models (see,  for example,  \citealp{bsb01,bp04,bs05,basu04}).  In this
subject, for now, it seems that `Better is worse.' 

With this  unclear situation regarding  heavy element abundances, what  is our
best  strategy   to  simulate  the  uncertainties  in   the  surface  chemical
composition? We hedge our bets. We simulate 5,000 solar models for both of the
following  cases: 1)  adopt AGS05  abundances using  the  perhaps 'optimistic'
uncertainties determined by \citet{ags05} and summarized in
Table~\ref{tab:abundanceuncertainties}  of  this  paper;  hereafter  AGS05-Opt
composition  choice;  and  2)  adopt  GS98  recommended  abundances  but  with
'conservative'  uncertainties given  in Table~\ref{tab:abundanceuncertainties}
of the present paper; hereafter  GS98-Cons composition choice. These two cases
represent our  primary Monte Carlo  simulation. In addition, we  compute 1,000
solar models for a third hybrid  case: 3) adopt the newer AGS05 abundances but
with conservative uncertainties; we denote this option AGS05-Cons.

\subsection{How this paper was written}
\label{subsec:howwritten}

Our greatest  fear in  carrying out  this project was  that we  would discover
something that  we wanted to change  after we had calculated  the 10,000 Monte
Carlo models. In order to avoid  this disaster, we went carefully over all the
details by examining  the outputs of many sets of small  numbers of models (10
to  100) that  were ultimately  discarded,  but which  we used  to refine  the
technical details  of how we handled  the simulations of  input parameters and
the calculations and analysis of solar models.

Based  upon the preliminary  calculations, we  wrote a  complete draft  of the
paper  that described  all the  technical details  and the  results (including
tables and  figures). We  decided we needed  to complete this  exercise before
running the  10,000 models to make  sure that the  results were understandable
and self-consistent and that the  simulations were indeed doing what we wanted
them to do.   Although this is an  unorthodox way to write a  paper, it turned
out to be essential for this project.

We  discovered  using  the preliminary  models  that  we  had to  adjust  some
important technical  aspects of our simulation.  For example, we  had to shift
the  mean  of the  log-normal  distribution  of  the simulated  heavy  element
composition  variables so as  to give  the observed  best-estimate composition
value (see eq.~[\ref{eq:zprimeshift}]). We  had to make several adjustments in
the size and distribution of the mesh points in our final models so as to make
possible robust  and automated helioseismological  inversions. As we  wrote up
the results, we  realized that there were additional things  that we needed to
print out and analyze or save.

A paper  that is written  in this  unusual way should  probably be read  in an
unusual way (see \S~\ref{subsec:readpaper}-\S~\ref{subsec:read}).

\subsection{Outline of this paper} \label{subec:outline}

We  present in  \S~\ref{sec:standardvalues} the  best estimates  and $1\sigma$
uncertainties  for 19  of the  21  input parameters,  including all  7 of  the
critical nuclear 
parameters,  as  well   as  the  solar  age  and   luminosity,  the  diffusion
coefficient,  and, perhaps  more  importantly, the  9  most significant  heavy
element  abundances. The  equation of  state and  the radiative  opacities are
treated separately in \S\ref{sec:opacityeos}.  In this section we describe how
we compute the 
effective $1\sigma$  uncertainties for the radiative opacity  and the equation
of state  for all of  the measurable quantities  that we calculate  with solar
models. We also give in  this section the computed $1\sigma$ uncertainties due
to  opacity  and equation  of  state  for all  of  the  predicted solar  model
quantities. We then describe in \S~\ref{sec:calcdetails} the stellar evolution
code  used  in  the  calculations  and  some  numerical  issues,  particularly
regarding the  precision with  which each solar  model has been  calculated in
order that the numerical error for every model is less than $0.1\sigma$ of the
estimated  uncertainty  in  each  of  the  calculated  helioseismological  and
neutrino  predictions.   In  \S~\ref{sec:standardsolarmodel}  we  present  and
discuss the best-estimate predictions of our standard (preferred) solar models
for 23 output  parameters. We also present in  this section the best-estimates
for the production  profile versus radius of each solar  neutrino flux and the
profiles  of  the  electron  and  neutron number  densities.   We  present  in
\S~\ref{sec:rczysurf}  our Monte  Carlo results  for  the depth  of the  solar
convective zone  and for the surface  helium abundance. We  also compare these
results      with      the      helioseismologically     measured      values.
\S~\ref{sec:soundsppeddensityprofiles}  compares  the  calculated solar  model
sound  speed   profiles  and  the   density  profiles  with  the   results  of
helioseismological  measurements. We  describe  in \S~\ref{sec:neutrinofluxes}
the Monte  Carlo results  for the distributions  of individual  solar neutrino
fluxes and  also illustrate the  important correlations between  the different
fluxes. In  \S~\ref{sec:correlationcoefficients}, we tabulate  and discuss the
correlation coefficients among the  predicted neutrino fluxes.  We present and
discuss  in  \S~\ref{sec:fusionfractions} the  fractions  of  the total  solar
nuclear energy  generation that occur via different  fusion pathways. Finally,
we   summarize  our   main   results  and   discuss   their  implications   in
\S~\ref{sec:discussion}.

\subsection{How should this paper be read?}
\label{subsec:readpaper}

We think  most readers  will be product  oriented. They  will want to  see the
results and,  will not be  as interested in  the technical details of  how the
calculations were done. We describe  the technical details in this paper; they
are necessary  in order  for the experts  to evaluate  our results and  may be
useful in  other contexts.  But, we  do not expect  that anyone  but dedicated
experts to read these descriptions.

Therefore, most  readers should begin  by leafing through  the paper to  get a
general impression of what is included, paying particular attention to figures
and tables. Very few readers need to go through the paper in the logical order
in which it is written.

\subsection{What to skip}
\label{sub:skip}

The average reader can easily  skip essential aspects of our presentation like
the choice  of the 19 best-estimate  values for the input  parameters that are
given in  \S~\ref{sec:standardvalues} and the  technical way that  we simulate
composition uncertainties (also  described in \S~\ref{sec:standardvalues}). To
use the results, it is also  not necessary to understand how we have evaluated
uncertainties due to the input  functions that represent the radiative opacity
and the equation of state (\S~\ref{sec:opacityeos}). Only aficionados of solar
modeling will be interested in \S~\ref{sec:calcdetails} on the precision with
which we have  calculated different parameters and technical  details like the
number of radial shells used in the evaluations.

\subsection{What to read}
\label{subsec:read}

We give here some examples of sections that may be of interest to readers with
expertise in different areas.

If you teach a course that touches on solar energy or on stellar evolution, or
if you  are an  astronomer working  in a specialty  not connected  to stellar
evolution or to the Sun, you may  find it interesting to peruse the section on
the standard  solar model, \S~\ref{sec:standardsolarmodel}.  This perusal will
give you a feel for what we can  calculate about the Sun. Then you can jump to
the final  summary and discussion, \S~\ref{sec:discussion}, to  get an overall
picture of the agreement between the solar model and different experiments and
to appreciate the outstanding challenges.

If you are interested in  helioseismology or astroseismology, you will want to
look      the     results      given     in      \S~\ref{sec:rczysurf}     and
\S~\ref{sec:soundsppeddensityprofiles}.  In  these  sections, we  present  the
uncertainties   in    predicting   quantities   that    have   been   measured
helioseismologically: the  distribution of  sound speeds, the  distribution of
the matter  density, the  surface abundance  of helium, and  the depth  of the
convective  zone.  We also  compare  the  measured  and predicted  values  for
helioseismological variables  and discuss the extent to  which the theoretical
and observed values agree or disagree.

If you  are interested in  neutrinos, you will  want to look carefully  at the
results presented in \S~\ref{sec:neutrinofluxes}.  We describe in this section
the uncertainties  in the predicted  neutrino fluxes and compare  the best-fit
values with the inferences from  solar neutrino experiments.  We also describe
the correlations  that are  potentially observable between  the pep,  p-p, and
$^7$Be  solar neutrino  fluxes.   In \S~\ref{sec:correlationcoefficients}  the
correlation coefficients  between all the computed neutrino  fluxes are given.
In   addition,   we   describe   in   \S~\ref{subsec:fluxesversusradius}   the
distribution of the production probability  of each of the neutrino fluxes and
in  \S~\ref{subsec:electronneutrondensities}  we   present  the  electron  and
neutron  number  densities versus  radius  (quantities  that  are required  to
discuss aspects of neutrino oscillations).

Of course, stellar model theorists may  be interested in some of the technical
details regarding the calculation of  our solar models, details that are given
in \S~\ref{sec:calcdetails}  (brief description of the  stellar evolution code
and  precision  of  the  models)  and  \S~\ref{sec:standardsolarmodel}  (input
parameters and their accuracy).  

Nuclear  astrophysicists  may  like  to  know the  fraction  of  solar  energy
generation that takes place in  different reaction paths.  This information is
given             in             Table~\ref{tab:fusionfractions}            of
\S~\ref{sec:fusionfractions}.  Nuclear physicists  in particular  may  find it
useful to  look at Table~\ref{tab:10parameters}  to see the current  status of
the  most  important nuclear  fusion  cross  sections  and the  discussion  in
\S~\ref{sec:neutrinofluxes} to understand how the nuclear uncertainties affect
the predicted neutrino fluxes.

We hope that most readers will be interested in the conclusions and discussion
presented in \S~\ref{sec:discussion}.  

\section{BEST-ESTIMATES AND UNCERTAINTIES FOR INPUT PARAMETERS}
\label{sec:standardvalues}

In this  section, we present the  best-estimate, or standard,  values we adopt
for each  of the  input parameters of  the solar  models. We also  present the
$1\sigma$ uncertainties of the best-estimate parameters.

In \S~\ref{subsec:10inputparameters}, we tabulate and discuss for 10 important
input  parameters the best-fit  values and  $1\sigma$ uncertainties.  These 10
parameters include  all the critical nuclear  parameters as well  as the solar
age  and   luminosity  and  the   diffusion  coefficient  for   heavy  element
diffusion.  In \S~\ref{subsec:compositionparameters},  we describe  for  the 9
most important  surface heavy element  abundances the best-estimate  values we
adopt and  their associated uncertainties.  The simulation of  the composition
uncertainties is less straightforward than the simulation of the uncertainties
for      the     10      non-composition      parameters     discussed      in
\S~\ref{subsec:10inputparameters}.           We           present           in
\S~\ref{subsec:compositionuncertainties}  the  equations   that  are  used  to
simulate     the     composition     uncertainties.     We     describe     in
\S~\ref{subsec:simulationsoftware} how the software works that produces the 19
simulated input parameters discussed above for each solar model.

The final 2 input  parameters that we consider, out of a  total of 21, are the
radiative  opacity and the  equation of  state.  The  opacity and  equation of
state  are complicated  functions,  unlike the  parameters  discussed in  this
section which are  all scalar numbers. Therefore we defer  a discussion of the
radiative  opacity and  the  equation of  state  to a  separate discussion  in
\S~\ref{sec:opacityeos}.

\subsection{Ten important input parameters}
\label{subsec:10inputparameters}

Table~\ref{tab:10parameters}  presents the  best-estimates and  the associated
$1\sigma$ uncertainties  that we have adopted  for each of  10 important input
parameters to the solar models. The most recent references on which we rely
for these data are given in the last column of the table.
\begin{table}[!t]
\begin{center}
\caption{Best-estimates and $1\sigma$
uncertainties for 10 important input parameters for solar models.
\label{tab:10parameters} }
\begin{tabular}{cccc}
\noalign{\smallskip} \tableline\tableline \noalign{\smallskip}
Quantity&Best&$1\sigma$&Ref.\\
&Estimate&Uncertainty&\\
 \noalign{\smallskip} \tableline
p-p & 3.94$\times 10^{-25}$ MeV b & 0.4\% & 1 \\
$^3$He+$^3$He & 5.4 MeV b & 6.0\% & 2,3 \\
$^3$He+$^4$He & 0.53 keV b & 9.4\% & 3,4 \\
$^7$Be+$e^-$ & Eq. (26), ref. 3 & 2\% & 3,5 \\
$^7$Be+p & 20.6 eV b & 3.8\% & 6 \\
hep & 8.6$\times 10^{-20}$ keV b & 15.1\% & 1 \\
$^{14}$N+p & 1.69 keV b & 8.4\% & 7,8 \\
age & $4.57 \times 10^{9}$ yr & 0.44\% & 9 \\
diffusion & 1.0 & 15.0\% & 10 \\
luminosity & $3.842\times 10^{33}{\rm ~erg\, s^{-1}}$ & 0.4\% & 9,11,12 \\
 \noalign{\smallskip} \tableline
\end{tabular}
\end{center}
\tablecomments{Some comments on the input  parameters are given in the text in
\S~\ref{subsec:10inputparameters}.   The  first  seven  quantities  listed  in
column~1 of  the table  refer to the  rates of  the low energy  nuclear fusion
reactions. The last three quantities represent the current age of the Sun, the
element  diffusion coefficient,  and  the present-day  luminosity  of the  Sun
measured with photons. The best-estimate  of 1.0 for the diffusion coefficient
corresponds to the  value calculated by Thoul et al.   (1994).  The second and
third columns give, respectively, the  best-estimate of each of the quantities
and the $1\sigma$ uncertainty, expressed  in percent of the best-estimate. The
references  are listed  in the  last column  and correspond  to  the following
articles: 
(1) \citet{park03}; (2) \citet{jun98};  (3) \citet{ade98}; (4) \citet{nara04};
(5) \citet{gru97};  (6) \citet{jun03};  (7) \citet{for04};  (8) \citet{run05};
(9) \citet{bp95};    (10)     \citet{tho94};    (11)    \citet{fro98};    (12)
\citet{bah05a}.}  
\end{table}

The    reader    may   find    useful    some    brief   comments    regarding
Table~\ref{tab:10parameters}.  The  first seven rows of the  table refer, with
the exception  of the row for the  $^7$Be + $e^-$ reaction,  to the low-energy
cross  section factors  for the  indicated nuclear  fusion reactions  (see for
example Chapter~3 of \citealp{bah89}). The entries for the p-p
reaction (low energy cross section  factor $S_{11}$) and the hep reaction have
recently  been  recalculated  with  a  rather  high  precision  \citep{park03}
culminating more than six decades of theoretical work 
on the p-p reaction. The $^3$He-$^3$He reaction ($S_{3,3}$) has been measured,
in an experimental  tour de force, down to the energies  at which solar fusion
occurs \citep{jun98}.

The rate of  the $^3$He($^4$He,$\gamma$)$^7$Be reaction ($S_{3,4}$) represents
the  most important  nuclear physics  uncertainty in  the prediction  of solar
neutrino fluxes  \citep[see][]{bp04}. We continue
to use  the estimated uncertainty  given by \citet{ade98}.  However,  a recent
measurement by \citet{nara04} gives a best-estimate that agrees exactly with
the Adelberger et al. recommended value but with a much smaller error bar. The
important  result Nara Singh  et al.  measurement should  be checked  by other
experimental groups before it can be used to reduce the error estimate for the
$^3$He +  $^4$He reaction. The measurements  should also be  extended to lower
energies; the Nara Singh measurement goes down to 420 keV.

The reaction $^7$Be($e^-$,$\nu$)$^7$Li is,  unlike the other nuclear reactions
listed in  Table~\ref{tab:10parameters}, an  electron capture reaction,  not a
nucleon-nucleon  fusion reaction.   The electron  is attracted  to  the $^7$Be
nucleus,  not repelled  by Coulomb  forces as  in a  nucleon-nucleon reaction.
Therefore, the  $^7$Be +  $e^-$ reaction cannot  be described by  a low-energy
cross  section factor  in the  way that  nucleon-nucleon fusion  reactions are
described.  The reaction rate  must be calculated theoretically, not measured.
We use  formula (26) of  \citet{ade98} Adelberger et  al. for the rate  of the
$^7$Be +  $e^-$ reaction.  This  formula has a  coefficient that is  about 1\%
higher than was obtained in the previous theoretical calculations that go back
more than 40 years.  The reason is that \citet{ade98} use the recalculation by
\citet{bah94} of the capture rate from  states of $^7$Be that are bound in the
Sun.  In his recalculation, Bahcall used profiles of the temperature, density,
and chemical composition obtained from modern solar models.

In    recent    years,    reevaluations    of    the    rate    of    the
$^{14}$N(p,$\gamma$)$^{15}$O reaction have yielded  values much smaller for the
astrophysical   factor   $S_{1,14}$    than   the   previous   adopted   value
(\citealt{ang01}; \linebreak \citealt{muk03}).  More importantly, the reaction
rate has been measured 
recently  by two beautiful,  independent experiments  \citep{for04,run05}.  We
use  the  weighted  average  cross  section  ($S_{1,14}$)  obtained  from  the
measurements  of Formicola  and Runkle  for this  reaction and  the associated
$1\sigma$ uncertainty. 

The   rates   of   other   important   nuclear   reactions   not   listed   in
Table~\ref{tab:10parameters} are taken from \citet{ade98}. 

We adopt  the
solar  age, and  the associated  uncertainty, determined  by G.  Wasserberg by
detailed analysis of meteoritic data (see discussion in \citealp{bp95}). 
The  solar  luminosity  is the  same  as  adopted  in \citet{bah05a}  and  its
uncertainty is discussed in \citet{bp95}.

We use the  diffusion subroutine that is described  in \citet{tho94} and which
is  publicly   available  at  \hbox{$  \rm   www.sns.ias.edu/\sim  jnb$}.  Our
best-estimate for the diffusion rate assumes 
that  the  results  from  this  subroutine  are  exactly  correct  (hence  the
best-estimate value  of 1.0 in Table~\ref{tab:10parameters}).  A discussion of
the   adopted  uncertainty   is  also   given  in   \citet{tho94} 
\citep[see also][]{prof94}.

\subsection{Composition parameters}
\label{subsec:compositionparameters}

In recent years, determinations of the solar abundances of heavy elements have
become   more   refined   and   detailed   (\citealp{lod03}   and   especially
\citealp{asp00a,   asp00b,  all01,  all02,   asp04,  ags05}).    These  recent
determinations yield 
significantly   lower  values   than   were  previously   adopted  (e.g.,   by
\citealp{gs98}) for the abundances of the volatile heavy
elements: C, N, O, Ne,  and Ar. However, these recent abundance determinations
lead  to  solar  models  that disagree  with  helioseismological  measurements
\citep{bp04,basu04}. By contrast, solar models that use the older
determinations  of  element  abundances   by  \citet{gs98}  are  in  excellent
agreement with helioseismology \citep{bp04, bs05, basu04, ant05, tur04, guz05,
  mon04}. 

As of  this writing,  we do not  know the  reason for the  discrepancy between
helioseismological   measurements  and   the  predictions   of   solar  models
constructed  with the more  recently determined  heavy element  abundances. We
therefore  carry out  independent simulations  using the  older  heavy element
abundances recommended by \citet{gs98} (which we call GS98 abundances) and the
more recent  heavy element abundances  recommended by \citet{ags05}  (which we
call   AGS05   abundances).   In   Table~\ref{tab:compo}   we  summarize   the
best-estimate values for the GS98 and AGS05 compositions adopted in this
paper. Only  elements accounted for  in the Opacity Project  radiative opacity
calculations are given \citep{bad05}.

We  follow the  compilers  of heavy  element  abundances in  regarding as  the
appropriate quantity on which to focus attention the logarithmic ratio
\begin{equation}
{\rm abundance}_i ~=~ \log \ninh + 12.0.
\label{eq:abundancedefn}
\end{equation}
The quantity  {\it abundance}$_i$  is the logarithmic  ratio of the  number of
atoms of type $i$ divided by the number of hydrogen atoms ($N_{\rm H}$) on the
scale  in   which  the   logarithm  of  the   number  of  hydrogen   atoms  is
12.0. 

We  vary  the  heavy  element  abundances for  the  following  nine  important
elements: C, N,  O, Ne, Mg, Si, S,  Ar, and Fe. We have  carried out numerical
experiments with different solar models to verify that the nine heavy elements
considered here  are overwhelmingly the  most significant for  solar modeling.
The remaining elements listed in  Table~\ref{tab:compo}, i.e.  Na, Al, Ca, Cr,
Mn, and  Ni, are kept  equal to their  best-estimate value in the  Monte Carlo
simulations.

\begin{table}[!t]
\begin{center}
\caption{Adopted Abundances}\label{tab:compo}
\begin{tabular}{lcc|lcc}
\noalign{\smallskip}\tableline \tableline \noalign{\smallskip}
Element & GS98 & AGS05 & Element & GS98 & AGS05 \\
\tableline
C & 8.52 & 8.39 & S & 7.20 & 7.16 \\
N & 7.92 & 7.78 & Ar & 6.40 & 6.18 \\
O & 8.83 & 8.66 & Ca & 6.35 & 6.29 \\
Ne & 8.08 & 7.84 & Cr & 5.69 & 5.63 \\
Na & 6.32 & 6.27 & Mn & 5.53 & 5.47 \\
Mg & 7.58 & 7.53 & Fe & 7.50 & 7.45 \\
Al & 6.49 & 6.43 & Ni & 6.25 & 6.19 \\
Si & 7.56 & 7.51 & & & \\
\tableline
\end{tabular}
\end{center}
\tablecomments{Best-estimate element abundances for the
two abundance compilations adopted in this work 
\citep[][GS98]{gs98} and  \citep[][AGS05]{ags05}. Only elements  accounted for
in the radiative opacity calculations by the Opacity Project group are given.} 
\end{table}

We  define  in the  next  two  subsections  abundances uncertainties  that  we
caricature as `conservative' uncertainties and `optimistic' uncertainties.

\subsubsection{Conservative Uncertainties}
\label{subsubsec:conservative}

We first  define ``Conservative  [Historical] Uncertainties" (see  column~2 of
Table~4  of  \citealp{bah05}).  We  calculate  conservative  uncertainties  by
assuming that the differences between the \citet{ags05} recommended abundances
and the  \citet{gs98} recommended  abundances represent the  current $1\sigma$
uncertainties. Thus
\begin{table}[!t]
\begin{center}
\caption{Adopted  $1\sigma$ uncertainties  for individual  heavy  elements (in
  dex).  
\label{tab:abundanceuncertainties} }
\begin{tabular}{ccc}
\noalign{\smallskip} \tableline\tableline \noalign{\smallskip}
Heavy& `Conservative'& `Optimistic' \\
Element&[historical]  (dex) & [Asplund et al. 2005] (dex) \\
\noalign{\smallskip} \tableline
C & 0.13 & 0.05\\
N & 0.14 & 0.06\\
O & 0.17 & 0.05\\
Ne & 0.24 & 0.06\\
Mg & 0.05 & 0.03\\
Si & 0.05 & 0.02\\
S & 0.04 & 0.04\\
Ar & 0.22 & 0.08\\
Fe & 0.05 & 0.03\\
\noalign{\smallskip} \tableline
\end{tabular}
\end{center}
\tablecomments{We give in column (2), under the heading
`Conservative,'  our preferred  estimated errors,  the differences
between the recent abundance determinations \citep{ags05}
and the previously standard values \citep{gs98} (see
eq.~[\ref{eq:sigmaabundance}]). Column (3), under the heading 
Asplund et al. 2005, lists our  'optimistic uncertainties'; these
uncertainties are quoted in the recent paper by \citep{ags05}. 
We use meteoritic abundances and uncertainties for the
non-volatile elements Mg, Si, S, and Fe.}
\end{table}

\begin{equation}
\sigma({\rm abundance_i})=\left|\,{\rm abundance}_i{\rm
(GS98)}-{\rm abundance}_i{\rm (AGS05)}
\label{eq:sigmaabundance}\right| \, ,
\end{equation}
where  in equation~(\ref{eq:sigmaabundance}) GS98  stands for  the composition
recommended by \citet{gs98} and AGS05 stands for
the composition recommended by \citet{ags05}.

\subsubsection{Optimistic Uncertainties}
\label{subsubsec:optimistic}

The primary uncertainties in the determination of heavy element abundances are
generally  not the measurement  errors. The  most important  uncertainties are
usually the systematic uncertainties that arise from the detailed modeling of
the solar  atmosphere that is necessary  in order to  infer element abundances
from the  measurements of line strengths.  It is very difficult  to assess the
systematic uncertainties that arise from the modeling. We cite as evidence of
this difficulty the fact that when compilers of element abundances list errors
they usually  do not specify  whether they intend  their errors to be  used as
$1\sigma$  uncertainties,  $3\sigma$  uncertainties,  or to  have  some  other
significance.

 We  define  here  as   `optimistic  $1\sigma$  uncertainties'  the  abundance
uncertainties recommended by \citet{ags05}.  We use the characterization
`optimistic'  in  contrast  to  the `conservative'  uncertainties  defined  in
\S~\ref{subsubsec:conservative}. The optimistic  uncertainties are a factor of
two or more smaller than  the conservative uncertainties for the most abundant
elements (see Table~\ref{tab:abundanceuncertainties}).

\subsection{Simulating Composition Uncertainties}
\label{subsec:compositionuncertainties} 
We  describe  in   this  subsection  how  we  simulate   the  distribution  of
uncertainties for each of the heavy element abundances. This question deserves
special  attention since  people working  in the  field of  element abundances
almost  universally  quote  best-estimates   and  uncertainties  in  terms  of
logarithms   of   the   number   abundances.   Since   symmetric   logarithmic
uncertainties result in asymmetric  errors on the abundances ([$10^{+\epsilon}
-1.0$]  is different from  $|10^{-\epsilon} -  1.0|$ ),  special care  must be
taken to make sure that logarithmic uncertainties translate into uncertainties
for the abundances that have the desired properties (e.g., the correct average
value).

Let
\begin{equation}
y = \log_{10}\left[\ninh/\ninh_0\right] \, ,
\end{equation}
where $\log \ninh_0$ is the tabulated (recommended) value of
the abundance. Let $\sigma$ be the  uncertainty in $\log
\ninh_0 $ that is listed in
Table~\ref{tab:abundanceuncertainties}. We assume a normal
distribution for $y$ with the tabulated value of $\sigma$ (in
dex). Thus
\begin{equation}
P(y)dy = [\sqrt(2\pi)\sigma]^{-1}\exp[-y^2/(2*\sigma^2)] dy .
\label{eq:normaldistribution}
\end{equation}

The normal distribution in the  logarithm of $\ninh/\ninh_0$ translates into a
log-normal distribution 
of $\ninh/\ninh_0$. This translation is exhibited by letting
\begin{equation}
z \equiv \frac{\ninh}{\ninh_0}\, .
\label{eq:zdefinition}
\end{equation}
Then
\begin{equation}
P(z) = [z\sqrt(2\pi)\sigma_{\rm l}]^{-1}\exp[-(\ln
z)^2/(2*\sigma_{\rm l}^2)]\, ,
\label{eq:zprimeshift}
\end{equation}
where $\sigma_{\rm l}$ is $\ln 10 \times \sigma$. The variable $z$
is log-normal distributed with an average value
\begin{equation}
\left<z\right> = \exp({\sigma_{\rm l}^2/2}).
\label{eq:averagesigmarelation}
\end{equation}

We want $\left< z\right>$ to be equal to 1.0.  To accomplish this we shift the
whole distribution by considering instead of $z$ the variable $z'$ where
\begin{equation}
z' \equiv z - \left(\exp({\sigma_1^2/2})- 1.0\right)\equiv \frac{\rm
\ninh'}{\ninh_0}\,. \label{eq:zprimedefn}
\end{equation}
Then, because of  the relation between the average  and the standard deviation
in  a  log-normal  distribution, equation~(\ref{eq:averagesigmarelation}),  we
have \hbox{$<z'> = 1.0$.}

We  calculate $z'$,  which is  used in  the stellar  evolution program  and in
evaluating     opacities,     from     equation~(\ref{eq:zdefinition})     and
equation~(\ref{eq:zprimedefn}). Thus
\begin{equation}
z' = \frac{\ninh'}{\ninh_0} = 10^y -
\left(\exp({\sigma_1^2/2})- 1.0\right) .
\end{equation}
In general,  the standard  deviation of  $z'$ can be  related to  the standard
 deviation of $y$ by 
\begin{equation}
\sigma(z') = \sigma(10^y) =\sqrt{\exp{(\sigma_1^2(y))}
\times \left(\exp{(\sigma_1^2(y))} -1 \right)}, 
\label{eq:sigmazprime}
\end{equation}
where, as before, $\sigma_1(y)= \ln{10} \ \sigma(y)$. For small values if 
$10^y = \ninh/\ninh_0$, they are related by a simple factor
\begin{equation}
\sigma(z') = \sigma(10^y) \simeq \ln10 \,\sigma(y)\, .
\label{eq:sigmazprimeII}
\end{equation}

\subsection{Simulation software}
\label{subsec:simulationsoftware}

The essence of  our Monte Carlo simulations is software  that chooses for each
solar model a randomly selected value  for each of the 19 parameters discussed
in   this   section.   For   each   of  the   10   parameters   discussed   in
\S~\ref{subsec:10inputparameters},  the software  chooses  a particular  value
from a Gaussian probability distribution  with the mean and standard deviation
given in  Table~\ref{tab:10parameters}. For  the 9 composition  variables, the
software  chooses for  each  solar model  particular  values from  probability
distributions        with       the       uncertainties        listed       in
Table~\ref{tab:abundanceuncertainties}   and,   as   appropriate,   with   the
best-estimate   heavy  element   abundances  as   given  in   \citet{gs98}  or
\citet{ags05}. 

Since  we consider very  large numbers  of models,  5,000 in  each simulation,
there is a small  chance that a simulated value for one  of the variables will
be non-physical,  if we accept without thinking  the probability distributions
discussed                 in                \S~\ref{subsec:10inputparameters},
\S~\ref{subsec:compositionparameters}                                       and
\S~\ref{subsec:compositionuncertainties}.  The problematic  cases could be the
neon 
and  argon  composition  variables  when the  conservative  uncertainties  are
adopted,  in which  case a  strict application  of the  log-normal probability
distribution would yield a few models  with neon or argon abundances less than
zero due to  the shift in mean value  (Eq.~\ref{eq:zprimedefn}).  To deal with
this  situation, the  software  rejects non-physical  values,  i. e.  negative
simulated values  for positive definite  quantities such as cross  sections or
compositions,  and repeats  the random  selection  until a  positive value  is
found. Given that negative, non-physical values of neon and argon are expected
to occur at  the $3.3$ and $3.9\,\sigma$ level respectively,  we do not expect
this procedure will introduce any bias in the simulated data.

\section{UNCERTAINTIES DUE TO OPACITY AND EOS}
\label{sec:opacityeos}

We  describe   in  \S~\ref{subsec:defnsigmaopacityeos}  how   we  compute  the
effective  $1\sigma$  uncertainties  that  arise  from  uncertainties  in  the
radiative opacity and  in the equation of state  (EOS). Since these quantities
are   not   single   numbers   like   the  input   parameters   discussed   in
\S~\ref{sec:standardvalues},  the  estimate of  the  effective  errors of  the
opacity and  the EOS  have to be  computed separately  for each
output quantity of  interest, depending upon the sensitivity  of each quantity
to the radiative opacity and the EOS.

We then  present and  discuss in \S~\ref{subsec:opacity1sigma}  the calculated
effective  $1\sigma$  uncertainties  due  to  the  radiative  opacity  and  in
\S~\ref{subsec:eos1sigma}   we   present  the   results   for  the   $1\sigma$
uncertainties due to the equation of state.

\subsection{Definition of Effective $1\sigma$ Uncertainties for
Opacity and Equation of State} \label{subsec:defnsigmaopacityeos}

We  begin this  subsection by  defining how  we compute  the  uncertainties in
different solar model  predictions that are caused by  our imperfect knowledge
of the radiative  opacity and the equation of state.  Then we illustrate these
definitions by  showing explicitly how  we calculate the uncertainties  in the
rms sound speed profile.

Let us denote by  $X$ the solar model quantity for which  we want to determine
the uncertainty introduced by the  opacity uncertainties.  First we evolve two
solar models that are identical except that one model uses the recent OP
opacity calculations \citep{bad05, sea04, sea05}  and the other model uses the
OPAL opacity \citep{igl96}. From this pair  of matched solar models we get two
values for $X$ 
we call $X_i({\rm  OP})$ and $X_i({\rm OPAL})$ for the models  with the OP and
the OPAL opacities respectively (the subscript $i$ denotes a given 
pair of matched models). The unbiased estimator $s^2_i$ for the variance 
is, 
\begin{equation}
s^2_i(X({\rm     opacity}))=    \frac{\left(X_i({\rm    OP})     -    X_i({\rm
    OPAL})\right)^2}{2}.  
\end{equation}
We define the fractional standard deviation squared
\begin{equation}
\sigma^2_i(X({\rm  opacity})) =  \frac{s^2_i(X({\rm opacity}))}{\mu^2_i(X({\rm
    opacity}))}     =    2     \frac{\left(X_i({\rm     OP})    -     X_i({\rm
    OPAL})\right)^2}{\left(X_i({\rm OP}) + X_i({\rm OPAL})\right)^2} 
\label{eqn:sigmaopacity}
\end{equation}
where $\mu_i(X({\rm opacity}))$ is the  mean value between $X_i({\rm OP})$ and
$X_i({\rm OPAL})$.  

In  order  to   obtain  a  more  representative  value   for  $\sigma(  X({\rm
opacity}))$,    we   decided    to   average    the   difference    shown   in
equation~(\ref{eqn:sigmaopacity})  over a  matched set  of $N  = 20$  pairs of
solar  models. In  all  $N$  cases, one  member  of each  pair  of models  was
constructed using the OP opacity and one member was constructed using the OPAL
opacity.  For  each pair  of  models, the  19  input  parameters discussed  in
\S~\ref{sec:standardvalues}     were     simulated     as     described     in
\S~\ref{subsec:simulationsoftware}. The  19 parameters were the  same for both
members of each  pair, but different parameters were simulated  for all of the
$N$ pairs. The OPAL 2001 equation of state was used in all cases.

In practice, we calculated $\sigma( X({\rm opacity}))$ from the equation
\begin{equation}
\sigma(    X({\rm   opacity}))   ~=~    \sqrt{N^{-1}\sum_i   \sigma^2_i(X({\rm
    opacity}))}\ ,
\label{eqn:inpracticesigmaopacity}
\end{equation}
where as  before $i$ denotes a pair  of matched solar models  (same values for
the 19 input parameters discussed in \S~\ref{sec:standardvalues} and EOS).  

The effective $1\sigma$  fractional uncertainty in $X$ due  to the equation of
state is calculated  in an analogous fashion. In this  case, the matched pairs
of solar models  are computed by changing  only the EOS. We use  the 2001 OPAL
equation of  state \citep{rog01} and the  earlier 1996 OPAL  equation of state
\citep{rog96}.  Thus 
\begin{equation}
\sigma( X({\rm EOS})) ~=~ \sqrt{N^{-1}\sum_i \sigma^2_i(X({\rm EOS}))}\ ,
\label{eqn:sigmaeos}
\end{equation}
where  $\sigma^2_i(X({\rm EOS}))$  for each  pair of  matched solar  models is
computed as 
\begin{equation}
\sigma^2_i(X({\rm EOS})) = 2 \frac{\left(X_i({\rm EOS 2001}) - X_i({\rm
EOS 1996})\right)^2}{\left(X_i({\rm 2001}) + X_i({\rm 1996})\right)^2} 
\end{equation}


One of  the quantities  that is  of greatest interest  is the  distribution of
sound speeds predicted  by the solar model. We  characterize this distribution
by the  root-mean-squared (rms) difference between the  sound speeds predicted
by  a given  solar  model and  the  sound speeds  inferred  from the  measured
helioseismological frequencies. Thus
\begin{equation}
\delta{\rm c} ~=~ \sqrt{{\rm M^{-1}}\sum_{i=1}^M{\left[\frac{(c_\odot -
c_{\rm model})^2}{c_\odot^2}\right]}} \, ,\label{eqn:deltac}
\end{equation}
where the  summation is carried  out over $M$  shells in the solar  model. For
consistency  and greatest  accuracy, the  inversion of  the helioseismological
frequencies  to obtain  the  solar sound  speeds  is accomplished  using as  a
reference model  the same  solar model whose  sound speed is  being considered
(see, e.g., \citealp{basu00}).  We define the rms difference in
densities, $\delta{\rho}$,  analogous to the definition of  $\delta{\rm c}$ in
equation~(\ref{eqn:deltac}) by
\begin{equation}
\delta{\rho} ~=~ \sqrt{{\rm M^{-1}}\sum_{i=1}^M{\left[\frac{(\rho_\odot
- \rho_{\rm model})^2}{\rho_\odot^2}\right]}} \,. 
\label{eqn:deltarho}
\end{equation}

We  perform  the   summation  indicated  in  equations~(\ref{eqn:deltac})  and
(\ref{eqn:deltarho})  over three  separate regions:  (1) the  interior region:
$0.07 R_\odot \leq R \leq 0.45 R_\odot$; (2) the exterior region: $0.45
R_\odot \leq R  \leq 0.95 R_\odot$; and (3) the  entire measured region: $0.07
R_\odot \leq R \leq 0.95 R_\odot$.   We have broken up the measured domains of
sound speeds and of 
densities into  these three  regions because the  region just below  the solar
convective zone  is relatively poorly  described by the standard  solar models
(see, e.g., Fig.~13 of \citealp{bsb01} or Fig.~1 of \citealp{bs05}).

We  illustrate   the  use  of   equation~(\ref{eqn:sigmaopacity})  by  showing
explicitly how we  calculate the effective $1\sigma$ uncertainty  of the sound
speed distribution due to uncertainties in the radiative opacity. We have
\begin{equation}
\sigma(  \delta  c({\rm  opacity})) ~=~  \sqrt{N^{-1}\sum_i  \sigma^2_i(\delta
  c({\rm opacity}))}\ ,
\label{eqn:sigmadeltacopac}
\end{equation}
where, analogously to eq.~\ref{eqn:sigmaopacity}, 
\begin{equation}
\sigma^2_i(\delta c({\rm opacity})) = 2 \frac{\left(\delta c_i({\rm OP}) -
\delta c_i({\rm OPAL})\right)^2}{\left(\delta c_i({\rm OP}) + \delta c_i({\rm
OPAL})\right)^2}.  
\end{equation}
%


Similarly,  for the  $1  \sigma$ uncertainty  in  $\delta{\rm c}$  due to  the
equation of state uncertainties, we have: 
\begin{equation}
\sigma(  \delta  c({\rm  EOS})) ~=~  \sqrt{N^{-1}\sum_i  \sigma^2_i(\delta
  c({\rm EOS}))}\ ,
\label{eqn:sigmadeltaceos}
\end{equation}
where  for  each  individual pair  of  matched  solar  models (same  19  input
parameters and radiative opacities), 
\begin{equation}
\sigma^2_i(\delta c({\rm EOS})) = 2 \frac{\left(\delta c_i({\rm EOS 2001}) -
\delta c_i({\rm EOS 1996})\right)^2}{\left(\delta c_i({\rm EOS 2001}) + \delta
  c_i({\rm EOS
1996})\right)^2}.  
\end{equation}


\begin{table*}[!t]
\caption{$\sigma({\rm opacity}_{\rm eff})$: Effective Standard
Deviations due to Uncertainties in the Radiative Opacity.
\label{tab:sigmasopacity} }
\begin{center}
\begin{tabular}{lccccc}
\noalign{\smallskip} \tableline\tableline \noalign{\smallskip}
Neutrino & Effective & Helioseismological & Effective & Nuclear & Effective \\
Flux & $1\sigma$  (\%) & Quantity & $1\sigma$(\%) &  Fusion Branch & $1\sigma$
(\%)\\ 
\noalign{\smallskip} \tableline
pp & 0.07 (0.04) & $Y_{\rm surf}$ & 0.32 (0.29) & p-p & 0.01 ($<0.01$)\\
pep & 0.17 (0.10) & $R_{\rm cz}$ & 0.17 (0.10) & CNO & 1.29 (0.97)\\
hep &  0.23 (0.18) &  $\delta {\rm c_{  all}}$ & 29.0  (12.6) & p-p(I)  & 0.10
(0.07)\\ 
$^7$Be & 0.78 (0.62) & $\delta {\rm c_{inner}}$ & 19.4 (7.2) & p-p(II) & 0.78
(0.61)\\ 
$^8$B & 1.87 (1.36) & $\delta {\rm c_{outer}}$ & 32.0 (13.5) & p-p(III) & 0.79
(0.61)\\ 
$^{13}$N & 1.14 (0.86) & $\delta {\rho_{\rm all}}$ & 26.8 (8.7) & &\\
$^{15}$O & 1.49 (1.12) & $\delta {\rho_{\rm inner}} $ & 17.4 (13.3) & &\\
$^{17}$F & 1.65 (1.24) & $\delta {\rho_{\rm outer}}$ & 29.2 (8.7) & &\\
\noalign{\smallskip} \tableline
\end{tabular}
\end{center}
\tablecomments{The standard deviations were computed with the aid
of equation~(\ref{eqn:inpracticesigmaopacity}), using solar models evolved
separately for the OPAL and OP radiative opacity determinations.
The values without parentheses were computed using solar models
that incorporate the \citet{gs98} heavy element
abundances; the values in parentheses were computed using \citet{ags05}
abundances. The first two columns of the table refer
to solar neutrino fluxes. The third and fourth columns give
results for helioseismological quantities: the surface helium
abundance, the depth of the convective zone, the rms difference
between the solar sound speed and the model sound speed (for the
total measured range; the inner region: $R \leq 0.45R_\odot$; and the
outer region: $R \geq 0.45R_\odot$, see eq.~[\ref{eqn:deltac}] and
eq.~[\ref{eqn:sigmadeltacopac}]), as well as the analogous rms
differences between the solar density and the model density. The
last two columns present results for percentages of the solar
energy generation that involves different nuclear paths: all p-p
reactions; all CNO reactions; p-p(I) (terminated by $^3$He-$^3$He
or p + $^2$H); p-p(II) (terminated through $e^- + ^7$Be) and
p-p(III) (terminated through $p + ^7$Be). The $1\sigma$
uncertainty is given in percent of the relevant quantity. }
\end{table*}

\subsection{Effective $1\sigma$ Uncertainties due to Radiative Opacity}
\label{subsec:opacity1sigma}

In  this   subsection,  we  describe  and  discuss   the  effective  $1\sigma$
uncertainties due to the radiative opacity.

Table~\ref{tab:sigmasopacity}  presents the effective  $1\sigma$ uncertainties
due  to  radiative opacity  for  individual  solar  neutrino fluxes,  measured
helioseismological  parameters,  and  the  parameters  that  characterize  the
different  nuclear fusion  reactions  that are  responsible  for solar  energy
generation.        The       results        were        calculated       using
equation~(\ref{eqn:inpracticesigmaopacity}).  The   numerical  values  without
parentheses were computed using solar models that incorporate the \citet{gs98}
heavy element abundances; the values in parentheses were computed using 
the  \citet{ags05} abundances.  The uncertainties  are given  in all  cases in
fractional percent.  

For all  the solar  neutrino fluxes, the  radiative opacity  introduces errors
that are small compared to the previously estimated total uncertainties in the
predicted and the measured solar neutrino fluxes \citep{bp04,bah05,bp-g04}.
This statement  is correct  for solar  models computed
with  both  the  \citet{gs98}  heavy   element  abundances,  as  well  as  the
\citet{ags05} abundances. However, the $\sim 2\%$ ($\sim 1\%$)
uncertainty in  the predicted $^8$B solar  neutrino flux due  to the radiative
opacity  is comparable to  some of  the other  commonly-calculated theoretical
uncertainties for this important  flux. Nevertheless, even for $^8$B neutrinos
the radiative opacity  contributes an uncertainty that is  a factor of several
below the total theoretical uncertainty for this important neutrino flux.

For the  surface helium abundance  and the depth  of the convective  zone, the
radiative opacity contributes uncertainties that are comparable to the claimed
accuracy  in  the  helioseismological  measurements. For  the  surface  helium
abundance,   the  quoted   measurement   error  is   0.0034   or  1.4\%   (see
eq.~[\ref{eq:Yhelio}] and \citealp{basu04}), which should be
compared  with the  smaller 0.3\% uncertainty  due to  the radiative
opacity (Table~\ref{tab:sigmasopacity}).  For
the depth  of the convective zone,  the spread among  accurate measurements is
about 0.001 or 0.14\% (see eq.~[\ref{eq:radiuscz}] and \citealp{basu04}
see also \citealp{kos91,jcd91}),
while the  radiative opacity  causes an uncertainty  of 0.17\%  (0.10\%, AGS05
abundances) that is comparable or larger (Table~\ref{tab:sigmasopacity}).

The  radiative opacity  causes  a huge  uncertainty,  $\sim 20\%-32$\%  ($\sim
7\%-14$\%, AGS05 heavy  element abundances), in the calculated  profile of the
sound  speed. The  uncertainty in  the density  profile due  to  the radiative
opacity varies from about 17\% (13\% for AGS05 abundances) in the inner region
($R \leq 0.45R_\odot$) to 30\% ($\sim$9\% for AGS05 abundances) in the
outer region of the Sun ($R \geq 0.45 R_\odot$).

The calculated fractions  of the nuclear fusion reactions  that take different
paths  in the  Sun are  practically independent  of uncertainties  due  to the
radiative opacity (last two columns of Table~\ref{tab:sigmasopacity}). In all
cases, the fractional uncertainties are $\lesssim 1$\% in the frequencies that
different nuclear fusion  paths are taken, with the only  exception of CNO for
the GS98 composition for which we get 1.3\%.

\subsection{Effective $1\sigma$ uncertainties due to equation of
state} \label{subsec:eos1sigma}

\begin{table*}[!t]
\caption{$\sigma({\rm Equation~ of~ State}_{\rm eff})$: Effective
Fractional Standard Deviations due to Uncertainties in the
Equation of State. \label{tab:sigmaseos} }
\begin{center}
\begin{tabular}{lccccc}
\noalign{\smallskip} \tableline\tableline \noalign{\smallskip}
Neutrino & Effective & Helioseismological & Effective & Nuclear & Effective \\
Flux  &  $1\sigma$(\%)  &  Quantity  &  $1\sigma$  (1\%)  &  Fusion  Branch  &
$1\sigma$(\%)\\ 
\noalign{\smallskip} \tableline
pp & 0.02 (0.02) & $Y_{\rm surf}$ & 0.12 (0.14) & p-p & 0.00 ($<0.01$)\\
pep & 0.01 (0.01) & $R_{\rm cz}$ & $<0.01$ ($<0.01$) & CNO & 0.22 (0.24)\\
hep &  0.05 (0.05) &  $\delta {\rm c_{  all}}$ & 11.6  (5.2) & p-p(I)  & 0.02
(0.02)\\ 
$^7$Be & 0.18 (0.20) & $\delta {\rm c_{inner}}$ & 16.2 (11.3) & p-p(II) & 0.18
(0.20)\\ 
$^8$B & 0.30 (0.33) & $\delta {\rm c_{outer}}$ & 13.7 (4.6) & p-p(III) & 0.18
(0.20)\\ 
$^{13}$N & 0.20 (0.21) & $\delta {\rho_{\rm all}}$ & 15.7 (4.2) & &\\
$^{15}$O & 0.24 (0.26) & $\delta {\rho_{\rm inner}}$ & 10.4 (13.0) & &\\
$^{17}$F & 0.26 (0.29) & $\delta {\rho_{\rm outer}}$ & 17.8 (4.1) & &\\
\noalign{\smallskip} \tableline
\end{tabular}
\end{center}
\tablecomments{The standard deviations were computed with the aid
of equation~(\ref{eqn:sigmaeos}), using solar models evolved
separately for the OPAL 1996 and OPAL 2001 equations of state. The
notation is the same as for Table~\ref{tab:sigmasopacity}.}
\end{table*}

In this subsection, we present  and discuss the calculated effective $1\sigma$
uncertainties due to the equation of state. We determine the uncertainties for
the EOS from equation~(\ref{eqn:sigmaeos}).

Table~\ref{tab:sigmaseos} summarizes the  effective uncertainties that are due
to  a lack of  knowledge of  the equation  of state.  We see  immediately from
Table~\ref{tab:sigmaseos}   that  the   uncertainty  in   the  EOS   does  not
significantly affect the  calculation of the neutrino fluxes  (see column~2 of
Table~\ref{tab:sigmaseos}) nor  the fraction of the  nuclear energy generation
that    occurs   via    different   fusion    pathways   (see    column~6   of
Table~\ref{tab:sigmaseos}).  For  both  the  neutrino fluxes  and  the  fusion
fractions, the fractional uncertainties are in all cases less than 0.5\%.

Also, the surface helium abundance is only affected by 0.2\%, and the depth of
the convective zone  by less than 0.01\%, by the uncertainty  in the EOS. Both
of  these   uncertainties  are   small  compared  to   the  helioseismological
measurement errors.

The  situation  is different  for  the sound  speed  profile  and the  density
profile.  For these profiles,  the uncertainty  in the  equation of  state can
cause  a  $1  \sigma$  difference  that  ranges from  about  12\%  (6\%  AGS05
abundances)  to  21\% (15\%  AGS05  abundances),  depending  upon whether  one
considers the  sound or the density  profile and whether one  is considers the
total profile or the inner or outer profile.

\section{SOLAR MODEL CALCULATIONS}
\label{sec:calcdetails}

In \S~\ref{subsec:code}  we briefly describe  the stellar evolution  code used
for  computing  the   solar  models  of  our  Monte   Carlo  simulations.   In
\S~\ref{subsec:accuracy} we describe the precision with which the solar models
were computed.  In particular, we summarize  the results of  tests carried out
using  different  numbers  of  radial  zones, time  steps,  and  criteria  for
convergence to the adopted solar luminosity, radius, and chemical composition.

\subsection{Stellar Evolution Code}
\label{subsec:code}

The stellar  evolution code used for  computing the solar models  in our Monte
Carlo  simulations is  the  Garching  stellar evolution  code  which has  been
described in some detail in \citet{wei00} with the
updates/modifications mentioned in \citet{bah05a}. Crucial to this
work is  the calculation of  appropriate radiative opacities, that  depend not
only  on the  total metallicity  assumed  for the  Sun but  on the  individual
element abundances.  For  this reason, we compute for each  solar model in our
simulations a  complete new set  of radiative opacity tables  corresponding to
the simulated composition. This has been performed using the data and software
tools provided by the Opacity Project group \citep{sea05}.

\subsection{Precision of Solar Model Calculations}
\label{subsec:accuracy}

In general, we have set the numerical parameters of our stellar evolution code
such   that  the   errors   we   make  in   calculating   the  desired   solar
parameters--neutrino related quantities and helioseismological parameters--are
less than $0.1\sigma$ of the current uncertainty in predicting each parameter.

Our best standard solar models \citep{bs05} have
approximately  2000 radial mesh  points. The  base of  the convective  zone is
particularly   well  resolved  by   using  a   grid  spacing   $\Delta  R/{\rm
R_\odot}  \approx 4\times10^{-5}$  in a  region centered  at the  base  of the
convective 
zone and extending by $0.002 {\rm R_\odot}$ both outwards and inwards. Because
the depth of  the convective zone evolves very  slowly during solar evolution,
redistributing mesh  points in each  evolutionary step is enough  to guarantee
that the depth of the convective zone  is very well defined at all times. This
high  density of  mesh points  near  the boundary  of the  convective zone  is
necessary   in  order   to  compute   a  precise   depth  of   the  convective
zone.  Evolution  from  the  Zero  Age  Main Sequence  to  the  solar  age  is
accomplished  with evolutionary time-steps  that are  not longer  than 10~Myr.
Convergence  of the  model to  the measured  values of  the  solar luminosity,
radius,  and  surface  $Z/X$  is  considered satisfactory  when  the  relative
differences  between the  computed and  the  adopted values  are smaller  than
$10^{-6}$  for  each of  the  three  quantities.  With these  conditions,  the
computational  time  required  to  calculate  a solar  model  is  kept  within
reasonable limits if only a few solar models have to be computed; however, the
computational time  becomes prohibitively large  when thousands of  models are
required.

The computational time can be reduced by relaxing the constraints on the model
accuracy.  However, when a  less stringent  convergence criterion  is adopted,
e.g.,  fewer mesh  points  are used,  or  a longer  evolutionary time-step  is
permitted, the solar model predictions deviate slightly from those of the more
accurate models. As  a practical compromise, we allow  small deviations of the
predicted solar model quantities from  the results of our most precise models,
deviations  that  are  less  than  or  equal to  $0.1\sigma$  of  the  current
uncertainty  in  the  predictions  of  each parameter.  Among  the  quantities
discussed in this  paper, the predicted values that are  most sensitive to the
numerical  accuracy  of  the  solar  models are,  given  their  small  current
theoretical uncertainties,  the depth of the  convective zone and  the p-p and
pep neutrino fluxes. The calculated  depth of the convective zone is sensitive
to the  radial mesh density while  the neutrino fluxes are  mostly affected by
the evolutionary time-step. 

Guided by trial and error, we  performed a series of numerical tests and found
an acceptable  set of  constraints that preserves  the desired  accuracy while
significantly  reducing  the  required  computational time.  There  are  three
important sets  of requirements that we  have used in evolving  models for the
Monte  Carlo calculations  discussed  in this  paper.  First, the  convergence
accuracy  is   $10^{-4}$  in  the   solar  radius,  luminosity,   and  surface
$Z/X$. Second,  the total number of mesh  points in each solar  model is about
1200 during the initial 3.5~\hbox{Gyr}  of evolution and is smoothly increased
from that moment on  until the model has about 1800 mesh  points at the end of
the  evolution. At  all times,  the high  mesh density  near the  base  of the
convective  zone  is same  as  in  our most  precise  models.  This fine  mesh
distribution is necessary for the  solar sound speed and density inversions to
have  a similar  level  of accuracy  as  our best  solar  models described  in
\S~\ref{sec:standardsolarmodel}.  Third,  evolutionary  time-steps  of  up  to
15~Myr (50\% longer than in our most precise models) are allowed.

The computational time is  reduced by more than a factor of  3 relative to our
standard  models   (see  \S~\ref{sec:standardsolarmodel})  for   solar  models
computed with these precision requirements.  However, the calculated values of
all 
the neutrino fluxes and nuclear fusion rates and all of the helioseismological
parameters we  discuss in  this paper are  the same  as in our  most precisely
calculated  models  to  within  an  accuracy of  $0.1\sigma$  of  the  current
theoretical uncertainty. In
particular, the  p-p and pep neutrino  fluxes and the depth  of the convective
zone of our best standard solar models \citep{bs05}
are reproduced  with the  less precise models  considered here to  better than
$0.07\sigma(\hbox{p-p})$,   $0.05\sigma({\rm   pep})$,  and   $0.07\sigma({\rm
convective~zone})$.   Other quantities  have larger  theoretical uncertainties
and  thus  the  errors  introduced   by  using  less  accurate  models  become
negligible. For example,  for the important $^8B$ neutrino  flux the error due
to the  reduced requirements  for the  precision of the  solar models  is only
$0.005\sigma({\rm ^8B})$.

\section{THE STANDARD SOLAR MODEL}
\label{sec:standardsolarmodel}

We present in this section the best-estimate predictions of our standard solar
models. The most important input parameters, aside from composition variables,
are listed in Table~\ref{tab:10parameters}. Any input quantities not discussed
explicitly in \S~\ref{sec:standardvalues} are the same as described in 
\citet{bs05,bp04} or \citet{bsb01} with the latest description taking
precedence. The best-estimate heavy element abundances are given in 
\citet{gs98} (GS98) and \citet{ags05} (AGS05). For short, we will
sometimes refer to these standard solar models as, respectively, the BSB(GS98)
and the BSB(AGS05) standard models.

The  only difference  between the  models discussed  in this  section  and the
models discussed in \citet{bs05} is that for the models
presented here (and throughout this paper) we use the improved low-temperature
opacities of \citet{fer05} rather than the
previously-available opacities of  \citet{af94}.  The improved low-temperature
opacities make 
no significant difference in any of the quantities we consider here except for
the depth of  the convective zone. For the BS05(OP)  model, the agreement with
helioseismology  is  slightly  improved  by  using  the  new  opacities.   The
\citet{fer05} opacities decrease the  depth of the convective  zone by 0.07\%
(or $0.0005  R_\odot$) relative to  the values obtained with  the \citet{af94}
values.  

The  free parameters in  our solar  models are:  the initial  helium abundance
$Y_{\rm init}$, the  initial metallicity $Z_{\rm init}$ and  the mixing lenght
parameter  $\alpha$.  Our  Zero  Age  Main Sequence  model  is  a  1~M$_\odot$
homogeneous star. An acceptable solar  model has to have the present-day solar
luminosity, radius and surface metallicity at the present solar age within a
precision already discussed in \S~\ref{subsec:accuracy}.

\subsection{Predictions for 23 Measurable Quantities}
\label{subsec:bestestimatepredictions}

Table~\ref{tab:standardmodelpredictions} gives,  for 23 measurable quantities,
the  calculated best-estimate  predictions  for our  preferred standard  solar
models, BSB(GS98) and BSB(AGS05). The  values that are not in parentheses were
calculated using the \citet{gs98} solar heavy
element abundance  (BSB(GS98) model); these  values are very similar  to those
obtained with the solar model BS05(OP) of \citet{bs05}. 
The values that are in parentheses were calculated using
the \citet{ags05} recommended solar
heavy  element abundances  (BSB(AGS05)  model); these  values correspond  most
closely to the values obtained from the solar model BS05(AGS, OP) of 
\citet{bs05}.

\begin{table*}[!t] \caption{Standard Solar Model Predictions:
measurable quantities. \label{tab:standardmodelpredictions} }
\begin{center}
\begin{tabular}{lccccc}
\noalign{\smallskip} \tableline\tableline \noalign{\smallskip}
Neutrino  &  Neutrino &  Helioseismological  &  Helioseismological  & Other  &
Calculated \\
Source & Flux & Quantity & Value & Quantities & Value \\
\noalign{\smallskip} \tableline
p-p  & 5.99 (6.06)  & $Y_{\rm  surface}$ &  0.2426 (0.2291)  & Cl(SNU)  & 8.12
(6.58) \\ 
pep  & 1.42  (1.45) &  $R_{\rm cz}$  &  0.7132 (0.7279)  & Ga  (SNU) &  126.08
(118.88) \\ 
hep  & 7.93  (8.25) &  $\delta {\rm  c_{ all}}$  & 0.00099  (0.00488) &  p-p &
99.2\%( 99.5\%) \\
$^7$Be &  4.84 (4.34) & $\delta {\rm  c_{inner}}$ & 0.00077 (0.00239)  & CNO &
0.78\% (0.50\%) \\
$^8$B & 5.69 (4.51) & $\delta  {\rm c_{outer}}$ & 0.00114 (0.00606) & p-p(I) &
88.3\% (89.6\%) \\
$^{13}$N & 3.05 (2.00) & $\delta {\rho_{\rm all}}$ & 0.0113 (0.0442) & p-p(II)
& 10.8 \% (9.6\%) \\
$^{15}$O  & 2.31  (1.44) &  $\delta {\rho_{\rm  inner}}$ &  0.0054  (0.0070) &
p-p(III) & 0.91\% (0.81\%) \\
$^{17}$F & 5.83 (3.25) & $\delta {\rho_{\rm outer}}$ & 0.0143 (0.0591) & & \\
\noalign{\smallskip} \tableline
\end{tabular}
\end{center}
\tablecomments{The  values without  parentheses
were calculated using the \citet{gs98} heavy element abundances
and represent our  preferred model BSB(GS98). The values  that are enclosed in
parentheses were obtained with a solar model that uses the \citet{ags05}
solar  heavy   element  abundances   and  represent   the  model
BSB(AGS05). The table presents the  predicted neutrino fluxes in the first two
columns,  in units  of $10^{10}$(p-p),  $10^{9}({\rm \,  ^7Be})$, $10^{8}({\rm
pep},  {\rm   ^{13}N,  ^{15}O})$,  $10^{6}   ({\rm  \,  ^8B,   ^{17}F})$,  and
$10^{3}({\rm hep})$  ${\rm cm^{-2}s^{-1}}$. The third and  fourth columns give
the  calculated  quantities  that are  measured helioseismologically:  the
surface helium abundance, the depth of the convective zone, and the fractional
uncertainties in the rms profiles ($\delta  c$ and $\delta \rho$) of the sound
speed and density (all measured points, as well as the inner and outer regions
of the Sun) (see eq.~[\ref{eqn:deltac}] for a definition of the rms fractional
differences). The last two columns  give the solar model predictions, assuming
no  neutrino  oscillations,  for  the  chlorine  and  gallium  solar  neutrino
experiments, and the percentage of  nuclear fusion energy that is generated by
different paths. The  quantities p-p and CNO refer,  respectively, to the full
collection of p-p and CNO  fusion reactions. The percentages for the different
p-p branches  are denoted by  p-p (I), p-p  (II), and p-p  (III), respectively
(see also the caption to Table~\ref{tab:sigmasopacity}).}
\end{table*}

We   now  comment   on   some   of  the   measurable   quantities  listed   in
Table~\ref{tab:standardmodelpredictions}.  We  first  consider  the  predicted
quantities  that have  been  measured with  helioseismology  and then  discuss
briefly the quantities that have been measured by solar neutrino experiments.

\subsubsection{Measured Helioseismological Quantities}
\label{subsubsec:measuredhelioseismology}

For        comparison       with        the        value       given        in
Table~\ref{tab:standardmodelpredictions},  the helioseismologically determined
depth of the convective zone is \citep{kos91,jcd91,guz93,basu97,basu04,basu98}:

\begin{equation}
R_{\rm CZ} = 0.713 \pm 0.001 R_\odot \, . \label{eq:radiuscz}
\end{equation}
The surface helium abundance of the Sun has recently been
redetermined by \citet{basu04}.  They find
\begin{equation}
Y_{\rm surf} = 0.2485 \pm 0.0034. \label{eq:Yhelio}
\end{equation}
The interpretation of the errors given in equation~(\ref{eq:radiuscz})
and equation~(\ref{eq:Yhelio}) is not simple since systematic
uncertainties are dominant.  However, it is clear from
Table~\ref{tab:standardmodelpredictions} that the best-estimates for
$R_{\rm CZ}$ and $Y_{\rm surface}$ computed with the \citet{gs98} 
abundances are in agreement with the
measured values while the best-estimate values computed with the
\citet{ags05} differ noticeably from the measured values.

We will compare in \S~\ref{sec:rczysurf} the Monte Carlo
distributions for $R_{\rm CZ}$ and $Y_{\rm surface}$ with the
observed values given above. The profiles, $\delta c$ and $\delta
\rho$, of the fractional differences, solar $-$ model, of the
sound speed and density are discussed in
\S~\ref{sec:soundsppeddensityprofiles} and compared with
helioseismological measurements. For completeness, we present in
\S~\ref{subsec:soundspeeddensity} the absolute values of the sound
speed and density at different radii in the Sun in our standard
models.

\subsubsection{Measured Solar Neutrino Quantities}
\label{subsubsec:measuredneutrino}

The measured event rate in the chlorine solar neutrino experiment,
expressed in solar neutrino units (SNU), is \citep{cle98}. 
\begin{equation}
\Sigma \phi(i) \sigma(i)|_{\rm Cl}~=~ 2.56 \pm 0.16 ~({\rm
statistical}) \pm 0.16 ~({\rm systematic}) ~~{\rm SNU} \,
,\label{eq:clratemeasured}
\end{equation}
where  the  summation  is  over  all   8  of  the  neutrino  fluxes  shown  in
Table~\ref{tab:standardmodelpredictions}. The difference between the predicted
standard model  value of the chlorine  event rate and the  measured event rate
created the `solar neutrino problem' in 1968 \citep{bah68,dav68}. The
predicted  rates  given  in Table~\ref{tab:standardmodelpredictions}  for  the
BSB(GS98) and BSB(AGS05) solar models bracket the predicted value estimated in
1968.

The  neutrino  absorption  cross  sections  and their  uncertainties  used  to
calculate   the  predicted  rate   for  the   chlorine  experiment   shown  in
Table~\ref{tab:standardmodelpredictions} are taken from \citet{bah88}
except  for the  $^8$B absorption  cross section,  which is  taken from
\citet{bah97}. The  uncertainties from the high energy
neutrinos (hep and $^8$B) are calculated separately and combined quadratically
with the  uncertainties from  the lower energy  neutrinos (all  other neutrino
sources). The reason is that the lower energy neutrinos essentially cause only
ground-state  to ground-state nuclear  transitions whereas  the hep  and $^8$B
neutrinos predominantly cause transitions to excited states.


The weighted average rate measured by the SAGE, GALLEX, and GNO solar neutrino
experiments is \citep{gallex99,sage02,sage03,gno05}
\begin{equation}
\Sigma   \phi(i)   \sigma(i)|_{\rm   Ga}~=~   68.1   \pm   3.85~~{\rm   SNU}\,
.\label{eq:garatemeasured}
\end{equation}
The  neutrino  absorption  cross  sections  and their  uncertainties  used  to
calculate   the    predicted   rate   in   the    gallium   experiments   (see
Table~\ref{tab:standardmodelpredictions}) are taken from \citet{bah97}. 
The uncertainties from the high energy and low
energy  neutrinos  are combined  quadratically,  as  explained  above for  the
chlorine experiment.

The  flux of  electron  neutrinos from  $^8$B  neutrino flux  measured in  the
Kamiokande, Super-Kamiokande,  and SNO experiments, assuming  no distortion of
the neutrino energy spectrum  (no neutrino oscillations), is \citep{snosalt05,
  snosalt04, kamiokande, superk1} 
\begin{equation}
\phi(^8{\rm B})_{\rm e}~=~ (1.68 \pm 0.10)\times 10^6 ~{\rm
cm^{-2}\,s^{-1}}. \label{eq:b8electrontypeflux}
\end{equation}

The  measured  rates  of  electron  type solar  neutrinos  determined  in  the
chlorine, gallium,  Kamiokande, Super-Kamiokande,  and SNO experiments  is, in
all cases, much less than the rate predicted by the standard solar models. The
discrepancies  can   be  seen  easily   by  comparing  the  values   given  in
Table~\ref{tab:standardmodelpredictions}    with   the    values    given   in
equation~(\ref{eq:clratemeasured}),       (\ref{eq:garatemeasured}),       and
(\ref{eq:b8electrontypeflux}).

The differences  between the predicted  standard model rates and  the measured
rates  in  the  chlorine  and  gallium solar  neutrino  experiments  are  well
explained by  the hypothesis  of solar neutrino  oscillations (\citealp{gri69,
  wol78,mik85,mik86}; see, for example, \citealp{bp-g04}).
The electron type neutrinos that  are produced in
the  Sun and  that  have been  measured  directly on  earth  have mostly  been
converted to  muon and tau  neutrinos by the  time they reach  the terrestrial
detectors. The quantitative  disagreements between solar neutrino measurements
and  the  predictions  of   the  standard  solar  model,  neglecting  neutrino
oscillations, are presented and discussed in \S~\ref{sec:neutrinofluxes}.

By  contrast, the  total  flux of  $^8$B  neutrinos (electron,  muon, and  tau
neutrinos) determined by  the SNO experiment (Aharmin et  al. 2005, average of
Phase I and Phase II measurements) is
\begin{equation}
\phi(^8{\rm  B}) ~=~  (4.99  \pm 0.33)\times  10^6  ~{\rm cm^{-2}\,s^{-1}}  \,
,\label{eq:total8bratemeasured}
\end{equation}
which is  in excellent agreement with  the predicted $^8$B  neutrino flux (see
Table~\ref{tab:standardmodelpredictions}).  In fact,  the  measured flux  lies
approximately halfway  between the values  predicted by the BSB(GS98)  and the
BSB(AGS05) solar models.

Given the reluctance  to accept the solar model results  by many physicists in
the 1980's  and 1990's  (which led to  the solar  neutrino problem), it  is of
interest to  compare the  present best-estimate rates  for the  standard solar
model predictions with the values in the systematic study by 
\citet{bah88}.   Despite two  decades  of refinements  in nuclear  parameters,
opacity, 
equation  of  state,  and the  inclusion  of  element  diffusion, as  well  as
intensive  studies  of the  surface  heavy  element  abundances, the  neutrino
predictions from  the standard solar  model remain almost unchanged.  The 1988
prediction for  the rate in the  chlorine experiment (then  the only available
solar neutrino experiment) was 7.9 SNU \citep{bah88},  which is
intermediate  between   the  values  of   8.1  SNU  and  6.6   SNU  predicted,
respectively,  by  the current  BSB(GS98)  and  BSB(AGS05)  solar models.  The
predicted gallium rate in 1988 was 132 SNU which is 5\% (10\%) higher than the
rate  currently  predicted  with  the  BSB(GS98) and  BSB(AGS05)  models.  The
best-estimate value  for the $^8$B neutrino  flux was $5.76  \times 10^6 {~\rm
cm^{-2} \, s^{-1}}$, within 2\%  of the current prediction using the BSB(GS98)
model.  In  all cases,  the  changes in  the  predictions  for solar  neutrino
experiments have  been less than the  quoted theoretical errors  given in 1988
(or now).

\subsection{Some Characteristics of the Standard Solar Models}
\label{subsec:somecharacteristics}

In  this subsection,  we present  some characteristics  of the  standard solar
model that  are important  and of general  interest, but which--unlike  the 23
quantities  discussed  in  \S~\ref{subsec:bestestimatepredictions}--cannot  be
measured directly.

Table~\ref{tab:somecharacteristicparameters}  lists in  the second  column the
central values of the temperature,  density, pressure, as well as the hydrogen
mass fraction and  the helium mass fraction. The values  that are not enclosed
in parentheses refer  to the BSB(GS98) standard solar model  and the values in
parentheses refer to  the BSB(AGS05) solar model. Column 4  of the table gives
the values at the base of the convective zone of the temperature, density, and
pressure,  as  well as  the  mass  enclosed in  the  convective  zone and  the
magnitude of  the radiative opacity at the  base of the zone.  The last column
gives the initial helium and  heavy element abundance, the present-day surface
abundance of Z/X, and the mixing length parameter.

\begin{table*}[!t]   \caption{Some  Characteristics   of  the   Standard  Solar
    Models\label{tab:somecharacteristicparameters} } 
\begin{center}
\begin{tabular}{cccccc}
\tableline\tableline \multicolumn{2}{c}{Center} &
\multicolumn{2}{c}{Base of convective zone}
& \multicolumn{2}{c}{Other quantities} \\
\tableline \tableline
$T_{\rm C}$ & 15.67 (15.48) & $T_{\rm  CZ}$ & 2.184 (2.006) & $Y_{\rm init}$ &
0.27250 (0.26001) \\ 
 $\rho_{\rm C}$ &  152.9 (150.4)& $\rho_{\rm CZ}$ &  0.1862 (0.1555) & $Z_{\rm
  init}$ & 0.01884 (0.01405) \\ 
$P_{\rm C}$  & 235.7 (233.8) &  $P_{\rm CZ}$ &  0.05584 (0.04341)& (Z/X)$_{\rm
  surf}$ & 0.02292 (0.01655)\\ 
$X_{\rm C}$  & 0.3461  (0.3647) &$M_{\rm CZ}$  & 0.02403  (0.01974) &$\alpha$&
2.2097 (2.1531) \\ 
$Y_{\rm C}$ & 0.6337 (0.6202) &$\kappa_{\rm CZ}$ & 20.62 (19.03)  & \\
\tableline
\end{tabular}
\end{center}
\tablecomments{Some characteristic solar  model quantities. The table presents
values calculated with the BSB(GS98) (no parentheses) standard solar model and
the  BSB(AGS05)  standard   model  (in  parentheses).   Present-epoch  central
quantities are the temperature $T_{\rm  C}$ (in units of $10^6$~\hbox{K}), the
density $\rho_{\rm C}$ (in units of~\hbox{g cm $^{-3}$}), the pressure $P_{\rm
C}$  in units  of ($10^{15}$~\hbox{erg  cm$^{-3}$}), as  well as  the hydrogen
$X_{\rm C}$ and  helium $Y_{\rm C}$ mass fractions. Conditions  at the base of
the  convective  zone are  given  by  the  temperature $T_{\rm  CZ}$,  density
$\rho_{\rm  CZ}$, pressure  $P_{\rm CZ}$  (same units  as before)  and opacity
$\kappa_{\rm CZ}$  (in units of~\hbox{cm$^2$  g$^{-1}$}). $M_{\rm CZ}$  is the
mass of the convective zone in units of solar masses. Finally, $Y_{\rm init}$,
$Z_{\rm  init}$, $(Z/X)_{\rm surface}$,  and $\alpha$  are the  initial helium
mass fraction  and metallicity, the  present surface heavy metals  to hydrogen
mass fraction of the models, and the mixing length parameter.}
\end{table*}

At the  present-epoch, the  solar core in  our standard models  is contracting
while the  outer layers are  expanding. The net  effect is an increase  in the
gravitational binding energy of the Sun  that releases energy at rate equal to
0.04\% of the  present solar luminosity, half of which  is radiated away while
the other half is stored as internal energy.

It is of interest to see  how the characteristic parameters of the solar model
have evolved  over the last two  decades, in which  important refinements have
been introduced into the calculations.  The refinements include taking account
of  the diffusion of  elements, using  a more  accurate radiative  opacity and
equation of state, and revising and refining the input nuclear cross sections.
As a reference model, we use the \citet{bah88} standard solar
model, which  represented the first  systematic combined investigation  of the
solar  neutrino problem and  of helioseismology  and which  was also  the most
comprehensive solar model study prior to the inclusion of element diffusion.

The central  values of $T_{\rm C}$,  $\rho_{\rm C}$, $P_{\rm  C}$, $X_{\rm C}$
and $Y_{\rm C}$ for the \citet{bah88} model were 15.6, 148, 229,
0.3411,       and      0.639       (same      units       as       in      the
Table~\ref{tab:somecharacteristicparameters}). We see by comparing the earlier
values with  the values given  in Table~\ref{tab:somecharacteristicparameters}
that the important  improvements over the past two decades  in the solar model
physics have left the central parameters of the model almost unchanged.

On the  other hand,  the quantities at  the base  of the convective  zone have
changed considerably  over the past two  decades. The depth  of the convective
zone  has  moved   deeper  as  the  result  of   including  element  diffusion
\citep{bp95}.  In 1988, the estimated depth of the
convective zone was  $0.74 R_\odot$, whereas the BSB(GS98)  and the BSB(AGS05)
solar models  locate the base  of the convective  zone at $0.713  R_\odot$ and
$0.728 R_\odot$ respectively, in much better agreement with helioseismological
measurements of  the convective zone depth  (see eq.~[\ref{eq:radiuscz}]). All
of the current best-estimate parameters  for the solar convective zone reflect
the fact that the transition between radiative and convective energy transport
occurs in  a deeper part  of the solar  model than it  did for the  Bahcall \&
Ulrich solar model.

In 1988, the best-estimate for  the initial helium abundance was $Y_{\rm init}
=  0.271$,  which  is  essentially  identical to  the  current  best-estimated
obtained  with  the  BSB(GS98)  solar   model  but  is  4\%  larger  than  the
best-estimate  obtained with the  BSB(AGS05) model.  The biggest  change since
1988 is in the adopted ratio of $Z/X_{\rm surf}$. In 1988, we used the value of
$Z/X_{\rm surf}= 0.02765$ from \citet{gre84}, which is
21\% larger than the \citet{gs98} value and 67\% larger than the \citet{ags05}
ratio. 

\subsection{Sound Speed and Density Versus Radius}
\label{subsec:soundspeeddensity}

\begin{table}[!t]
\caption{The sound speed in the Sun as a function of radius for
the standard solar models BSB(GS98) and BSB(AGS05).
\label{tab:csound}}
\begin{center}
\begin{tabular}{cccccc}
\tableline\tableline $R/{\rm R_\odot}$ & BSB(GS98) & BSB(AGS05) &
$R/{\rm R_\odot}$ & BSB(GS98) &
BSB(AGS05) \\
\tableline
0.000 & 5.0666e+02 & 5.0873e+02 & 0.675 & 2.3923e+02 & 2.3707e+02 \\
0.025 & 5.0803e+02 & 5.0996e+02 & 0.700 & 2.2940e+02 & 2.2767e+02 \\
0.050 & 5.1074e+02 & 5.1236e+02 & 0.725 & 2.1702e+02 & 2.1699e+02 \\
0.075 & 5.1167e+02 & 5.1304e+02 & 0.750 & 2.0345e+02 & 2.0358e+02 \\
0.100 & 5.0838e+02 & 5.0963e+02 & 0.800 & 1.7603e+02 & 1.7613e+02 \\
0.150 & 4.8748e+02 & 4.8852e+02 & 0.850 & 1.4748e+02 & 1.4755e+02 \\
0.200 & 4.5498e+02 & 4.5569e+02 & 0.900 & 1.1609e+02 & 1.1614e+02 \\
0.250 & 4.2068e+02 & 4.2100e+02 & 0.920 & 1.0201e+02 & 1.0207e+02 \\
0.300 & 3.8941e+02 & 3.8933e+02 & 0.930 & 9.4440e+01 & 9.4501e+01 \\
0.350 & 3.6228e+02 & 3.6178e+02 & 0.940 & 8.6375e+01 & 8.6441e+01 \\
0.400 & 3.3866e+02 & 3.3773e+02 & 0.950 & 7.7661e+01 & 7.7731e+01 \\
0.450 & 3.1782e+02 & 3.1647e+02 & 0.960 & 6.8017e+01 & 6.8093e+01 \\
0.500 & 2.9912e+02 & 2.9737e+02 & 0.970 & 5.6939e+01 & 5.7020e+01 \\
0.550 & 2.8188e+02 & 2.7976e+02 & 0.980 & 4.4299e+01 & 4.4401e+01 \\
0.600 & 2.6531e+02 & 2.6295e+02 & 0.990 & 2.8017e+01 & 2.7985e+01 \\
0.650 & 2.4831e+02 & 2.4599e+02 & 1.000 & 7.9193e+00 & 7.9889e+00 \\
\tableline
\end{tabular}
\end{center}
\tablecomments{The tabulated values of $c$ are the sound speed in
${\rm km~s}^{-1}$. More extensive numerical tables of $c$  are
available at http://www.sns.ias.edu/$\sim$jnb.}
\end{table}
In this  subsection, we present  and discuss the  sound speed profile  and the
density profile  in the Sun. These  profiles are not  directly measurable, but
are     nevertheless    of     considerable    theoretical     interest.    In
\S~\ref{sec:soundsppeddensityprofiles}, we compare the sound speed and density
profiles in  Monte Carlo solar models  with the corresponding  profiles in the
Sun.  Helioseismological  inversions of  solar observations determine  not the
absolute  values of  the sound  speed and  density that  are discussed  in the
present subsection,  but rather  the differences between  the model  and solar
profiles that are discussed in \S~\ref{sec:soundsppeddensityprofiles}.

Table~\ref{tab:csound} presents the sound speeds as a function of solar radius
for both  the BSB(GS98)  and the BSB(AGS05)  standard solar models.  The sound
speed in  the solar  models varies  from about $500~  {\rm km~s}^{-1}$  in the
solar  center to  about $8~  {\rm km~s}^{-1}$  on the  solar surface.  The two
standard solar  models have  very similar sound  speed profiles.  The relative
difference  of the  solar sound  speed  between the  BSB(GS98) and  BSB(AGS05)
models  in the  convective envelope  is about  $-0.05\%$. At  the base  of the
convective zone the  sound speed of BSB(GS98) becomes  larger and the relative
difference  has a  maximum of  about 1\%  at 0.65~$R_\odot$.  From  that point
inwards, the  difference decreases,  becoming negative again  at 0.3~$R_\odot$
and reaches the value of $-0.04\%$ at the solar center.

\begin{table}[!t]
\caption{The logarithm of the total density in the Sun as a
function of radius for the standard solar models BSB(GS98) and
BSB(AGS05). \label{tab:rho}}
\begin{center}
\begin{tabular}{cccccc}
\tableline\tableline $R/{\rm R_\odot}$ & BSB(GS98) & BSB(AGS05) &
$R/{\rm R_\odot}$ & BSB(GS98)&
BSB(AGS05) \\
\tableline
0.000 &   2.185 &   2.177 & 0.675 &  -0.591 &  -0.610 \\
0.025 &   2.164 &   2.158 & 0.700 &  -0.684 &  -0.706 \\
0.050 &   2.109 &   2.104 & 0.725 &  -0.768 &  -0.798 \\
0.075 &   2.032 &   2.028 & 0.750 &  -0.853 &  -0.883 \\
0.100 &   1.943 &   1.941 & 0.800 &  -1.042 &  -1.073 \\
0.150 &   1.752 &   1.753 & 0.850 &  -1.274 &  -1.304 \\
0.200 &   1.546 &   1.548 & 0.900 &  -1.586 &  -1.616 \\
0.250 &   1.321 &   1.325 & 0.920 &  -1.754 &  -1.784 \\
0.300 &   1.082 &   1.085 & 0.930 &  -1.854 &  -1.884 \\
0.350 &   0.836 &   0.839 & 0.940 &  -1.970 &  -2.000 \\
0.400 &   0.592 &   0.593 & 0.950 &  -2.108 &  -2.138 \\
0.450 &   0.355 &   0.354 & 0.960 &  -2.280 &  -2.309 \\
0.500 &   0.127 &   0.124 & 0.970 &  -2.508 &  -2.537 \\
0.550 &  -0.091 &  -0.097 & 0.980 &  -2.852 &  -2.880 \\
0.600 &  -0.299 &  -0.309 & 0.990 &  -3.506 &  -3.533 \\
0.650 &  -0.496 &  -0.512 & 1.000 &  -6.783 &  -6.774 \\
\tableline
\end{tabular}
\end{center}
\tablecomments{The tabulated values are $\log_{10}{\rho}$, where $\rho$
is the total density in \hbox{g/cm$^3$}. 
More extensive numerical tables of $\rho$ and are
available at http://www.sns.ias.edu/$\sim$jnb.}
\end{table}

Quadratic interpolation within Table~\ref{tab:csound} accurately
reproduces the numerical values from the solar models. The sound
speed can be interpolated from the table to an accuracy that is
typically much better than 0.1\% from the center of the Sun up to
$0.98~R_\odot$. Only in the region $0.15R_\odot<R<0.2R_\odot$ is the
accuracy of the quadratic interpolation degraded to about 0.15\%.

Table~\ref{tab:rho} presents the density profiles for the BSB(GS98)
and BSB(AGS05) standard solar models. The density varies by 9
orders of magnitude from the solar interior to the solar surface,
from $153 {\rm ~gm~cm}^{-3}$ ($150 {\rm ~gm~cm}^{-3}$ ) at the
center of the Sun to $1.65 \times 10^{-7} {\rm ~gm~cm}^{-3}$
($1.68 \times 10^{-7} {\rm ~gm~cm}^{-3}$) at the solar surface
(optical depth equal 0.312). Quadratic interpolation in
Table~\ref{tab:rho} reproduces the solar model densities to an
accuracy better than 0.2\% in the inner $0.8~R_\odot$ and is more
accurate than 0.5\% up to $0.97~R_\odot$.

The fractional differences between the densities for the BSB(GS98)
and the BSB(AGS05) solar models are larger than the fractional
differences of the sound speeds. The
density in the convective envelope is about 7\% larger in the
BSB(GS98) model than in the BSB(AGS05) model. This difference
smoothly drops to zero at 0.45~$R_\odot$ where it becomes negative
and at 0.25~$R_\odot$ the BSB(AGS05) density is about 1\% larger
than that of the BSB(GS98) model. Towards
the center the BSB(GS98) model again has higher density 
than the BSB(AGS05) model, the difference being close to 2\% in the center.

\begin{figure*}[!t]
\begin{center}
\includegraphics[bb=50 70 545 720,angle=0.0,scale=0.6]{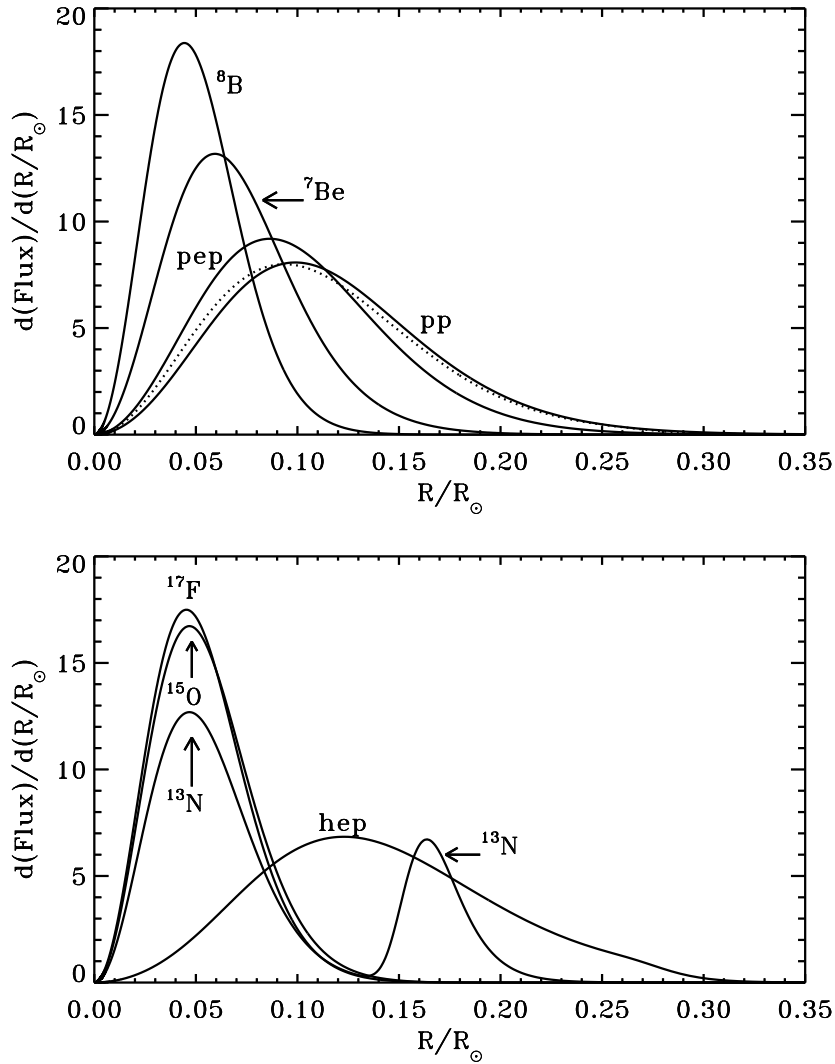}
\caption{Neutrinos  fluxes  versus radius.  The  figure  shows the  production
profiles of the principal neutrino fluxes versus radius for our standard solar
model BSB(GS98). The dotted line in  the upper panel, close to the profile for
the  p-p  neutrino  flux,  represents  the production  profile  of  the  solar
luminosity.  The  production profiles are normalized to  unity when integrated
over  $dR/R_\odot$. The  double-peaked distribution  of the  $^{13}$N neutrino
flux is explained in the text.\label{fig:fluxesvsradius}}
\end{center}
\end{figure*}

\subsection{Neutrino Fluxes Versus Radius}
\label{subsec:fluxesversusradius}

Figure~\ref{fig:fluxesvsradius} shows the production profiles versus radius of
each of the important solar neutrino  fluxes, as well as the solar luminosity.
The  $^8$B, $^7$Be, $^{15}$O,  and $^{17}$F  neutrino fluxes  are concentrated
toward the  center of the  Sun, while  the p-p, pep,  and hep fluxes  are more
broadly distributed.

The  $^{13}$N neutrino production  profile is  double-peaked.  The  inner peak
(small radii) represents neutrinos produced where the CN reactions are 
approximately in  steady state.  The outer peak  (large radii)  represents the
residual 
burning          of         $^{12}$C          by          the         reaction
$^{12}$C(p,$\gamma$)$^{13}$N($\beta^+$)$^{13}$C   at   radii   at  which   the
temperature is too low to permit the subsequent burning of nitrogen.

Table~\ref{tab:fluxpeaks}  gives  the  locations  of  the  peak  in  the  flux
distribution per  unit radius  for each  solar neutrino flux,  as well  as the
locations  below and  above the  peak  radius within  which 34.1\%  (effective
$\sigma/2$)  of the  flux is  produced. The  table presents  values  that were
computed using the \citet{gs98}  recommended heavy element abundances and also
values that were computed  using the \citet{ags05} recommended abundances. One
can see immediately from the table that the flux distributions are practically
independent of  which of  the two recommended  compositions is used,  which is
another indication  that solar  neutrino fluxes are  not much affected  by the
choice of heavy element composition (within the currently fashionable range of
surface heavy element abundances).

\subsection{Electron and Neutron Number Densities Versus Radius}
\label{subsec:electronneutrondensities}

We present in this subsection the electron and neutron number
densities that are required to calculate the propagation of
neutrinos through  solar material.

\begin{table}[!t]       \caption{Neutrino       Flux       Dependence       on
    Radius\label{tab:fluxpeaks} } 
\begin{center}
\begin{tabular}{lccccccc}
\noalign{\smallskip} \tableline\tableline \noalign{\smallskip}
Neutrino & \multicolumn{3}{c}{BSB(GS98)} & & \multicolumn{3}{c}{BSB(AGS05)} \\
Flux & $R_{\rm peak}$ & $R_{\rm inner}$ & $R_{\rm outer}$ & & $R_{\rm peak}$ &
$R_{\rm inner}$ & $R_{\rm outer}$ \\
\noalign{\smallskip} \tableline
p-p & 0.0990 & 0.0470 & 0.1471 & & 0.0990 & 0.0471 & 0.1472 \\
pep & 0.0864 & 0.0410 & 0.1286 & & 0.0866 & 0.0411 & 0.1290 \\
hep & 0.1230 & 0.0616 & 0.1796 & & 0.1230 & 0.0618 & 0.1797 \\
$^7$Be & 0.0594 & 0.0276 & 0.0889 & & 0.0592 & 0.0274 & 0.0887 \\
$^8$B & 0.0443 & 0.0220 & 0.0654 & & 0.0442 & 0.0219 & 0.0653 \\
$^{13}$N & 0.0468 & 0.0221 & 0.0698 & & 0.0470 & 0.0224 & 0.0701 \\
$^{13}$N$_2$ & 0.1637 & 0.1473 & 0.1781 & & 0.1615 & 0.1450 & 0.1758 \\
$^{15}$O & 0.0468 & 0.0220 & 0.0700 & & 0.0470 & 0.0222 & 0.0703 \\
$^{17}$F & 0.0454 & 0.0217 & 0.0675 & & 0.0455 & 0.0218 & 0.0677 \\
\noalign{\smallskip} \tableline
\end{tabular}
\end{center}
\tablecomments{The table presents the peak radius, $R_{\rm peak}$,
and the inner and outer radii, $R_{\rm inner}$ and $R_{\rm outer}$,
for each neutrino flux. The peak radius corresponds to the maximum
of the flux production per unit radius, while the inner and outer
radii correspond to the points at which 34.1\% of the flux is
produced on either side of the peak.  The flux distributions are
shown in Figure~\ref{fig:fluxesvsradius}.  The second $^{13}$N
peak, $^{13}$N$_2$, can be seen clearly in
Figure~\ref{fig:fluxesvsradius}  and is explained in the text.The
second, third, and fourth columns were computed using \citet[GS98]{gs98}
heavy element abundances; the fifth, sixth,
and seventh columns were computed using \citet[AGS05]{ags05}
heavy element abundances. }
\end{table}

The dominant effect for converting an electron type neutrino to a
muon or tau neutrino in the Sun is proportional to the profile of
the electron number density minus one-half the neutron number
density, $n_e - 0.5\times n_{\rm n}$, as a function of solar
radius \citep{wol78,mik85,lim88}. This
matter-induced change of neutrino flavors is known as the MSW
effect, after its discoverers Mikheyev, Smirnov, and Wolfenstein.
The probability for matter to induce transformations of other
active neutrinos, $\nu_\mu$ or $\nu_\tau$, is proportional to $-
0.5\times n_{\rm n}$. If one only considers active neutrinos,
$\nu_e$, $\nu_\mu$, and $\nu_\tau$, then it is not necessary to
know $n_{\rm n}$ because the common phase induced by $n_{\rm n}$
does not affect the oscillation probability. However, in order to
calculate the propagation of sterile neutrinos, one requires the
profile of $n_e - 0.5\times n_{\rm n}$ \citep{bar91}.

\begin{table*}[!t]
\caption{ The electron number density and the neutron number
density versus radius for the standard solar models (GS98 and
AGS05, see \S~\ref{subsec:bestestimatepredictions}.).
\label{tab:ne}}
\begin{center}
\begin{tabular}{ccccccccccc}
\tableline\tableline $R/{\rm R_\odot}$ &
\multicolumn{2}{c}{$\log_{10}(n_e/{\rm N_A})$} &
\multicolumn{2}{c}{$\log_{10}
 (n_n/N_{\rm       A})$}       &       &       $R/{\rm       R_\odot}$       &
\multicolumn{2}{c}{$\log_{10}(n_e/{\rm                N_A})$}                &
\multicolumn{2}{c}{$\log_{10}(n_n/N_{\rm A})$} \\ 
& GS98 & AGS05 & GS98 & AGS05 &&& GS98 & AGS05 & GS98 & AGS05\\
\tableline 
0.000 &  2.0125 &  2.0114 & 1.6990  & 1.6795  && 0.600 &  -0.3649 &  -0.3708 &
-1.1501 & -1.1878 \\ 
0.025  & 1.9981 &  1.9974 &  1.6665 &  1.6476 &&  0.650 &  -0.5621 &  -0.5732 &
-1.3489 & -1.3919\\ 
0.050 &  1.9581 &  1.9585 & 1.5770  & 1.5593  && 0.700 &  -0.7460 &  -0.7651 &
-1.5582 & -1.6008 \\ 
 0.075 &  1.8998 & 1.9015  & 1.4495 &  1.4325 && 0.750  & -0.9130 &  -0.9394 &
-1.7395 & -1.8012 \\ 
0.100  &1.8295 &  1.8326 &  1.3037 &  1.2862 &&  0.800 &  -1.1024 &  -1.1287 &
-1.9289 & -1.9906 \\ 
0.150  & 1.6648  & 1.6703  & 1.0102  &0.9901 &&  0.850 &  -1.3341 &  -1.3603 &
-2.1605 & -2.2221 \\ 
0.200 &  1.4711 &  1.4783 & 0.7440  & 0.7223  && 0.900 &  -1.6462 &  -1.6721 &
-2.4726 & -2.5340 \\ 
0.250 &  1.2509 &  1.2591 & 0.4950  & 0.4729  && 0.940 &  -2.0306 &  -2.0561 &
-2.8571 & -2.9180 \\ 
0.300  &1.0129 &  1.0211 &  0.2462 &  0.2236 &&  0.950 &  -2.1685 &  -2.1938 &
-2.9950 & -3.0557 \\ 
0.350 &  0.7678 & 0.7752  & -0.0050  &-0.0285 && 0.960  & -2.3399 &  -2.3648 &
-3.1663 & -3.2267 \\ 
0.400 &  0.5244 & 0.5303  & -0.2519  &-0.2769 && 0.970  & -2.5683 &  -2.5926 &
-3.3947 & -3.4545 \\
0.450 &  0.2877 & 0.2916  & -0.4911  &-0.5183 && 0.980  & -2.9124 &  -2.9355 &
-3.7388 & -3.7973 \\ 
0.500  & 0.0602 &  0.0614 &  -0.7209&-0.7509 &&  0.990 &  -3.5662 &  -3.5890 &
-4.3927 & -4.4508 \\ 
0.550 & -0.1574  & -0.1594 & -0.9407  &-0.9740 && 1.000 & -6.8436  & -6.8296 &
-7.6700 & -7.6915 \\
\tableline
\end{tabular}
\end{center}
\tablecomments{The tabulated  values are $\log_{10}  (n_e/N_A)$ and $\log_{10}
  (n_n/N_A)$, 
where $n_e (n_n) $ is the electron (neutron) number density measured in number
per  ${\rm cm^3}$  and $N_A$  is Avogadro's  number. More  extensive numerical
files      of       $n_e$      and      $n_n$       are      available      at
http://www.sns.ias.edu/$\sim$jnb.  The numerical  values were  calculated with
our  preferred solar  models  constructed  using either  GS98  or AGS05  heavy
element  abundances and are  described in  \S~\ref{sec:standardsolarmodel} and
Table~\ref{tab:standardmodelpredictions}. }
\end{table*}

We present here the separate distributions  for $n_e$ and $n_n$.  The user can
easily form  the combined density $n_e  -0.5\times n_n$ from  the values given
here. In addition, since we give separately $n_e$ and $n_n$ the user can study
more exotic, non-standard interactions that may require different combinations
of $n_e$ and $n_n$ (see, e.g., \citealp{wol78} and \citealp{fri04}).

Table~\ref{tab:ne} gives,  at representative points  in the Sun,  the electron
and neutron  number densities as a  function of solar radius  for our standard
solar   model.   For   the  electron   distribution,  the   values   given  in
Table~\ref{tab:ne} can be used (with quadratic interpolation) to reproduce the
actual electron distribution in the given solar models to better than 0.1\% in
all the  solar interior from  the center up  to $0.7 R_\odot$ and,  for larger
radii, to better than 2\% up  to $0.99 R_\odot$. For the neutron distribution,
quadratic  interpolation  in Table~\ref{tab:ne}  reproduces  the solar  models
distribution to an accuracy of 0.5\%  (usually much better) from the center up
to $0.7  R_\odot$ and, for  larger radii,  to an accuracy  of 2\% up  to $0.99
R_\odot$. More extensive numerical files of the number densities are available
at http://www.sns.ias.edu/$\sim$jnb.

As  can  be seen  in  Table~\ref{tab:ne},  the  electron and  neutron  density
distributions are  different for different assumed solar  compositions. In the
case of the electron density  distribution, the differences between the models
are smaller than 2\% for $R<0.6 R_\odot$  and rise up to 6\% in the convective
envelope  ($R>  0.7  R_\odot$)  where  the higher  values  correspond  to  the
BSB(GS98) model.  On the other hand, the neutron density is 5\% higher for the
BSB(GS98) model  at the  center and this  difference smoothly increases  up to
10\% at the base of the  convective envelope, while in the convective envelope
itself the difference is about 15\%.

\begin{figure*}[!t]
\begin{center}
\includegraphics[bb=50 50 545 360,scale=0.65]{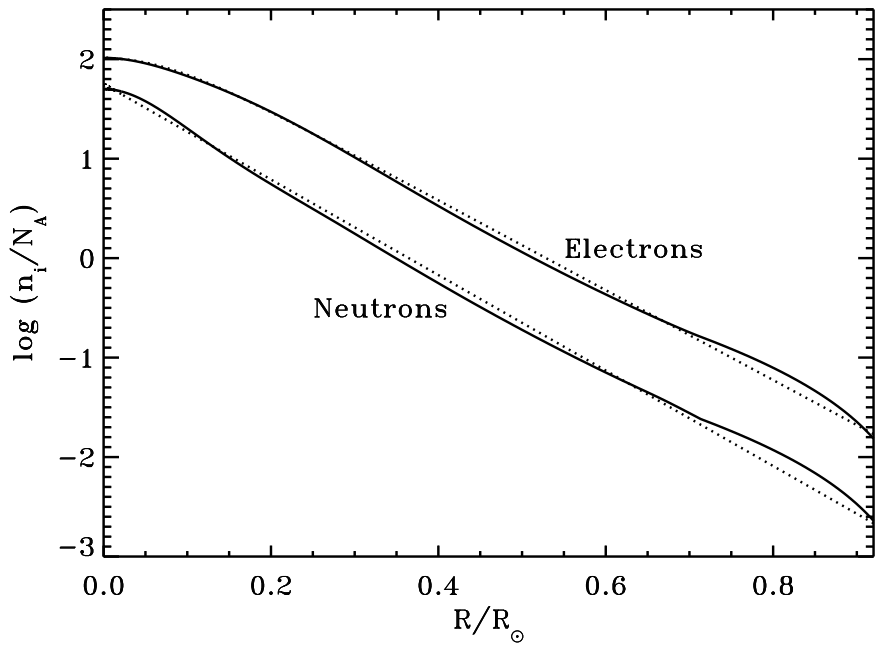} 
\medskip\medskip\smallskip
\caption[]{Electron  and Neutron  Number Densities  Versus Radius.  The figure
shows as solid lines the logarithm of the number densities of electrons and of
neutrons, divided by  Avogadro's number ${\rm N_{\rm A}}$,  versus radius. The
dotted   lines  represent   the   analytic  fits   presented   in  the   text,
equation~(\ref{eq:enumberdensity}) and equation~(\ref{eq:nnumberdensity}). For
radii  $R <  0.3 R_\odot$,  the analytic  fits are  sufficiently close  to the
actual model distributions that it is  difficult to see them as separate lines
in  the figure.  The electron  and neutron  number densities  plotted  in this
figure  were calculated  using  a  solar model  that  incorporated GS98  heavy
element abundances.  The number  densities are practically  the same  if AGS05
heavy         element         abundances         are         used,         see
text and Table~\ref{tab:ne}. \label{fig:electronneutrondensity}}
\end{center}
\end{figure*}

Figure~\ref{fig:electronneutrondensity}  shows   the  calculated  solar  model
values of  the electron and neutron  number densities versus  solar radius for
the standard solar  model constructed with the GS98  heavy element abundances.
In the inner  region of the Sun,  $R < 0.3 R_\odot$, where  matter effects are
most  relevant for  neutrino  oscillations, the  electron  and neutron  number
densities can be  approximated by analytic formulae. We  find for the electron
number density
\begin{equation}
\log{n_e/{\rm N_A}}= 2.36 - 4.52x - 0.33\
\exp{\left[(-x/0.075)^{1.1} \right]} \, ,
\label{eq:enumberdensity}
\end{equation}
and for the neutron number density
\begin{equation}
\log{n_n/{\rm N_A}}= 1.72 - 4.80x \, , \label{eq:nnumberdensity}
\end{equation}
where $x=R/{\rm R_\odot}$. The analytic fits given in
equation~(\ref{eq:enumberdensity}) and
equation~(\ref{eq:nnumberdensity}) are shown as dotted lines in
Figure~\ref{fig:electronneutrondensity}.

The  first  two  terms  in equation~(\ref{eq:enumberdensity})  (with  slightly
different     constants),      originally     derived     by     \citet{bah88}
\citep[see][]{bah89}, have
been  used by  many authors  in  calculations of  MSW survival  probabilities.
However, these two terms  alone significantly overestimate the electron number
density for radii  less than $0.12R_\odot$ (see Fig.~6d  of \citealp{bah88} or
Fig.~4d of \citealp{bah89}).  We have therefore added the third term in
equation~(\ref{eq:enumberdensity}), which leads to satisfactory agreement with
the numerical values for  the solar model in the inner region  of the Sun. The
rms difference between  the values given by equation~(\ref{eq:enumberdensity})
is  7\%  for  $R <  0.7  R_\odot$.   The  agreement  of  the values  given  in
equation~(\ref{eq:nnumberdensity}) with the solar model values is 12\% rms for
$R < 0.7 R_\odot$.  In the outer  region, for $R > 0.8 R_\odot$ (which is less
important for standard MSW  transformations), the analytic fits represented by
equation~(\ref{eq:enumberdensity})  and equation~(\ref{eq:nnumberdensity}) are
not accurate and  one must use the more  precise numerical values extrapolated
from Table~\ref{tab:ne} or take the values directly from the solar model.

For our standard solar model  constructed with AGS05 heavy element abundances,
the      coefficients     for      the      corresponding     versions      of
equation~(\ref{eq:enumberdensity})  and equation~(\ref{eq:nnumberdensity}) are
practically  the  same.  The  three  coefficients  in  the  AGS05  version  of
equation~(\ref{eq:enumberdensity})the  numerical constants  are, respectively,
2.38, 4.56, and -0.36 (instead of the values of 2.36, 4.52, and -0.33 that are
optimal    for    GS98    abundances).    For    the    AGS05    version    of
equation~(\ref{eq:nnumberdensity}), the numerical constants are 1.72 and -4.84 
(instead of 1.72 and -4.80 for GS98 abundances). 

The   analytic  formulae   given  in   equation~(\ref{eq:enumberdensity})  and
equation~(\ref{eq:nnumberdensity})  can  be used  in  analytic discussions  of
solar neutrino  oscillations (just as they  have been used by  many authors in
previous  decades) and  for  most  purposes these  formulae  are adequate  for
numerical calculations of neutrino oscillations. However, for the most precise
work, quadratic interpolation in Table~\ref{tab:ne} should be used.

\section{MONTE  CARLO RESULTS  FOR CONVECTIVE  ZONE DEPTH  AND  SURFACE HELIUM
  ABUNDANCE} 
\label{sec:rczysurf}

We present in this section our Monte  Carlo results for the depth of the solar
convective zone and the surface  helium abundance. We compare the results with
the  helioseismologically   measured  values.   We  begin   by  discussing  in
\S~\ref{subsec:convectivezone}  the  comparison  between  the  calculated  and
measured  values for  the depth  of the  convective zone.  We then  discuss in
\S~\ref{subsec:helium}  the  comparison between  the  calculated and  measured
values of the surface helium abundance.

Our major results are  summarized in Figure~\ref{fig:rczsurfacey}.  All of the
panels in the figure  show the number of solar models that  were found to have
the depth of  the convective zone (or, for the  right-hand panels, the surface
helium abundance) in a given bin. The top two rows of the figure were obtained
from  5,000 solar  models  each and  the  bottom (lowest)  row summarizes  the
results for  1,000 solar models.   All of the  distributions are well  fit (as
judged by a reduced $\chi^2$ calculation)  by a Gaussian shape, which is shown
in each panel as a smooth curve.

\subsection{Depth of The Convective Zone}
\label{subsec:convectivezone}

\begin{figure*}
\begin{center}
\includegraphics[bb=0 40 566 730,angle=0.0,scale=0.65]{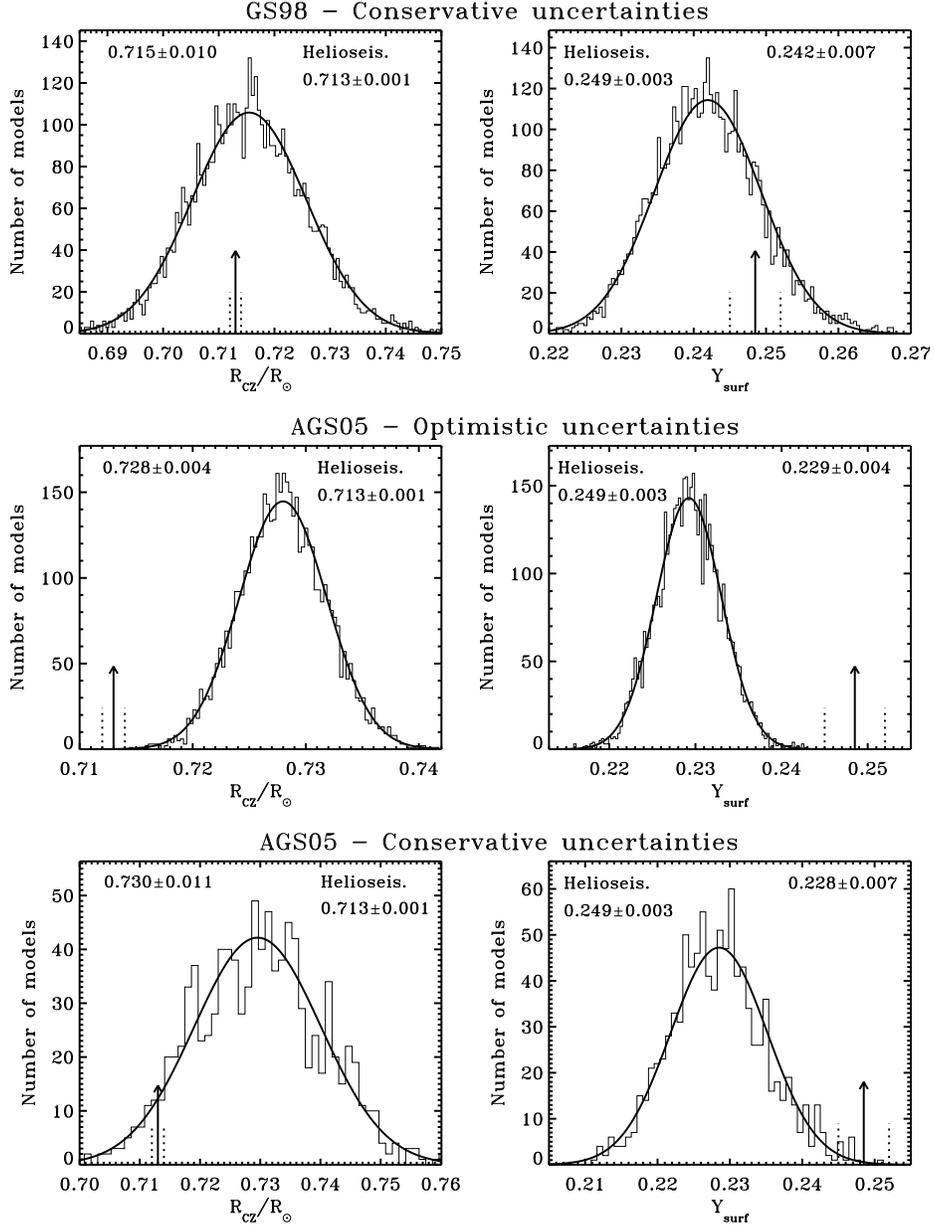}
\caption[]{Depth  of the  convective  zone, $R_{\rm  cz}$  and surface  helium
abundance,  $Y_{\rm surf}$.  The three  left panels  present the  number of
models  with different  values of  the  depth of  the convective  zone for  a)
conservative composition errors and \citet{gs98} 
recommended abundances; b) the recommended errors and abundances given by 
\citet{ags05}; and c) conservative composition errors and the \citet{ags05}
recommended  abundances.   The  composition  uncertainties are  discussed  in
\S~\ref{subsec:compositionparameters}.  The right three panels show the number
of solar models with different values  of the surface helium abundance for the
same three cases as  for the depth of the convective zone.  The arrows and the
dotted  lines  represent the  measured  values  and  their uncertainties.  The
distribution    of    models    is,    in    all   six    cases    shown    in
Figure~\ref{fig:rczsurfacey}, well described with a Gaussian distribution, the
smooth curves in each panel. The confidence level in the top two plots is, for
example, 96.3\% and 80.0\%. \label{fig:rczsurfacey}}
\end{center}
\end{figure*}

The top two left panels of Figure~\ref{fig:rczsurfacey} summarize
the results for the depth of the convective zone that were
obtained with our two standard composition choices: 1) GS98
abundances and `conservative' uncertainties (GS98-Cons); 2) AGS05 abundances
and `optimistic' uncertainties (AGS05-Opt). The third (lowest) row was
computed with a hybrid choice of AGS05 abundances and conservative
uncertainties (AGS05-Cons). 

The GS98 abundances and the `conservative' composition uncertainties listed in
the  second   row  of  Table~\ref{tab:abundanceuncertainties}   were  used  in
calculating the solar models to which the top panel refers. The mean value and
$1 \sigma$ uncertainty for the calculated values of the convective zone are
\begin{equation}
R_{\rm CZ} = 0.7154 \pm 0.0102 R_\odot; ~{\rm GS98-Cons}. 
\label{eq:gs98conservradiuscz}
\end{equation}
In this case, the solar model results are  in good agreement with
the helioseismologically determined depth of the convective zone
(see eq.~[\ref{eq:radiuscz}]).

However, the situation is different if the AGS05 abundances and
the AGS05 uncertainties (`optimistic' uncertainties, see column~3
of Table~\ref{tab:abundanceuncertainties}) are both used. The
second row of Figure~\ref{fig:rczsurfacey} shows that the
disagreement in this case is  significant.  We find
\begin{equation}
R_{\rm CZ} = 0.7280 \pm 0.0037 R_\odot;~{\rm AGS05-Opt}. 
\label{eq:ags05optimisticradiuscz}
\end{equation}
Thus the composition and composition uncertainties recommended by
\citet{ags05} lead to
solar models with values for $R_{\rm CZ}$ that differ from the
helioseismologically measured value by $3.7 \sigma$, where the
$\sigma$ used here is the quadratically-combined solar model and
helioseismological errors.

If the AGS05 \citep{ags05} abundances and the
`conservative' composition uncertainties are used to calculate the
depth of the convective zone, we find (see second row of
Fig.~\ref{fig:rczsurfacey})
\begin{equation}
R_{\rm CZ} = 0.7296 \pm 0.0105 R_\odot;~{\rm AGS05-Cons}.
\label{eq:ags05conservradiuscz}
\end{equation}
There is no strong disagreement  between the AGS05 abundances and the measured
depth  of the convective  zone if  conservative composition  uncertainties are
assumed.  Note,  however,  that  this  a  result  of  the  large  conservative
composition uncertainties.

\subsection{Surface Helium Abundance}
\label{subsec:helium}

The right hand panels  of Figure~\ref{fig:rczsurfacey} compare the solar model
calculations   of  the   present-day   surface  helium   abundance  with   the
helioseismologically measured values.

Using the GS98 abundances and conservative uncertainties, we find
(see the top right panel of Fig.~\ref{fig:rczsurfacey}):
\begin{equation}
Y_{\rm surf} = 0.2420 \pm 0.0072; ~{\rm GS98-Cons},
\label{eq:GS98conservYhelio}
\end{equation}
which is in very good agreement with the helioseismologically
determined value of $Y_{\rm surf} = 0.2485$ (see
eq.~[\ref{eq:Yhelio}]).

The AGS05 abundances and uncertainties lead to solar model
predictions,
\begin{equation}
Y_{\rm surf} = 0.2292 \pm 0.0037; ~{\rm AGS05-Opt}, 
\label{eq:AGS05optimYhelio}
\end{equation}
that differ from the helioseismologically measured value by $3.8
\sigma$ (combined solar model and helioseismological error).

The agreement is still not  good if we use AGS05 abundance and
conservative uncertainties. In this case, we find from the solar
models that
\begin{equation}
Y_{\rm surf} = 0.2285 \pm 0.0067; ~{\rm AGS05-Cons}. 
\label{eq:AGS05conservYhelio}
\end{equation}
The best-estimate with the AGS05 abundances and conservative
uncertainties differs from the helioseismologically measured value
(eq.~[\ref{eq:Yhelio}]) by about $2.7\sigma$.

\section{MONTE CARLO RESULTS FOR SOUND SPEED AND DENSITY PROFILES}
\label{sec:soundsppeddensityprofiles}

In the previous section we examined the distribution of the position of the CZ
base 
and the helium abundance of the models. While the CZ base and helium abundance
are two  very important quantities  obtained from helioseismic studies  of the
Sun, 
there are by no means the only two. Helioseismic inversions have allowed us to
determine the solar sound-speed and density profiles for most of the Sun,
the results are valid from roughly 5-7\% of the solar radius to 95\% of the
solar radius. Thus we have additional information about the Sun with
which to compare our models.
We present in this section the  Monte Carlo results for the comparison between
the solar model  and helioseismologically obtained solar sound-speed and
density profiles.  

Inversions to determine solar the solar sound-speed and density profiles are
done by inverting the frequency differences between the Sun and a solar
model to obtain the sound-speed and density between the Sun and the model
(called the ``reference model''). Thus inversions directly show us whether or
not the structure of a solar model agrees with that of the Sun.
For this work,  the frequency differences
between the Sun  and the models were inverted  using the Subtractive Optimally
Localized Averages (SOLA) technique \citep{pij92,pij94}. We set up the
inversion problem  and  determined  the various parameters of  the inversion in
the manner  described by \citet{rab99}.
  For the helioseismological data, we use solar oscillation frequencies
obtained  by  the  Michelson Doppler  Imager  (MDI)  on  board the  Solar  and
Heliospheric Observatory  (SOHO). In  particular, we use  frequencies obtained
from MDI  data that were collected for  the first 360 days  of its observation
\citep{schou98}.  This data set  was chosen because it was derived from a
long time  series when solar activity was  low. The length of  the time series
results in  reduced noise  and hence a  larger number  of modes for  which the
frequencies can  be determined  reliably. Mode sets  derived from  longer data
sets   are   available,   but   they   only  consist   of   low-degree   modes
\citep[e.g., ][]{ber00}. 
In addition,  a longer  time series  would have  meant adding
observations  from periods  of  increasing solar  activity,  which would  have
changed the frequencies. It is  a well-established fact that solar frequencies
increase with solar activity. However, it  is also known that increase is such
that  it  does  not reflect  a  change  in  structure  of the  solar  interior
\citep{basu02}. 
To quantize the difference between the structure of the Sun and the
models, we determine the rms difference between the sound-speed
and density profiles of the Sun and the models. A larger rms difference
would denote a worse model.
We note here that in order  to minimize systematic effects, the inversion of
the solar frequencies has been done independently with each of the models in
our simulation as the reference model.

We present the relusts discussion the sound-speed differences between the
Sun and the different models in \S~\ref{sec:sound}. The density differences
are described in \S~\ref{sec:density}. The results are
summarized by Figs.~\ref{fig:soundspeed} and \ref{fig:density}, as
well as Table~\ref{tab:rmscandrho}.

\subsection{Sound speed profiles}
\label{sec:sound}

\begin{figure*}[!t]
\begin{center}
\includegraphics[bb=0 40 566 730,angle=0.0,scale=0.65]{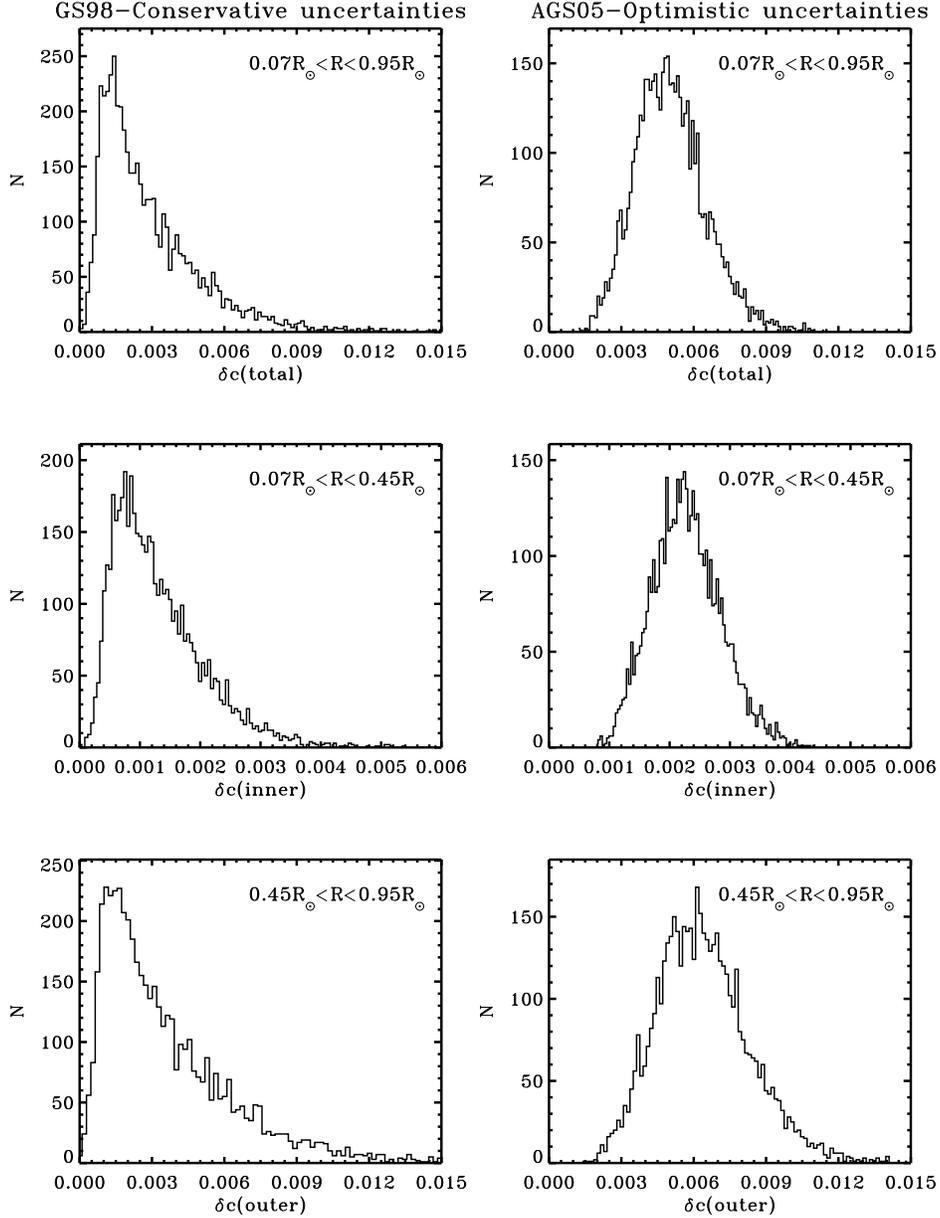}
\caption[]{\label{fig:soundspeed}
Distributions of the rms Sound-speed differences.  Panels on the left (right) 
correspond  to  the GS98-Cons  (AGS05-Opt)  composition  choice.  From top  to
bottom,  the total  $\delta c_{\rm  all}$, inner  $\delta c_{\rm  inner}$, and
outer  $\delta c_{\rm  outer}$ rms  differences  are shown.  The solar  region
involved in the  definition of each rms difference is  shown in the respective
panels.  }
\end{center}
\end{figure*}

\begin{figure*}[!t]
\begin{center}
\includegraphics[bb=0 40 566 730,angle=0.0,scale=0.65]{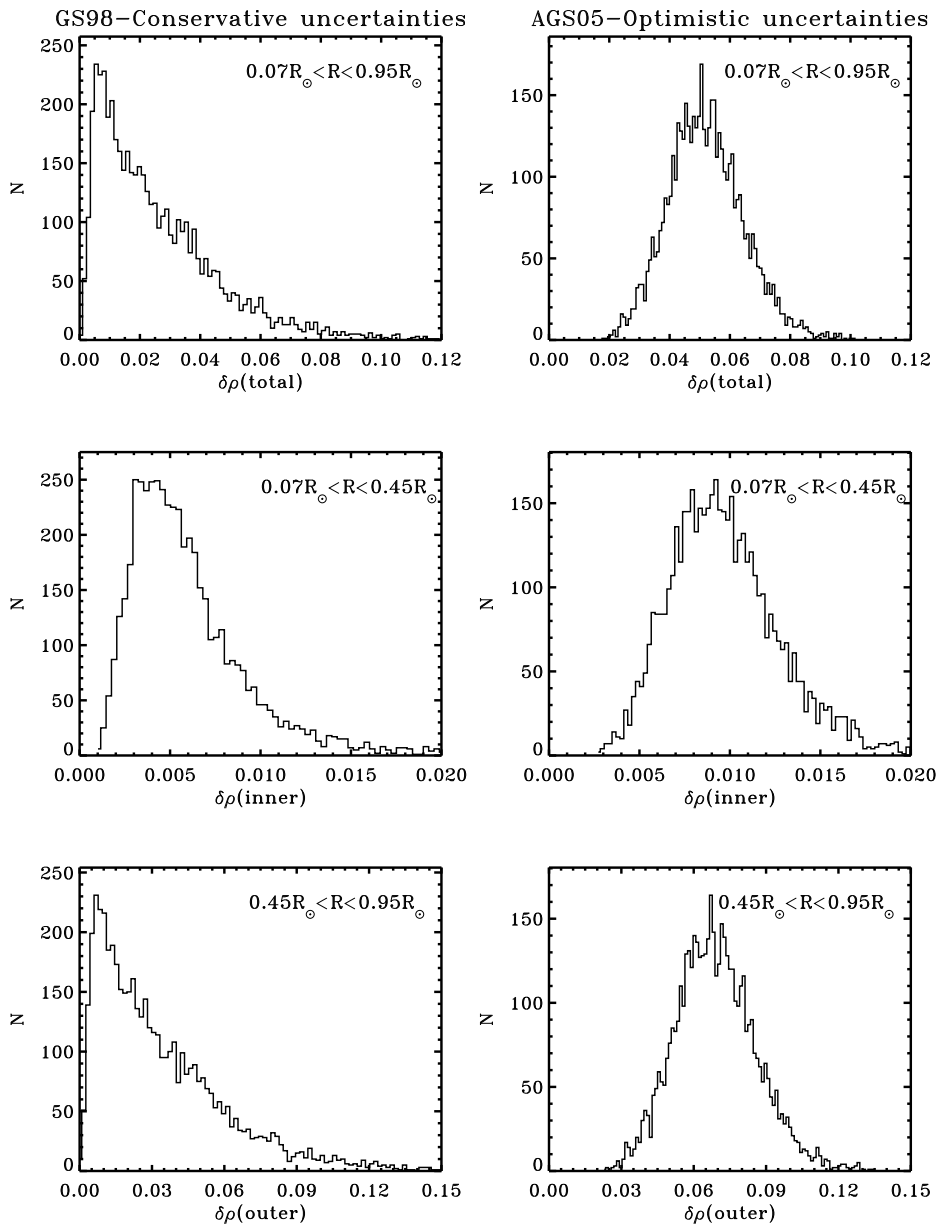}
\caption[]{\label{fig:density}
Density rms difference distributions.  Panels on the left (right)
correspond  to  the GS98-Cons  (AGS05-Opt)  composition  choice.  From top  to
bottom, the  total $\delta \rho_{\rm  all}$, inner $\delta  \rho_{\rm inner}$,
and  outer $\delta  \rho_{\rm outer}$  rms differences  are shown.   The solar
region  involved in  the definition  of each  rms difference  is shown  in the
respective panels.  }
\end{center}
\end{figure*}

The rms difference  between the sound-speed profile of the  Sun and the models
is shown in Fig.~\ref{fig:soundspeed}. The  differences are shown not only for
the entire range of radii over which the inversion results are valid, but also
a few  smaller radius ranges to check  for the sensitivity of  the profiles to
input physics.

The internal structure  of solar models is sensitive,  at different depths, to
different  physical inputs  in the  calculations.  For  instance,  the adopted
solar composition  has a more dramatic  effect on the structure  of the models
for  $R  \grtsim  0.45{\rm  R_\odot}$  \citep{bah05a,neon05}.   Of  particular
importance is  how composition affects the  location of the base  of the solar
convective envelope and how this affects the sound-speed profile in the region
$R  \grtsim 0.45{\rm  R_\odot}$.  In  the  temperature range  in this  region,
opacity is dominated by bound-free transitions and this largely determines the
temperature gradient.  Higher metal abundances  lead to higher opacities and a
steeper  radiative  temperature gradient.   As  a  result,  the condition  for
convective  stability (for  which  we adopt  the  Schwarzschild criterion)  is
satisfied  at  higher  temperatures  (i.e.,  a  larger  depth)  for  a  higher
metallicity.   Given the sharp  transition in  temperature at  the CZ  base, a
mismatch in its location in the  solar models with respect to the actual solar
location gives  rise to  a large  difference sound speed  in that  region.  At
inner regions, however, free-free  transitions and electron scattering opacity
sources  become gradually  more important,  diminishing the  influence  of the
details of the solar composition. This leads us to discuss our results for the
sound speed  profiles (and analogously for  the density profiles)  not only in
terms of the  rms sound speed difference $\delta c_{\rm  all}$ over the entire
range of validity  of the inversion, but also in two  smaller ranges, an inner
rms $\delta c_{\rm inner}$ defined for $0.07{\rm R_\odot} \leq R \leq 0.45{\rm
R_\odot}$,  and an  outer sound  speed difference  rms $\delta  c_{\rm outer}$
defined  for $0.45{\rm  R_\odot}  \leq  R \leq  0.95{\rm  R_\odot}$ (see  also
\$~\ref{subsec:defnsigmaopacityeos}).

The resulting  distributions for $\delta c_{\rm all}$,  $\delta c_{\rm inner}$
and  $\delta c_{\rm  outer}$  are shown  in Figure~\ref{fig:soundspeed}  (top,
middle and botton panels respectively).  The column on the left corresponds to
the models  obtained from the  Monte Carlo simulations adopting  the GS98-Cons
composition choice, while the right column corresponds to models obtained with
the  AGS05-Opt composition.   In the  case  of the  GS98-Cons simulation,  the
distribution of  each of the difference  rms is strongly peaked  very close to
zero, reinforcing  the well-known result  that solar models adopting  the GS98
solar abundances give in general very good agreement with helioseismic results
regardless  of the  uncertainties in  the other  input physics  (e.g.  nuclear
cross  section,  EOS,  radiative  opacities, element  diffusion).   The  tails
extending  to high  rms  values  result from  the  adopted large  conservative
uncertainties.   For the  AGS05-Opt composition  simulation  the distributions
peak at  much higher values  than in the  GS98-Cons case, reflecting  the fact
that the sound speed of the solar models constructed with this composition are
not  in   good  agreement   with  the  sound   speed  profile   inferred  from
helioseismology.  Additionally,  it is evident  that the distributions  do not
extend to such low rms values as those found with the GS98 composition.

\begin{table}[!t] 
\caption{Sound Speed and Density Profiles:
RMS Differences Between Solar Models and Helioseismological Measurements
\label{tab:rms}}
\begin{center}
\begin{tabular}{lcccccccccc}
\tableline\tableline         &         \multicolumn{2}{c}{GS98-Cons}        &&
\multicolumn{3}{c}{AGS05-Opt} && \multicolumn{3}{c}{AGS05-Cons} \\ \tableline 
& $Q$ & $\sigma_{0.68}$ && $Q$ & $\sigma_+$ & $\sigma_-$ && $Q$ & $\sigma_+$ & 
$\sigma_-$ \\
\tableline
$\delta c_{\rm  all}$ & 0.00143  & 0.00334 &&  0.00487 & 0.00136 &  0.00119 &&
0.00484 & 0.00316 & 0.00256 \\
$\delta c_{\rm inner}$  & 0.00075 & 0.00151 && 0.00216 &  0.00060 & 0.00052 &&
0.00238 & 0.00096 & 0.00095 \\
$\delta c_{\rm outer}$  & 0.00111 & 0.00423 && 0.00582 &  0.00206 & 0.00167 &&
0.00523 & 0.00453 & 0.00312 \\
$\delta  \rho_{\rm all}$ &  0.0055 &  0.0311 &&  0.0486 &  0.0135 &  0.0112 &&
0.0440 & 0.0311 & 0.0224 \\
$\delta \rho_{\rm  inner}$ & 0.0039  & 0.0067 &&  0.0086 & 0.0032 &  0.0025 &&
0.0076 & 0.0057 & 0.0033 \\
$\delta \rho_{\rm  outer}$ & 0.0069  & 0.0408 &&  0.0646 & 0.0180 &  0.0149 &&
0.0580 & 0.0416 & 0.0295 \\
\tableline \\
\end{tabular}
\end{center}
\tablecomments{The rms differences  $\delta c$ and $\delta \rho$  of the sound
speed  and  the  density  profiles from  the  helioseismologically  determined
profiles  are given  in the  table for  three regions  of the  solar interior.
Equations~(\ref{eqn:deltac})   and   (\ref{eqn:deltarho})   define   the   rms
differences.    In  the  case   of  the   GS98  abundances   and  conservative
uncertainties the distributions are highly asymmetric and we characterize them
by their most probable value (or mode) $Q$ (second column) and their one-sided
68.3\%  confidence  level  $\sigma_{0.68}$  (third  column).   For  the  AGS05
composition and optimistic uncertainties we  give for the distribution of each
rms difference the mode $Q$  (fourth column) and the $\sigma_+$ and $\sigma_-$
values  defining  the  68.3\%   confidence  level  (fifth  and  sixth  columns
respectively). The  same quantities  are given for  the AGS05  composition and
conservative  uncertainties  in columns  7  to  9.   Details on  the
definitions  of $\sigma_{0.68}$, $\sigma_+$  and $\sigma_-$  are given  in the
text.   The  three regions  are  (inner): $0.07  \leq  R  \leq 0.45  R_\odot$;
(outer):  $0.45 \leq  R  \leq 0.95  R_\odot$;  and (all):  $0.07  \leq R  \leq
0.95R_\odot$. \label{tab:rmscandrho}}
\end{table}

Table~\ref{tab:rmscandrho}   summarizes    our   results   by   quantitatively
characterizing  the  distributions  we   have  obtained  in  our  Monte  Carlo
simulations for the rms differences.  In the top 3 rows of Table~\ref{tab:rms}
we characterize  the 3 sound speed  difference rms distributions  for both the
GS98-Cons  and  AGS05-Opt  simulations   and,  for  completeness,  the  hybrid
AGS05-Cons case.   In the  case of the  GS98-Cons composition  we characterize
each of the rms distributions by  giving their most probable value (mode) that
we  denote by  $Q$, and  their one-sided  68.3\% ($1\sigma$)  confidence level
value,  $\sigma_{0.68}$ (for  each  quantity of  interest, $\sigma_{0.68}$  is
defined such that 68.3\% of all  the models in the Monte Carlo simulation have
this  quantity  in the  interval  $\left[0,  \sigma_{0.68}\right]$).  For  the
AGS05-Opt case, we find that the distributions are well described by lognormal
distributions  and in  Table~\ref{tab:rms}  we  give their  mode  $Q$ and  the
$1\sigma$ confidence level interval $\left[M-\sigma_-, M+\sigma_+\right]$ (see
Appendix for  details on  how $\sigma_-$ and  $\sigma_+$ are  defined).  Solar
models using the AGS05 composition show a worse agreement with helioseismology
results  than  models  using  the  GS98 composition.   This  is  evident  from
Figure~\ref{fig:soundspeed}. An indicative measure of the degradation is given
by  the  ratio  of the  most  probable  values  of  the $\delta  c_{\rm  all}$
distributions for each  composition choice, e.g.  $Q_{\delta c  \rm , all}{\rm
(AGS05)}/Q_{\delta c,\rm  all}{\rm (GS98)}$, and  we find that this  number is
$0.00487/0.00143=3.4$.

From  Figure~\ref{fig:soundspeed} and Table~\ref{tab:rms}  it is  evident that
the largest contribution to $\delta  c_{\rm all}$ originates in the outer half
of the solar  model.  Note in particular the smaller scale  of the abscissa of
the middle  panels as compared to the  top and bottom panels.  This shows that
the sound-speed difference between the Sun  and the models is quite low in the
region that includes  the core. Indeed, the low  sound-speed difference in the
core is  the basis of the  helioseimic solution of the  solar neutrino problem
(e.g., \citealp{bah97a}). The  larger values of $\delta c_{\rm  outer}$ can be
understood  by noting  that  the region  over  which the  quentity is  defined
includes,  in  addition  to  the  convective envelope,  the  radiative  layers
immediately below it.   The change in temperature gradient at  the base of the
convection zone causes a large change in sound-speed, and a mismatch of the CZ
base position  between the models and  the Sun translates  to relatively large
differences  in  the  inferred sound  speed  profiles.   In  the case  of  the
GS98-Cons composition, for instance, the  ratio of the most probable values of
the  $\delta  c_{\rm  outer}$  and  $\delta c_{\rm  inner}$  distributions  is
$Q_{\delta c, \rm outer}/Q_{\delta c , \rm inner}=1.5$.  The effect of a wrong
location  of $R_{\rm  CZ}$ is  more evident  in the  simulations  adopting the
AGS05-Opt composition, for which  the difference between the predicted $R_{\rm
CZ}$ and that measured by helioseismology  becomes very large. In this case we
get  $Q_{\delta c,\rm  outer}/Q_{\delta c,\rm  inner}=2.7$. This  reflects the
fact that $\delta c_{\rm outer}$  is more affected than $\delta c_{\rm inner}$
by the adoption the AGS05 solar  composition, and the underlying reason is the
effect of composition on opacities.

\subsection{Density profiles}
\label{sec:density}

We present the results for the differences in the density profiles in a manner 
similar  to those of the sound-speed differences.  The 
results are shown in Figure~\ref{fig:density}. 
As in  the case of sound-speed,  we define the rms  density difference between
the sun and the models --- $\rho_{\rm all}$, $\rho_{\rm
inner}$, and $\rho_{\rm outer}$ ---  with  definitions analogous to the
sound-speed difference case.   The last three rows of Table~\ref{tab:rms}  
shows the characteristics of the distributions of  $\rho_{\rm all}$,
$\rho_{\rm  inner}$, and  $\rho_{\rm  outer}$  that we find  for our
model obtained from the Monte  Carlo simulations.

Figure~\ref{fig:density} presents,  from top to bottom,  the distributions for
$\rho_{\rm  all}$, $\rho_{\rm  inner}$,  and $\rho_{\rm  outer}$  and for  the
GS98-Cons (left column)  and AGS05-Opt (right column) composition choices.  
Again, distributions from  the GS98-Cons simulation are 
 one-sided  distributions.  Although  this may not  be strictly true  in the
case of the $\rho_{\rm inner}$ distribution, it is nevertheless a highly
asymmetric  distribution  and  we  keep  this  description  for  the  sake  of
simplicity.   AGS05-Opt   results  are  again  well   described  by  lognormal
distributions.    Again,  the  hybrid  AGS05-Cons   results  are   included
in the table for the sake of completeness.

Qualitatively, results for the rms  density differences resemble those for the
sound-speed differences, although the density differences are generally larger
than  the  sound-speed  differences.  As  in  the  case  of sound  speed,  the
distributions for the GS98-Cons composition are strongly peaked close to zero,
showing  the  consistency  between  standard  solar  models  that  adopt  this
composition  and  the  helioseismological  inferences for  the  solar  density
profile.  On  the other  hand, the Monte  Carlo simulation with  the AGS05-Opt
composition give  distributions for $\rho_{\rm all}$,  $\rho_{\rm inner}$, and
$\rho_{\rm   outer}$    that   show    a   much   degraded    agreement   with
helioseismology. An indicative value of this degradation is given by the ratio
of the most probable  values $Q_{\delta \rho,\rm all}$(AGS05)/$Q_{\delta \rho,
\rm  all}$(GS98)$= 0.0486/0.0055\approx  8.8$.  This  seems to  indicate that,
despite the fact  that solar density profile is  somewhat less well determined
by inversions of the solar frequencies than the solar sound speed profile, the
density profile  in the solar  models is very  sensitive to the  input physics
adopted.   Besides, it  is  known that  envelope  models for  the  Sun can  be
constructed  with near  perfect sound  speed differences  even when  the AGS05
composition is adopted \citep{basu04,ant05}, these models, however, still have
a density  profile in disagreement with that  determined from helioseismology.
This points  in the  direction that, although  somewhat more limited  from the
observational   point  of   view,   density  profiles   can   be  a   powerful
helioseismological tool. This  appears to be particularly true  in the case of
the problem posed by the new determinations of the solar composition.

From the  results in this  and the previous  subsection, we conclude  that the
disagreement    between   the   standard    solar   model    predictions   and
helioseismological measurements of the  solar sound-speed and density profiles
introduced  by the  adoption of  the new  solar composition  \citep{ags05}, is
unlikely  to be  explained by  changing the  other input  physics  included in
standard  solar  models within  the  currently  accepted uncertainties.   This
result strengthens those described  in \S~\ref{sec:rczysurf} where we compared
the helium abundance and depth of  the convective zone of the models with that
of the Sun.

\section{MONTE CARLO RESULTS FOR INDIVIDUAL NEUTRINO FLUXES}
\label{sec:neutrinofluxes}

We present  in this section  the Monte Carlo  results for the  distribution of
each of the neutrino fluxes and their total uncertainties from all sources for
the predicted solar neutrino fluxes.  Here, results are presented, as in other
sections, for the three separate  cases which are distinguished by which heavy
element composition and by which set of composition uncertainties are adopted,
as       explained      in       \S~\ref{subsec:dilemmaheavyelements}      and
\S~\ref{subsec:compositionparameters}.

In  terms of  the  resulting shape  of  the distributions,  we  find that  the
neutrino fluxes can be separated  into 2 groups, regardless of the composition
choice. The  first group is  formed by the  p-p, pep, hep and  $^7$Be neutrino
fluxes.   For  these   fluxes,  we  find  by  $\chi^2$   analysis  that  their
distributions can be  described as normal distributions to  better than a 95\%
confidence level. The  only exception is the $^7$Be  flux distribution for the
GS98-Cons composition  choice, for  which the confidence  level is  70\%.  The
second group is formed by the  fluxes that have uncertainties dominated by the
composition uncertainties,  i.e.  the $^8$B, $^{13}$N,  $^{15}$O, and $^{17}$F
neutrino fluxes. The distributions of  these fluxes are very well described by
lognormal  distributions,  the $\chi^2$  analysis  yield
confidence levels better than 97\%  in all cases, that reflect our assumptions
regarding the solar composition uncertainties. 

\begin{figure*}
\begin{center}
\includegraphics[bb=0 40 566 730,angle=0.0,scale=0.65]{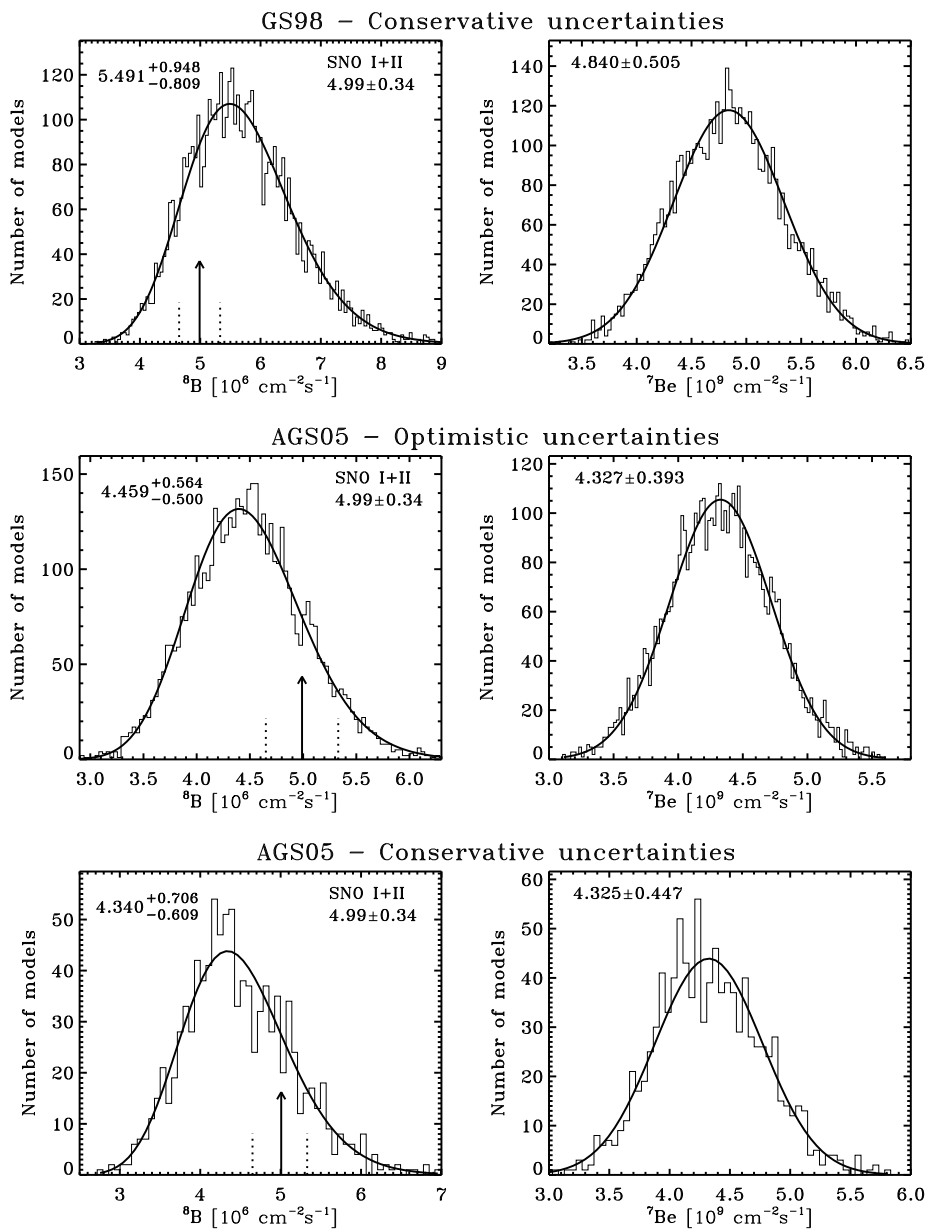}
\caption{The $^8$B and $^7$Be neutrino fluxes.  The figure shows the number of
solar models  from our Monte Carlo  simulations that have  $^8$B (left panels)
and $^7$Be (right panels) neutrino fluxes in the indicated ranges. From top to
bottom,  rows refer to  the GS98-Cons,  AGS05-Opt, and  AGS05-Cons composition
choices.  The conservative and optimistic abundance uncertainties are given in
Table~\ref{tab:abundanceuncertainties}.   The   smooth  curves  represent  the
lognormal  (normal)  distributions  inferred   for  the  $^8$B  ($^7$Be)  flux
distributions  from our  simulations.  For  the $^8$B  flux the  mode  $Q$ and
$\sigma_+$ and $\sigma_-$ as defined in  the text are given in each panel.  In
the case  of the $^7$Be flux, the  mean and the standard  deviation are given.
Fluxes units are the same as in Table~\ref{tab:standardmodelpredictions}. 
\label{fig:b8-be7}}
\end{center}
\end{figure*}

For  the four  most experimentally  important neutrino  fluxes, the  p-p, pep,
$^7$Be,  and  $^8$B neutrino  fluxes,  we  present  histograms for  all  three
composition  options. In Figure~\ref{fig:b8-be7}  and Figure~\ref{fig:pp-pep},
the top,  middle and bottom rows  correspond to the  GS98-Cons, AGS05-Opt, and
AGS05-Cons  composition  choices  respectively.   For the  more  difficult  to
measure hep, $^{13}$N, $^{15}$O, and $^{17}$F neutrinos, we present histograms
only for the GS98-Cons case.

Tables~\ref{tab:fluxgauss} and \ref{tab:fluxlognor} give the parameters needed
to    characterize   the    fluxes   distributions    of    our   simulations.
Table~\ref{tab:neutrinofluxuncertainties}    summarizes   the    Monte   Carlo
uncertainties  for all  eight neutrino  fluxes and  for all  three assumptions
regarding the composition.  

Our  Monte  Carlo  technique  only  provides  direct  results  for  the  total
uncertainties  of each  neutrino flux.   However, dominant  contributions from
individual sources to  the total uncertainty can generally  be identified with
the     aid    of     the     input    standard     deviations    given     in
Tables~\ref{tab:abundanceuncertainties},               \ref{tab:sigmasopacity},
\ref{tab:sigmaseos}.  In the following subsections, we comment on the dominant
individual sources of uncertainty for each flux where this seems relevant.

Section~\ref{subsec:b8}  describes  the histogram  of  results  for the  $^8$B
neutrinos   which   have   been    measured   directly   in   the   Kamiokande
\citep{kamiokande},    Super-Kamiokande   \citep{superk1,superk2}    and   SNO
experiments     \citep{snosalt04,snosalt05}.     We     then     discuss    in
Section~\ref{subsec:be7}  the $^7$Be neutrinos  which will  be studied  in the
Borexino  experiment  \citep{borexino}  and  perhaps  the  KamLAND  experiment
\citep{kamland}      or      the      LENA     experiment      \citep{ober05}.
Section~\ref{subsec:pppep}  presents results  for the  calculated  Monte Carlo
distribution of the fundamental p-p and pep neutrino fluxes. This section also
provides predictions  for the  anti-correlations of the  p-p and  pep neutrino
fluxes with the  $^7$Be neutrino flux (see figure~\ref{fig:corr-pepbe7ppbe7}),
as  well as  the predicted  correlation between  the p-p  and pep  fluxes (see
figure~\ref{fig:pp-pepcorrelation}).           We          present          in
Section~\ref{subsec:hepn13o15f17}  the  results  for the  difficult-to-measure
hep, $^{13}$N, $^{15}$O, and $^{17}$F neutrino fluxes.

\begin{table}[!t] 
  \caption{Neutrino fluxes with Gaussian distributions \label{tab:fluxgauss}}
  \begin{center}
    \begin{tabular}{ccccccccc}
      \tableline \tableline
      & \multicolumn{2}{c}{GS98-Cons} & & \multicolumn{2}{c}{AGS05-Opt} & &
      \multicolumn{2}{c}{AGS05-Cons}\\ 
 \tableline
      Flux & $\mu$ & $\sigma$ && $\mu$ & $\sigma$ && $\mu$ & $\sigma$ \\
      \tableline
      p-p & 5.987 & 0.056 && 6.055 & 0.042 && 6.054 & 0.050 \\
      pep & 1.419 & 0.022 && 1.451 & 0.016 && 1.451 & 0.018 \\
      hep & 7.970 & 1.236 && 8.251 & 1.276 && 8.281 & 1.264 \\
      $^7$Be & 4.840 & 0.505 && 4.327 & 0.393 && 4.325 & 0.447 \\
      \tableline
    \end{tabular}
  \end{center}
  \tablecomments{Parameters  of the Gaussian  distributions that  describe the
  neutrino fluxes distributions in our  Monte Carlo simulations. For each flux
  and each  composition choice,  the mean value  $\mu$ and  standard deviation
  $\sigma$   are   given.    Fluxes   are    in   the   same   units   as   in
  Table~\ref{tab:standardmodelpredictions}.}  
\end{table}

\begin{table*}[!t]
  \begin{center}
  \caption{Neutrino fluxes with lognormal distributions \label{tab:fluxlognor}}
  \begin{tabular}{ccccccccccccccc}
    \tableline \tableline
      & \multicolumn{4}{c}{GS98-Cons} & & \multicolumn{4}{c}{AGS05-Opt} & &
      \multicolumn{4}{c}{AGS05-Cons}\\ 
      \tableline
      Flux & $m$ & $s$ & $Q$ & $\mu$ && $m$ & $s$ & $Q$ & $\mu$ && $m$ & $s$ &
      $Q$ & $\mu$ \\
      \tableline
      $^8$B & 1.728 & 0.157 & 5.49 & 5.70 && 1.495 & 0.119 & 4.40 & 4.49
      && 1.490 & 0.148 & 4.34 & 4.49 \\
      $^{13}$N &  1.070 & 0.299  & 2.67 &  3.05 && 0.684  & 0.134 &  1.95 &
      2.00 && 0.644 & 0.292 & 1.75 & 1.99 \\ 
      $^{15}$O &  0.792 & 0.304  & 2.01 &  2.31 && 0.350  & 0.151 &  1.39 &
      1.44 && 0.312 & 0.295 & 1.25 & 1.43 \\
      $^{17}$F &  1.674 & 0.486  & 4.21 &  6.00 && 1.166  & 0.152 &  3.14 &
      3.25 && 1.105 & 0.466 & 2.43 & 3.37 \\
      \tableline
  \end{tabular}
  \end{center}
  \tablecomments{Characterization of the lognormal distributions that describe
the neutrino  fluxes distributions  in our Monte  Carlo simulations.  For each
flux  and each  composition  choice, the  scale  parameter $m$  and the  shape
parameter $s$  are given.  In addition,  the mode $Q=\exp{(m -  s^2)}$ and the
mean value $\mu= \exp{(m+s^2/2)}$ are also given.  Fluxes are in
the same units as in Table~\ref{tab:standardmodelpredictions}.}
\end{table*}

\subsection{The $^8$B  Neutrino Flux}
\label{subsec:b8}

The three panels on the left side of Figure~\ref{fig:b8-be7} show
the histograms of the number of computed models with $^8$B
neutrino fluxes in each flux bin. The assumed abundances and
abundance uncertainties are written above each of the three rows
of panels.

The weighted  average value of the  SNO neutral current  measurements from the
Neutral Current Phase I  and Phase II measurements \citep{snosalt04,snosalt05}
is $4.99 \pm 0.34 \times 10^6  {\rm cm^{-2}s^{-1}}$. This value is shown as an
arrow  perpendicular to  the  horizontal axis  of  each of  the $^8$B  panels,
together  with a  dotted  line that  shows  the $1\sigma$  uncertainty of  the
weighted average. The figure shows that adopting either of the recommended set
of heavy element  abundances, GS98 or AGS05, leads to  good agreement with the
total  $^8$B neutrino  flux measured  by the  SNO neutral  current experiments
\citep[see][]{bp04,bs05}. The measured value  of the $^8$B neutrino flux falls
slightly  below the  best-fit solar  model prediction  if GS98  abundances are
adopted  (upper left panel  of Fig.~\ref{fig:b8-be7})  and is  slightly higher
than  the best-fit value  if AGS05  abundances are  assumed.  The  $^8$B solar
neutrino  flux  is not  very  sensitive  to which  of  the  two heavy  element
compositions, GS98 or AGS05, is incorporated into the solar models.

The effect  of the  composition uncertainties is,  however, noticeable  in the
shape of the $^8$B flux distribution.  It is apparent, particularly in the top
and  bottom  left  panels  of  Figure~\ref{fig:b8-be7}, that  the  $^8$B  flux
distributions  are  not symmetric.   This  is  a  consequence of  the  assumed
distribution         for         the         composition         uncertainties
(\S~\ref{subsec:compositionuncertainties}). As  anticipated, we find  that the
$^8$B  flux distribution  of  our Monte  Carlo  simulations can  be very  well
described  by  lognormal distributions  (with  confidence  levels better  than
98.5\%) for  all composition choices.   The parameters characterizing  each of
these distributions are given in Table~\ref{tab:fluxlognor}.  A summary of the
properties of lognormal  distributions relevant to this paper  is given in the
Appendix.  

Table~\ref{tab:neutrinofluxuncertainties}   and  Figure~\ref{fig:b8-be7}  show
that  the  total $1\sigma$  theoretical  uncertainty  in  the predicted  $^8$B
neutrino flux  varies from  17\% to 11\%,  depending upon what  assumptions we
make  regarding  the  heavy   element  abundances.  The  SNO  neutral  current
measurements have an accuracy of $\pm 7$\% 
\citep{snosalt04,snosalt05}.  If one
includes  all  of the  solar  neutrino and  reactor  data  and the  luminosity
constraint,  then  the  $^8$B  neutrino   flux  is  determined  to  $\pm  5$\%
\citep{bp-g04}. The  theoretical uncertainty for the $^8$B solar
neutrino  flux is  more than  a  factor of  two larger  than the  experimental
measurement error.

The two largest  contributors to the total uncertainty  in the predicted $^8$B
neutrino  flux are  the  cross section  factor,  $S_{34 }$,  for the  reaction
$^3$He($^4$He,$\gamma$)$^7$Be  (which   contributes  about  7.5\%  uncertainty
$1\sigma$, \citealp{bp04}) and the heavy element
abundances (which contribute about $12$\% for GS98 abundances and conservative
uncertainties   and   about   5\%   for  AGS05   abundances   and   optimistic
uncertainties).

\subsection{The $^7$Be Neutrino Flux}
\label{subsec:be7}

The three  panels on the right  hand side of  Figure~\ref{fig:b8-be7} show the
histograms of  the computed $^7$Be  neutrino flux for  the three cases  we are
considering.   We find  that the  $^7$Be neutrino  flux distribution  for each
composition choice can  be described by a normal  distribution, the parameters
of   which    are   given   in   each    panel   in   the    figure   and   in
Table~\ref{tab:fluxgauss}.   The   $1\sigma$  deviation   is,   as  shown   in
Figure~\ref{fig:b8-be7}              and             summarized             in
Table~\ref{tab:neutrinofluxuncertainties}, practically the  same for all three
cases,  and  ranges  from  9.3\%  for the  AGS05  composition  and  optimistic
uncertainties to 10.5\% for the GS98 conservative case. 
The theoretical uncertainty  for the $^7$Be solar neutrino  flux is relatively
insensitive  to the assumptions  made regarding  heavy element  abundances and
their uncertainties.

The cross section factor $S_{34 }$ contributes the largest amount, $\sim 8$\%,
to  the  total  computed  uncertainty   of  the  $^7$Be  neutrino  flux.  This
uncertainty could  be reduced by improved laboratory  measurements (see, e.g.,
\citealp{nara04}). 

The  $^7$Be neutrino  flux will  be measured  by the  Borexino  solar neutrino
experiment \citep{borexino}  and hopefully also the
KamLAND experiment  \citep{kamland}.  In this connection,
it is useful  to analyze all of the available solar  and reactor neutrino data
treating  the solar  neutrino fluxes  as unknown  variables and  including the
effects of the luminosity constraint \citep{bah02,spi90}.
When this  is done, the  constraint on
the  $^7$Be  neutrino  flux  is   (see  Table~3  of \citealp{bp-g04}):

\begin{equation}
\phi(^7{\rm Be}) = 1.03^{+0.24}_{-1.03}~\phi(^7{\rm Be})_{\rm
BP04}, \,~~ {\rm exp.~data~+~luminosity~constraint} .
\label{eq:be7inferred}
\end{equation}
The ratio of the BP04 prediction for  the $^7$Be flux to that predicted by the
BSB(GS98) model in  this paper is $4.86/4.84 = 1.004$ (cf.  the $^7$Be flux in
Table~\ref{tab:standardmodelpredictions} of  this paper to the  $^7$Be flux in
Table~1   of   BP04).   The   coefficient   on  the   right   hand   side   of
equation~(\ref{eq:be7inferred})  should  be multiplied  by  1.004 when  the
basis for the rate calculation is  the BSB(GS98) solar model discussed in this
paper.

Unlike the  situation with  regard to  the $^8$B neutrino  flux for  which the
experimental  error is  less  than the  theoretical  uncertainty, the  current
experimental constraints on  the $^7$Be neutrino flux are  much less stringent
than the theoretical uncertainty in the predicted rate.  If all the
experimental evidence is  combined with the best solar  model prediction, then
the uncertainty in the predicted rate for the $\nu + e$ scattering experiments
is  $\pm 3\%$  \citep{bp-g04}.  This uncertainty  has  to be  combined with  a
realistic uncertainty of the solar model predictions of the $^7$Be flux, which
we show in this paper is of the order of $\pm 10\%$.


\begin{table}[!t]
\caption{Total Percent $1\sigma$ Deviations in Neutrino Fluxes due
to all Sources. \label{tab:neutrinofluxuncertainties}}
\begin{center}
\begin{tabular}{lcccccccc}
\noalign{\smallskip} \tableline\tableline \noalign{\smallskip}
 & \multicolumn{2}{c}{GS98-Cons} && \multicolumn{2}{c}{AGS05-Opt} &&
\multicolumn{2}{c}{AGS05-Cons} \\
\noalign{\smallskip} \tableline 
Flux & \multicolumn{2}{c}{$\sigma${\small [\%]}} &&
 \multicolumn{2}{c}{$\sigma${\small [\%]}} &&
\multicolumn{2}{c}{$\sigma${\small [\%]}} \\
\tableline
pp     &    \multicolumn{2}{c}{0.9}  &   &     \multicolumn{2}{c}{0.7}   &&
\multicolumn{2}{c}{0.8} \\ 
pep     &      \multicolumn{2}{c}{1.5}    & &     \multicolumn{2}{c}{1.1}  & &
\multicolumn{2}{c}{1.3} \\
hep     &     \multicolumn{2}{c}{15.5}     &&    \multicolumn{2}{c}{15.5}   &&
\multicolumn{2}{c}{15.3} \\ 
 $^7$Be   &   \multicolumn{2}{c}{10.5}  &   &   \multicolumn{2}{c}{9.3}  &   &
 \multicolumn{2}{c}{10.3} \\ 
\tableline
Flux & $\sigma_+${\small [\%]} & $\sigma_-${\small [\%]} &&
 $\sigma_+${\small [\%]} & $\sigma_-${\small [\%]}  &&
 $\sigma_+${\small [\%]} & 
$\sigma_-${\small [\%]} \\ 
\tableline
$^8$B & 17.3 & 14.7 && 12.7 & 11.3 && 16.1 & 14.1 \\
$^{13}$N & 36.6 & 26.8 && 14.5 & 12.7 && 35.5 & 26.2 \\ 
$^{15}$O & 37.4 & 27.2 && 16.5 & 14.2 && 36.1 & 26.5 \\ 
$^{17}$F & 72.4 & 42.0 && 16.6 & 14.2 && 67.6 & 40.4 \\ 
\noalign{\smallskip} \tableline
\end{tabular}
\end{center}
\tablecomments{For the neutrino flux p-p, pep, hep, and $^7$Be neutrino fluxes
the total $1\sigma$ uncertainty is given in \% of the mean values of each flux
distribution  listed  in  Table~\ref{tab:fluxgauss}.   The  same  results  are
obtained  if  the  best-estimate   neutrino  fluxes,  listed  in  column~2  of
Table~\ref{tab:standardmodelpredictions}, are used.  Columns 2-4 correspond to
the  different composition  choices described  in the  text.  For  fluxes with
lognormal  distributions,   $^8$B,  $^{13}$N,  $^{15}$O,   and  $^{17}$F,  the
uncertainties $\sigma_+$ and $\sigma_-$ that define the $1\sigma$ confidence
level are given separately. 
A detailed definition of $\sigma_+$ and $\sigma_-$
is given in the Appendix.
Relative uncertainties  are computed with  respect to the most  probable value
$Q$ of each distribution, given in Table~\ref{tab:fluxlognor}. 
Table~\ref{tab:abundanceuncertainties} gives the numerical values for
conservative and optimistic abundance uncertainties.} 
\end{table}

\subsection{The p-p and pep Neutrino Fluxes}
\label{subsec:pppep}

We begin  in \S~\ref{subsubsec:ppflux} by  discussing the distribution  of the
solar  model values  for  the fundamental  p-p  solar neutrino  flux and  then
describe in \S~\ref{subsubsec:pepflux} the distribution of the closely related
pep neutrino flux. The histograms of both the p-p and the pep fluxes are shown
in  Figure~\ref{fig:pp-pep}.   In  all  cases   we  find  the  fluxes  in  our
simulations to be normally distributed.

\subsubsection{The p-p Neutrino Flux}
\label{subsubsec:ppflux}

\begin{figure*}
\begin{center}
\includegraphics[bb= 10 40 566 650,angle=0.0,scale=0.65]{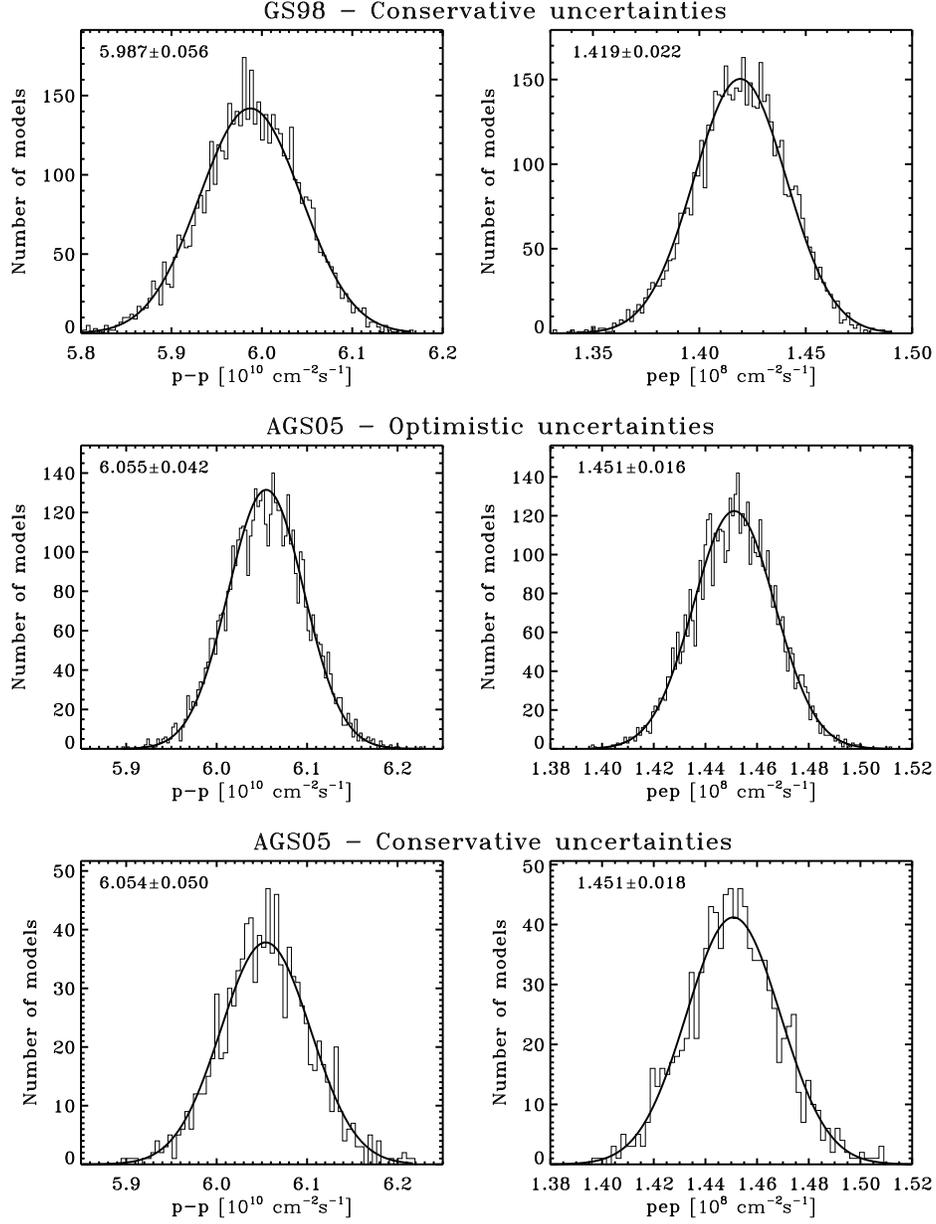}
\caption[]{The  p-p  and pep  neutrino  fluxes. This  figure  is  the same  as
Figure~\ref{fig:b8-be7} except that  the present figure refers to  p-p and pep
solar neutrinos  rather than  $^8$B and $^7$B  neutrinos.  In each  panel, the
smooth line shows the normal  distribution inferred from the data. Mean values
and  standard deviations are  also given  for each  flux and  each composition
case. \label{fig:pp-pep}}
\end{center}
\end{figure*}

The left-hand panels of Figure~\ref{fig:pp-pep} show the histograms of the p-p
neutrino fluxes  for the indicated  three assumptions regarding  heavy element
composition  and  their  uncertainties.  In  all  three  cases,  the  standard
deviation of  the theoretical prediction, $\sigma(\hbox{p-p})$,  is about 1\%.
Moreover,  the difference  between the  best-estimate flux  that  was computed
assuming the  GS98 composition  and the best-estimate  flux that  was computed
assuming      the     AGS05     composition      is     also      1\%     (see
Table~\ref{tab:standardmodelpredictions}).

The p-p  neutrino flux  is rather  well determined by  the existing  solar and
reactor experiments plus the luminosity constraint \citep{bah02}. The
available data constrain the p-p flux to (see Table~3 of \citealp{bp-g04}):
\begin{equation}
\phi(\hbox{p-p}) = 1.01^{+0.02}_{-0.02}~\phi(\hbox{p-p})_{\rm BP04},
\,~~{\rm exp.~data~+~luminosity~constraint}.
\label{eq:ppexperimental}
\end{equation}

The ratio  of the BP04 prediction  for the p-p  flux to that of  the BSB(GS98)
model  flux  in this  paper  is  $5.94/5.99 =  0.992$  (cf.  the  p-p flux  in
Table~\ref{tab:standardmodelpredictions} of this paper  to the flux in Table~1
of   BP04).   Hence,   the   coefficient   on   the   right   hand   side   of
equation~(\ref{eq:ppexperimental})  should  be multiplied  by  0.992 when  the
basis for the rate calculation is the currently preferred solar model with GS98
abundances. It should be stressed that the constraint on the p-p neutrino flux
that is given in  equation~(\ref{eq:ppexperimental}) is somewhat indirect.  Of
the   solar  neutrino  experiments   performed  so   far,  only   the  gallium
radiochemical  experiments (see,  e.g., \citealp{gallex99,sage02,gno05})
provide measurement constraints  on the p-p flux and  the gallium measurements
do not  give a unique flux since  neutrino energies are not  measured. For the
gallium   experiments,   all   neutrinos   above   a   fixed   threshold   are
counted. Moreover, the luminosity  constraint is critical for obtaining bounds
on the p-p  flux; without the luminosity constraint, the  allowed range of the
p-p  flux  is  very  large,  much  larger  than  the  theoretical  uncertainty
\citep{bp-g04}. At present, the accuracy of the experimental
determination  of   the  p-p  flux  when  supplemented   with  the  luminosity
constraint, is comparable to the  theoretical uncertainty in the prediction of
this flux.

The American  Physical Society  multi-divisional neutrino study  recommended a
precision measurement of the p-p  neutrino flux to, among other things, ``test
our  understanding  of how  neutrinos  change  flavor,  probe the  fundamental
question   of  whether   the  Sun   shines  only   through   nuclear  fusion''
\citep[see][]{free04}. Since  the  p-p reaction  initiates,
according to  the standard solar model,  more than 99\% of  the nuclear energy
generation  in  the  Sun  (see  Table~\ref{tab:standardmodelpredictions}),  an
accurate direct measurement of the p-p flux is of great importance for testing
the widely-used theory of stellar evolution.

A number  of promising approaches  to measuring the  p-p neutrino flux  are in
various  stages  of development  \citep{rag76,  rag01,  goro99, eji00,  suz00,
  mck00, mck05, nak01, mcd04, suz05, ober05, dol05, lan05}

\subsubsection{The pep Neutrino Flux}
\label{subsubsec:pepflux}

The three  right panels of  Figure~\ref{fig:pp-pep} present the  histograms of
the  calculated flux of  pep solar  neutrinos. The  standard deviation  of the
flux, $\sigma_{\rm pep}$, varies between 1.1\% and 1.5\%, depending upon which
heavy element abundances and uncertainties  are adopted.  There is no existing
significant experimental constraint on the  pep flux, which is about 400 times
smaller than the pp flux.

However, the monoenergetic  pep neutrinos have an energy  of 1.4 MeV, compared
to  the maximum energy  of 0.43  MeV of  the p-p  neutrinos. Therefore,  it is
possible that the  pep neutrinos could be measured in an  $\nu + e$ scattering
experiment like Borexino \citep{borexino} or KamLAND \citep{kamland}.

\subsubsection{The pep vs. $^7$Be and p-p vs. $^7$Be Correlations}
\label{subsubsec:pepppvsbe7}

We know from general considerations of the reactions in the p-p chain that the
p-p and  $^7$Be solar  neutrino fluxes are  inversely correlated.  If  the p-p
chain is terminated by the $^3$He-$^3$He reaction, then two p-p neutrinos, and
no  $^7$Be neutrinos,  are produced.  If the  p-p chain  is terminated  by the
$^3$He-$^4$He reaction, then one p-p  neutrino and one $^7$Be neutrino (nearly
always) is  produced (only rarely is  the $^7$Be neutrino replaced  by a $^8$B
neutrino).  Moreover, the pep flux is very nearly proportional to the p-p flux
\citep{bah69} and  can be used as  a surrogate for  the p-p flux in  the above
discussion.  

It is  possible that both the pep  neutrino flux and the  $^7$Be neutrino flux
will be measured in the next few years in the existing Borexino solar neutrino
experiment \citep{borexino}. For a discussion of this
possibility, the reader is referred to the paper by \citet{gal04}.
In  any event,  we  can  look  forward to  the
measurement of the p-p neutrino flux  in one of the solar neutrino experiments
currently being developed for this  purpose (see, for example, the discussions
by \citealp{rag76,  rag01, goro99, eji00,  suz00, mck00, mck05,  nak01, mcd04,
  suz05, ober05, lan05}). 

If either the pep or the p-p neutrino flux is measured and the $^7$Be neutrino
flux  is  also  determined  experimentally,  then  one  can  test  directly  a
fundamental  prediction of  stellar evolution  theory.  The  prediction  to be
tested is the anti-correlation between the  p-p (or pep) neutrino flux and the
$^7$Be neutrino flux. In what follows, we suppose for specificity that the pep
neutrino flux is  measured before the p-p flux and  therefore we first explore
the  anti-correlation  between  the   pep  and  $^7$Be  neutrino  fluxes.  The
calculational steps involved in determining the anti-correlation are identical
for pep  vs. $^7$Be  and p-p vs.   $^7$Be.  We  will present results  for both
cases.

How can  we determine quantitatively  the pep vs. $^7$Be  anti-correlation? To
answer  this  question it  is  convenient  to  define dimensionless  variables
$\delta({\rm pep})$ and $\delta({\rm ^7Be}) $ by the relations

\begin{equation}
\delta({\rm pep}) = \frac{\phi({\rm pep}) - \mu({\rm pep})}
{\mu({\rm pep})}\, ;~~\delta({\rm ^7Be}) =
\frac{\phi({\rm ^7Be}) - \mu({\rm ^7Be})}{\mu({\rm ^7Be})}\, ,
\label{eq:deltadefn}
\end{equation}
where  $\phi({\rm  pep})$ is  the  pep  flux from  a  single  solar model  and
$\mu({\rm  pep})$  is   the  mean  pep  flux  in   our  simulations  given  in
Table~\ref{tab:fluxgauss}   (with  analogous   definitions   for  $\delta({\rm
  ^7Be})$, $\phi({\rm ^7Be})$ and $\mu({\rm ^7Be})$). 

\begin{figure*}
\begin{center}
\includegraphics[bb=5 50 560 660,angle=0.0,scale=0.65]{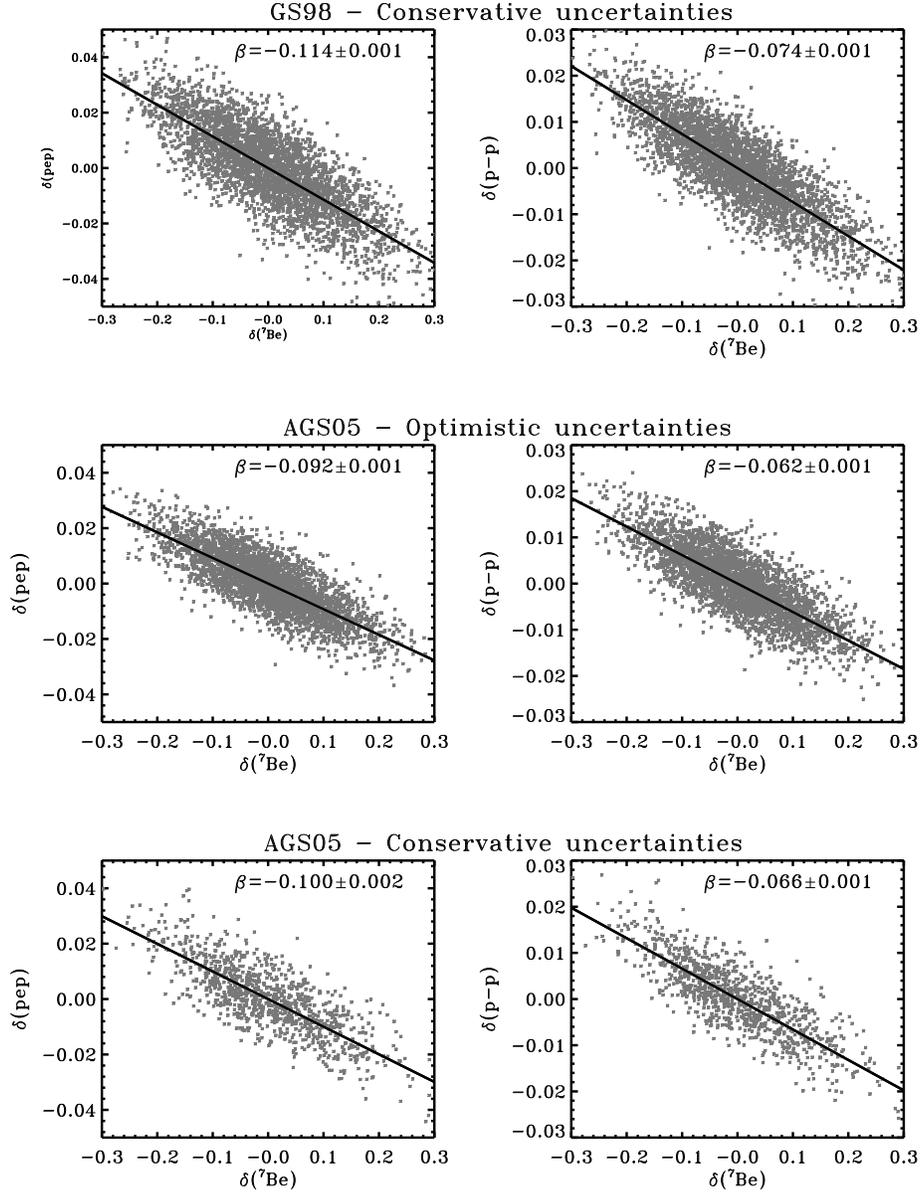}
\caption[]{The pep vs. $^7$Be and  the p-p vs.  $^7$Be anti-correlations.  The
two  fluxes  are anti-correlated  since  only one  p-p  (or  pep) neutrino  is
produced  if  hydrogen  burning  proceeds  through  the  $^3$He($^4$He,$\gamma
$)$^7$Be reaction,  whereas hydrogen burning  in which $^7$Be is  not involved
creates two p-p neutrinos (or occasionally pep neutrinos). 
\label{fig:corr-pepbe7ppbe7}}
\end{center}
\end{figure*}

Figure~\ref{fig:corr-pepbe7ppbe7}  shows in  the three  left panels  the solar
model  prediction for  the  anti-correlation between  $\delta({\rm pep})$  and
$\delta  ({\rm ^7Be})$.   For  the top  left  panel, the  GS98-Cons case,  the
best-fit straight line of the form 
\begin{equation}
\delta({\rm pep})~=~\beta({\rm pep, ^7Be})\delta({\rm ^7Be}) \,
\label{eq:deltapepvsdelta7Be}
\end{equation}
computed by a least squares fitting, has a slope
\begin{equation}
\beta({\rm pep, ^7Be})~=~-0.114 \pm 0.001\, . 
\label{eq:betapepbe7}
\end{equation}
The residuals 
\begin{equation}
R({\rm pep,^7Be})= \delta({\rm pep})-\beta({\rm pep, ^7Be})\delta({\rm ^7Be})
\end{equation}
are normally distributed with a standard deviation 
\begin{equation}
\sigma({\rm pep};\,^7{\rm Be})~=~0.0096\, . 
\label{eq:sigmapepfrombe7}
\end{equation}
If one measures precisely the $^7$Be neutrino flux, then our current knowledge
of solar model determines the expected pep neutrino flux to an accuracy of
0.96\% at  the $1\sigma$  level.  The  results are similar  for the  other two
cases shown  in Figure~\ref{fig:corr-pepbe7ppbe7}.  For the  AGS05-Opt case we
find   $\beta({\rm  pep,  ^7Be})   =  -0.092   \pm  0.001$   and  $\sigma({\rm
pep};\,^7{\rm  Be}) =  0.0066$ while  for AGS05-Cons  we get  $\beta({\rm pep,
^7Be})~=~-0.100 \pm 0.0024$ and $\sigma({\rm pep};\,^7{\rm Be}) = 0.0079$. 

Our two main Monte Carlo simulations yield values for $\beta({\rm pep, ^7Be})$
that differ by about 20\%.  This does not affect the predictive capability of
equation~\ref{eq:deltapepvsdelta7Be}. It can be easily shown that the relative
error in $\phi({\rm  pep})$ introduced by the uncertainty  in $\beta({\rm pep,
  ^7Be})$ is 
\begin{equation}
\frac{\Delta \phi({\rm pep})}{\phi({\rm  pep})} = \frac{\Delta \beta({\rm pep,
    ^7Be})}{\beta({\rm pep, ^7Be})} \delta(^7{\rm Be}) \beta({\rm pep, ^7Be}).
\label{eq:errorbeta}
\end{equation}
If  we  assume as  a  typical value  of  $\delta(^7{\rm  be})$ the  fractional
1$\sigma$  theoretical  uncertainty  of  the  $^7$Be  flux  (about  10\%;  see
Table~\ref{tab:neutrinofluxuncertainties}),  a $\beta({\rm pep,  ^7Be})$ value
equal to its average between the  GS98-Cons and AGS05-Opt values, and adopt as
its  uncertainty   the  relative  difference  between  the   values  from  the
simulations, we get 
\begin{equation}
\frac{\Delta \phi({\rm  pep})}{\phi({\rm pep})} \approx 0.2  \times 0.1 \times
0.103 \approx 0.002.
\end{equation}
We see that a typical uncertainty is  of the order of 0.2\%, i.e. much smaller
than       the      standard       deviation       of      the       residuals
(equation~\ref{eq:sigmapepfrombe7}).

The   right  hand   panels  of   Figure~\ref{fig:corr-pepbe7ppbe7}   show  the
anti-correlation between the p-p and  the $^7$Be neutrino fluxes.  For the top
right panel,  GS98-Cons composition choice,  the best-fit straight line  has a
slope 
\begin{equation}
\beta({\rm pp, ^7Be}) ~=~-0.0736 \pm 0.0007\, .
\label{eq:betappbe7}
\end{equation}
The corresponding $1\sigma$ uncertainty  in predicting $\delta({\rm pp})$ from
a known value of the $^7$Be neutrino flux is
\begin{equation}
\sigma(\hbox{p-p};\,^7{\rm Be})~=~0.0054.
\label{eq:sigmappfrombe7}
\end{equation}
Thus  one can  predict the  p-p flux  to  an accuracy  of about  0.54\% from  a
precisely  measured  value   of  the  $^7$Be  flux.  The   p-p  versus  $^7$Be
anti-correlation   is   somewhat   tighter   than  the   pep   versus   $^7$Be
anti-correlation.

Similar values are obtained for the other two cases.  We find, respectively, 
$\beta({\rm   pp,    ^7Be})   ~=~-0.0617    \pm    0.0006$   and
$\sigma(\hbox{p-p};\,^7{\rm Be})~=~0.0044$ and   $\beta({\rm  pp,   ^7Be})
~=~-0.0659 \pm 0.0014$ and $\sigma(\hbox{p-p};\,^7{\rm Be})~=~0.0046$ for the 
AGS05-Opt and AGS05-Cons cases respectively.

If one is interested in  the inverse correlations, e.g.  $\delta({\rm ^7 Be})=
\beta({\rm ^7Be,pep})\delta({\rm pep})$, they can be easily 
obtained by recalling that given two quantities $x$ and $y$, then if $x=
\beta(x,y)\, y $ and $y= \beta(y,x)\, x $, where $\beta(x,y)$ and $\beta(y,x)$
are computed from least squares fitting, then they satisfy the relation
\begin{equation}
\beta(x,y)\beta(y,x) = \rho^2(x,y)
\end{equation}
where $\rho(x,y)$ is the correlation coefficient between $x$ and $y$.  The
correlation coefficients of the neutrino fluxes are discussed in
\S~\ref{sec:correlationcoefficients}        and       summarized       in
Tables~\ref{tab:GS98correlationcoefficients}                                and
\ref{tab:AGS05correlationcoefficients}.

If our  general picture  of how  nuclear fusion reactions  occur in  the solar
interior is  correct, then  measurements of  the pep (or  p-p) and  the $^7$Be
neutrino  fluxes  must  lie  on  one  of  the  very  similar  lines  shown  in
Figure~\ref{fig:corr-pepbe7ppbe7}.  The values  given in  this  subsection for
$\beta({\rm pep,  ^7Be})$ represent a  fundamental and testable  prediction of
the  theory  of nuclear  energy  generation  in  stars. They  encapsulate  the
competition in the solar interior between the two primary branches, p-p(I) and
p-p(II), of the p-p chain.

\subsubsection{The pep vs. p-p Correlation}
\label{subsubsec:pepvspp}

What is the relation between the  production of the p-p and the pep neutrinos?
They share the same nuclear matrix 
element and differ in a  multiplicative factor that depends approximately upon
the  electron  number  density divided  by  the  square  root of  the  ambient
temperature \citep{bah69}. This factor does not
change very much from one solar model to the next. As a result, the pep rate
is very nearly proportional to the p-p rate. 

Suppose the pep flux is measured before the p-p flux is determined by a direct
experiment. How accurately can one infer the p-p neutrino flux if one measures
the  pep  flux?   To  answer  this  question,  we   plot  $\delta({\rm  pep})$
vs.  $\delta(\hbox{p-p})$,  where the  meaning  of  the  operator $\delta$  is
defined in equation~(\ref{eq:deltadefn}).

\begin{figure}[!t]
\begin{center}
\includegraphics[bb=100 80 530 500,angle=0.0,scale=0.6]{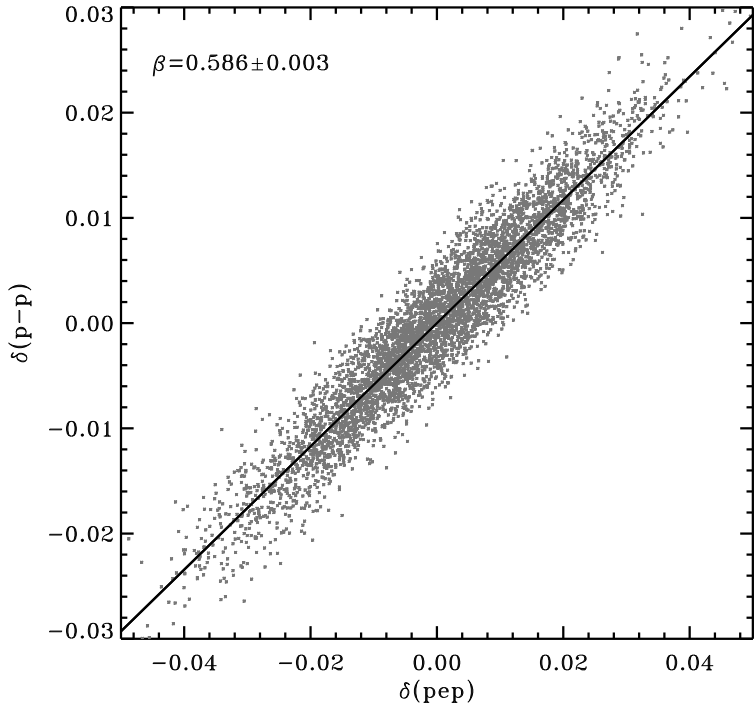}
\caption[]{The correlation between the p-p and pep
fluxes. The two fluxes are approximately proportional to each
other since they share the same nuclear matrix element (Bahcall \&
May 1969). \label{fig:pp-pepcorrelation}}
\end{center}
\end{figure}

Figure~\ref{fig:pp-pepcorrelation} shows  the rather-tight correlation between
$\delta({\rm pep})$ and $\delta (\hbox{p-p})$.  For the GS98-Cons case, the best-fit straight line of the form 
\begin{equation}
\delta(\hbox{p-p})~=~\beta({\rm\hbox{p-p},pep})\delta({\rm pep}) \,
\end{equation}
has a slope
\begin{equation}
\beta({\rm\hbox{p-p,pep}}) ~=~0.586 \pm 0.003\, ,
\end{equation}
and
\begin{equation}
\sigma(\hbox{p-p, pep})~=~0.0028\, .
\label{eq:sigmapppep}
\end{equation}
Thus  the p-p  neutrino flux  can be  inferred from  a precisely  measured pep
neutrino flux to  an accuracy of about 0.3\%. The results  are similar for the
other two cases we are considering.  For the AGS05-Opt case, we find 
$\beta({\hbox{p-p,pep}})   ~=~0.642    \pm   0.002$   and   $\sigma(\hbox{p-p,
pep})~=~0.0019$.  Similarly,  for AGS05-Cons we  find $\beta({\hbox{p-p,pep}})
~=~0.598 \pm 0.007$ while $\sigma(\hbox{p-p, pep})~=~0.0027$. 

Finally, we apply for the p-p  vs.  pep correlation the same reasoning leading
to   equation~\ref{eq:errorbeta}.    Adopting   a   value   for   $\delta({\rm
  pep})=0.015$,  the  relative uncertainty  in  $\phi({\rm  pp})$  due to  the
different values  of $\beta({\hbox{p-p,pep}})$ that  result from our  two main
simulations is only 0.1\%.  

In summary, the p-p flux can be inferred from a precisely measured pep flux to
an  accuracy  of $0.25\%  \pm  0.05\%$,  depending  upon which  heavy  element
abundances   are  adopted   and  which   abundance  uncertainties   are  used.
From the  inverse correlation, the pep  flux can be inferred  from a precisely
measured p-p flux to an accuracy of $0.39\% \pm 0.09\%$. 

\subsection{The hep, $^{13}$N, $^{15}$O, and $^{17}$F Neutrino Fluxes}
\label{subsec:hepn13o15f17}

\begin{figure*}[!t]
\begin{center}
\includegraphics[bb=0 275 566 720,angle=0.0,scale=0.65]{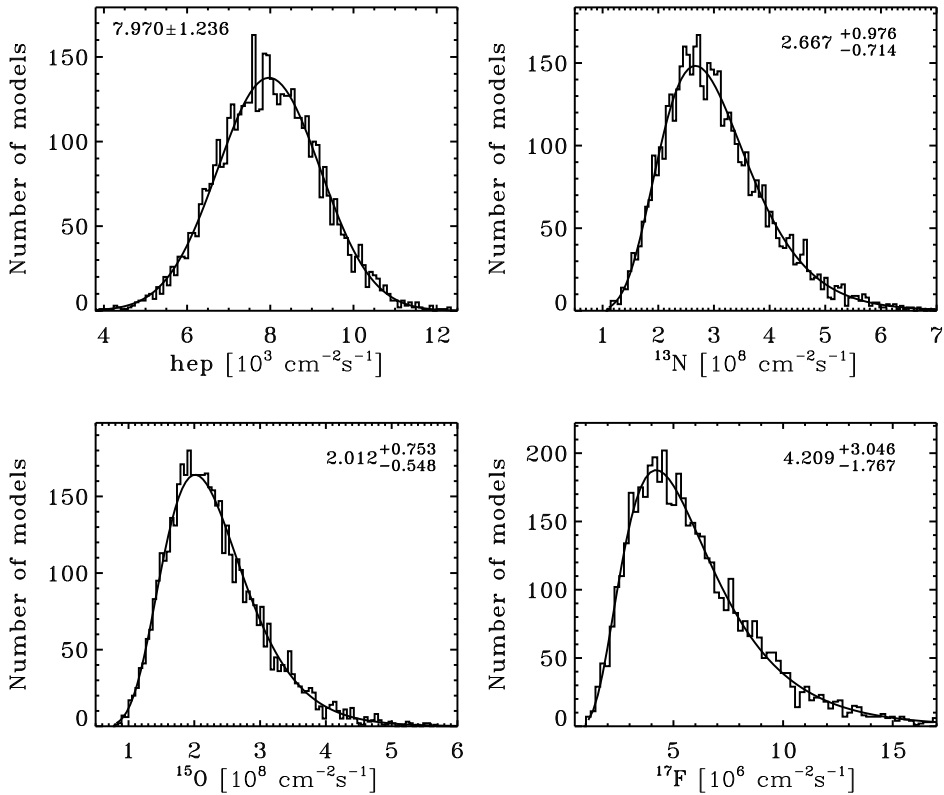}
\caption[]{The hep, $^{13}$N, $^{15}$O and $^{17}$F 
neutrino fluxes from out Monte Carlo simulation with the GS98-Cons composition
choice.  The hep distribution follows a normal distribution with the
parameters shown in the top-left panel. 
The distributions of the CNO fluxes are markedly
asymmetric as they reflect the lognormal distribution for the
composition        uncertainties       that       is        discussed       in
\S~\ref{subsec:compositionuncertainties}. Results are analogous for the 
AGS05-Opt and AGS05-Cons composition choices. 
\label{fig:hepn13o15f17}}
\end{center}
\end{figure*}

Figure~\ref{fig:hepn13o15f17} shows, for GS98 abundances and
conservative uncertainties, the histograms of the number of solar
models with different values of the hep, $^{13}$N, $^{15}$O, and
$^{17}$F solar  neutrino fluxes.  While  the hep flux is  normally distributed
(99.9\% C.L.),
the 3 CNO fluxes follow  lognormal distributions (better than 95\% C.L.).  The
results for all three 
composition cases that we are considering are summarized in
Tables~\ref{tab:fluxgauss},              \ref{tab:fluxlognor}              and
\ref{tab:neutrinofluxuncertainties}.  

The total uncertainty for the hep neutrino flux is dominated by
the 15.1\% uncertainty \citep{park03} from the calculation of
the nuclear matrix element. All other sources of uncertainty
contribute less than or of the order of 2\%. Therefore, the
calculated standard deviations for the hep neutrinos are
essentially independent of the adopted heavy element abundances
and their uncertainties.

For the $^{13}$N,  $^{15}$O, and $^{17}$F solar neutrino  fluxes, the standard
deviations are dominated by composition uncertainties if we adopt conservative
uncertainties. The marked  asymmetry of the distributions of  the CNO neutrino
fluxes is apparent  in Figure~\ref{fig:hepn13o15f17}.  This asymmetry reflects
the lognormal distribution for  composition uncertainties that we have adopted
and   discussed  in   \S~\ref{subsec:compositionuncertainties}.   If   we  use
optimistic composition uncertainties, the  cross section factor $S_{1,14}$ for
the   $^{14}$N(p,$\gamma$)$^{15}$O  reaction   (8.4\%  uncertainty)   and  the
composition uncertainties  make comparable  contributions to the  $^{13}$N and
$^{15}$O neutrino fluxes uncertainties.

There are no funded experiments for  which the detection of the hep, $^{13}$N,
$^{15}$O, and  $^{17}$F solar neutrino fluxes  seems likely if  the fluxes and
their uncertainties calculated in this paper are correct.

\section{Neutrino Flux Correlation Coefficients}
\label{sec:correlationcoefficients}

We have seen  in \S~\ref{sec:neutrinofluxes} that some of  the neutrino fluxes
are  highly correlated.   These correlations  are exhibited  in,  for example,
Figure~\ref{fig:corr-pepbe7ppbe7},  which illustrates the  anti-correlation of
the     pep    and    pp     fluxes    with     the    $^7$Be     flux,    and
Figure~\ref{fig:pp-pepcorrelation}, which shows the correlation of the pep and
the pp fluxes.   These correlations play an important  role in global analysis
of solar neutrino experiments (see, e.g., \citealp{fog95, bah01, fog02}).

The correlations of the fluxes arise  from two sources: 1) the solution of the
equations of  stellar evolution; and 2)  the effects of  changes in individual
input parameters.

In the  past, the correlations of  solar neutrino fluxes have  been taken into
account by  using the logarithmic  partial derivatives of  individual neutrino
fluxes with respect to 9  input parameters \citep{bah88, bah89}.  The standard
treatment of the flux-correlations, when represented by partial derivatives of
fluxes with respect  to input parameters, is contained  in the important paper
by \citet{fog95}.

In this section,  we use the results of our Monte  Carlo simulations to derive
the correlations  due to all  21 input parameters  and to the solution  of the
equations  of  stellar  evolution.  By directly  calculating  the  correlation
coefficients among the  Monte Carlo neutrino fluxes, we are  able to present a
simple and  complete summary  of the correlations.  These results  will enable
simpler and more accurate theoretical analysis of solar neutrino oscillations.

We  summarize  the correlations  in  terms  of  the correlation  coefficients,
$\rho(i,j)$, defined in the usual way by
\begin{equation}
\rho(i,j)~=~\frac{N^{-1}\sum_{n=1}^N \Delta \phi_i^n \Delta
\phi_j^n}{\sigma_i\times \sigma_{j}} \, ,
\label{eq:defncorrelation}
\end{equation}
where $\Delta \phi_i^n = \phi_i^n - \phi_{i,{\rm average}}$, $\sigma_i$ is the
standard deviation  of the ith flux type  ($i =$ pp, pep,  hep, $^7$Be, $^8$B,
$^{13}$N, $^{15}$O,  and $^{17}$F),  and $N  = 5,000$ is  the total  number of
cases  considered in the  separate Monte  Carlo simulations  that incorporated
GS98 or AGS05  heavy element abundances. The correlation  matrix is symmetric,
$\rho(i,j)~=~\rho(j,i)$, and has by definition unit diagonal elements.

\begin{table}[!t]
\caption{Correlation Coefficients for 5000 Sets of Neutrino
Fluxes: \citet{gs98} Heavy Element Abundances and
Conservative Uncertainties.
\label{tab:GS98correlationcoefficients} }
\begin{center}
\begin{tabular}{lcccccccc}
\noalign{\smallskip} \tableline\tableline \noalign{\smallskip}
Flux&pp&pep&hep&$^7$Be&$^8$B&$^{13}$N&$^{15 }$O&$^{17}$F\\
\noalign{\smallskip} \tableline 
pp & 1.000 & 0.954 & 0.082 &-0.819 &-0.720 &-0.349 &-0.381 &-0.319 \\
pep & 0.954 & 1.000 & 0.087 &-0.780 &-0.730 &-0.407 &-0.439 &-0.369 \\
hep & 0.082 & 0.087 & 1.000 &-0.062 &-0.086 &-0.052 &-0.058 &-0.076 \\
$^7$Be &-0.819 &-0.780 &-0.062 & 1.000 & 0.887 & 0.154 & 0.204 & 0.332 \\
$^8$B & -0.720 &-0.730 &-0.086 & 0.887 & 1.000 & 0.269 & 0.333 & 0.486 \\
$^{13}$N &-0.349 &-0.407 &-0.052 & 0.154 & 0.269 & 1.000 & 0.991 & 0.172 \\
$^{15}$O &-0.381 &-0.439 &-0.058 & 0.204 & 0.333 & 0.991 & 1.000 & 0.219 \\
$^{17}$F &-0.319 &-0.369 &-0.076 & 0.332 & 0.486 & 0.172 & 0.219 & 1.000 \\
\noalign{\smallskip} \tableline
\end{tabular}
\end{center}
\tablecomments{The  correlation  coefficients  in  the table  are  defined  by
equation~(\ref{eq:defncorrelation}).   The  fluxes   used   to  evaluate   the
coefficients were calculated using solar models that incorporated \citet{gs98}
surface    heavy   element   abundances   and   conservative
uncertainties.} 
\end{table}

Table~\ref{tab:GS98correlationcoefficients}                                and
Table~\ref{tab:AGS05correlationcoefficients}     present    the    correlation
coefficients  calculated   with  the   Monte  Carlo  simulations   that  used,
respectively,  the  \citet{gs98} heavy  element  abundances with  conservative
uncertainties and  the \citet{ags05} heavy element  abundances with optimistic
uncertainties.

We      see      from     Table~\ref{tab:GS98correlationcoefficients}      and
Table~\ref{tab:AGS05correlationcoefficients}  that  the  pp and  pep  neutrino
fluxes are strongly anti-correlated with  the $^7$Be and $^8$B neutrino fluxes
and  mildly anti-correlated  with the  CNO neutrino  fluxes.  The  pp  and pep
fluxes  are,   as  discussed  in   \S~\ref{subsubsec:pepvspp},  very  strongly
correlated.   The $^7$Be and  $^8$B neutrino  fluxes are  strongly correlated,
since  both  are  initiated   by  the  same  fusion  reaction,  $^3$He($^4$He,
$\gamma$)$^7$Be. Since the $^7$Be and $^8$B fluxes both occur predominantly in
higher  temperature regions  of the  Sun  (see Fig.~\ref{fig:fluxesvsradius}),
where the Gamow  penetration factor is more easily  overcome, these fluxes are
also  mildly correlated  with the  CNO neutrino  fluxes that  are  also mostly
produced  at  higher  temperatures.   Of  course, the  $^{13}$N  and  $^{15}$O
neutrino fluxes  are strongly correlated with  each other since  they are both
involved in  the CN  cycle that  operates close to  steady-state in  the inner
($R<0.12 R_\odot$) regions of the Sun.

Comparing       the        correlation       coefficients       given       in
Table~\ref{tab:GS98correlationcoefficients}                                and
Table~\ref{tab:AGS05correlationcoefficients},  we see  that  the same  general
trends  are obtained  independent of  which assumption  we make  regarding the
heavy  element  abundances  and   their  uncertainties.   However,  there  are
important quantitative differences.  In particular, the correlations involving
the CNO  neutrino fluxes  with fluxes from  the pp  chain are weaker  when the
\citet{ags05} abundances and optimistic uncertainties are adopted.

\begin{table}[!t]
\caption{Correlation Coefficients for Neutrino Fluxes: \citet{ags05}
Heavy Element Abundances and Optimistic Uncertainties.
\label{tab:AGS05correlationcoefficients} }
\begin{center}
\begin{tabular}{lcccccccc}
\noalign{\smallskip} \tableline\tableline \noalign{\smallskip}
Flux&pp&pep&hep&$^7$Be&$^8$B&$^{13}$N&$^{15 }$O&$^{17}$F\\
\noalign{\smallskip} \tableline
 pp &  1.000  &0.967 &-0.012 &-0.796 &-0.642 &-0.127 &-0.132 &-0.111 \\
pep &  0.967  &1.000  &0.001 &-0.793 &-0.667 &-0.162 &-0.171 &-0.137 \\
hep & -0.012  &0.001  &1.000 & 0.022 & 0.021 &-0.005 &-0.008 &-0.014 \\
$^7$Be &-0.796 &-0.793  &0.022 & 1.000  &0.878 & 0.125 & 0.155 & 0.237 \\
$^8$B & -0.642 &-0.667  &0.021 & 0.878  &1.000 & 0.257 & 0.296 & 0.412 \\
$^{13}$N &-0.127 &-0.162 &-0.005 &0.125 &0.257 &1.000 &0.984 &0.299 \\
$^{15}$O &-0.132 &-0.171 &-0.008 &0.155 &0.296 &0.984 &1.000 &0.338 \\
$^{17}$F &-0.111 &-0.137 &-0.014 &0.237 &0.412 &0.299 &0.338 &1.000 \\
\noalign{\smallskip} \tableline
\end{tabular}
\end{center}
\tablecomments{The correlation coefficients in the table are
defined by equation~(\ref{eq:defncorrelation}). The fluxes used to
create evaluate the coefficients were calculated using \citet{ags05}
surface heavy element abundances and optimistic
uncertainties.}
\end{table}

\section{Nuclear Fusion Fractions}
\label{sec:fusionfractions}

For pedagogical purposes  and in order to have a  conceptual overview of solar
energy generation, it is useful to calculate the fraction of the total nuclear
energy generation that occurs via each  of the most important fusion paths. We
present  in this  section  the best-estimates  (given  in the  last column  of
Table~\ref{tab:standardmodelpredictions} for  the best current  standard solar
models) and the $1\sigma$ uncertainties in the best-estimates of the fractions
that correspond to different ways of burning hydrogen.

Table~\ref{tab:fusionfractions}  gives  the  fractions  of the  total  nuclear
fusion  energy  generation in  standard  solar  models  that are  produced  by
different  fusion reaction paths.   We present  results for  all three  of the
composition options.

More than 99\%  of the total nuclear energy generation is  produced by the p-p
reactions 
in our solar models, while less than 1\% of the energy is generated by the CNO
reactions. These  fractions are robust to  all the input  uncertainties of our
standard solar models.   The total standard deviations from  variations in all
the 21 input  parameters in the Monte Carlo simulations  are between 0.07\% or
0.3\%.

About  88\% or 90\%  of the  energy is  derived, on  average for  our standard
models,  from reactions  that  begin  with the  fundamental  p-p reaction  and
terminate with the \hbox{$^3$He($^3$He, 2 p)$^4$He} reaction (p-p(I)). Nearly
all of the remainder of the nuclear  energy, 10\% or 11\%, is generated in our
models by  p-p reactions  that go through  the reaction  \hbox{$^3$He($^4$He,
$\gamma$)$^7$Be}  that  creates $^7$Be  solar  neutrinos  by electron  capture
(p-p(II)).  These fractions  also have  relatively small  variations (standard
deviations) due  to the  choice of different  input parameters.   The standard
deviations are typically 1\% for the p-p(I) and p-p(II) fractions.

Extraordinary as it seems, most of solar neutrino astronomy so far has focused
on  the  p-p(III)  reactions  that   involve  the  production  of  rare  $^8$B
neutrinos.  The  crucial reaction  sequence  terminating  these p-p  reactions
consists  of   the  reaction  $^3$He($^4$He,$\gamma$)$^7$Be  followed  by
$^7$Be(p,$\gamma$)$^8$B. Less than  1\% of the energy generation  in our solar
models corresponds to this rare reaction pathway and the standard deviation of
this fraction is only about 0.1\%.  The Kamiokande \citep{kamiokande}, 
Super-Kamiokande \citep{superk1,superk2},
and SNO solar neutrino experiments \citep{snosalt04,snosalt05} only
detect  neutrinos from  this rare  set  of reactions.  Moreover, the  original
chlorine  solar  neutrino experiment  by  R.  Davis,  Jr. and  his  colleagues
\citep{cle98} is primarily sensitive to $^8$B
neutrinos because  of a special superallowed transition  between $^{37}$Cl and
$^{37}$Ar \citep{bah64}.

\begin{table}[!t]
\caption{Fractions of Nuclear Energy Generation  That Are Produced
by Different Reaction Pathways \label{tab:fusionfractions}}
\begin{center}
\begin{tabular}{lccc}
\noalign{\smallskip} \tableline\tableline \noalign{\smallskip}
Fusion Branch & GS98-Cons &AGS05-Opt & AGS05-Cons\\
Fraction & & & \\
\noalign{\smallskip} \tableline
pp & $99.2 \pm 0.3$ & $99.5 \pm 0.1$ & $99.5 \pm 0.2$ \\
CNO & $0.8 \pm 0.2$ & $0.5 \pm 0.07$ & $0.5 \pm 0.1$ \\
p-p(I) & $88.3 \pm 1.3$ & $89.6 \pm 1.0$ & $89.6 \pm 1.1$ \\
p-p(II) & $10.8 \pm 1.1$ & $9.6 \pm 0.9$ & $9.6 \pm 1.0$ \\
p-p(III) & $0.9 \pm 0.1$ & $0.8 \pm 0.08$ & $0.8 \pm 0.08$ \\
\noalign{\smallskip} \tableline
\end{tabular}
\end{center}
\tablecomments{The  table presents  results  for percentages  of solar  energy
generation via  different nuclear  paths: all p-p  reactions (row~1);  all CNO
reactions (row~2); p-p(I) (terminated  by $^3$He($^3$He, 2p)$^4$He or p($^2$H,
$\gamma$)$^3$He,    row~3);   p-p(II)   (terminated    through   $e^-$($^7$Be,
$\nu_e$)$^7$Li,   row~4)   and    p-p(III)   (terminated   through   p($^7$Be,
$\gamma$)$^8$B,  row~5).  The  values  in column~2  correspond  to GS98  heavy
element  abundances and  conservative uncertainties,  column~3  corresponds to
AGS05  abundances and  optimistic uncertainties,  and column~4  corresponds to
AGS05         abundances        and         conservative        uncertainties.
Table~\ref{tab:abundanceuncertainties}   gives   the   numerical  values   for
conservative and optimistic abundance uncertainties.}
\end{table}

\section{CONCLUSIONS AND DISCUSSION}
\label{sec:discussion}

We provide  in this  paper quantitative estimates  of the accuracy  with which
standard solar models predict  measurable quantities.  These estimates provide
the most comprehensive summary of solar  model predictions and for many of the
predicted quantities, e. g., the 8 helioseismologically measured quantities
($Y_{\rm surf}$, $R_{\rm CZ}$, and the 6 difference rms)
the estimates given here provide the only consistent quantitative estimates of
the uncertainties of the predictions.

As  of this  writing, there  is  considerable uncertainty  regarding the  best
estimates  for  the  surface chemical  composition  of  the  Sun, one  of  the
important sets  of input parameters for  our Monte Carlo  simulations. We have
therefore carried  out parallel  sets of calculations  for two  very different
sets of heavy element abundances: the \citet[GS98]{gs98}
abundances and the \citet[AGS05]{ags05} recommended abundances. We
have used  conservative estimates  for the composition  uncertainties together
with  the  GS98 abundances  and  optimistic  uncertainties  together with  the
AGS05 abundances. Throughout this paper, and unless otherwise
noted,  we give  without  parentheses  the results  calculated  with the  GS98
abundances and conservative composition uncertainties and with parentheses the
results calculated with AGS05 abundances and optimistic uncertainties.

\begin{description}

\item[Input Parameters and Their Uncertainties.]  
In \S~\ref{sec:standardvalues}, we present  and discuss the best-estimates and
uncertainties for 19 important input  parameters that are used in constructing
the solar models  that are considered in this  paper. These parameters include
nuclear fusion  cross sections,  the solar age  and luminosity,  the diffusion
coefficient, and the 9 most  important heavy element abundances on the surface
of the Sun. Two additional input 'parameters', the radiative opacity and the
equation of  state, are  discussed separately in  \S~\ref{sec:opacityeos}. The
opacity  and the  equation of  state are  complicated functions  of  the local
conditions in the  star and must therefore be treated in  a different way than
the       single-valued        input       parameters       discussed       in
\S~\ref{sec:standardvalues}. For each standard solar model we simulate, all 21
input parameters are chosen from their separate probability distributions that
are described in \S~\ref{sec:standardvalues} and \S~\ref{sec:opacityeos}.

\item[Standard  Solar   Models:  23   Predicted  Quantities  and   Some  Model
  Characteristics.]  
We  present in  \S~\ref{sec:standardsolarmodel} the  best-estimate predictions
for  23 solar  quantities  that  are either  already  measured or  potentially
measurable. These quantities include the 8 dominant neutrino fluxes, the event
rates for  the chlorine and  gallium solar neutrino experiments,  8 quantities
that have been determined  precisely by helioseismological measurements, and 5
quantities (not  all independent) that characterize the  relative frequency of
different nuclear fusion  reactions in the Sun. We also  summarize some of the
main characteristics  (not directly measurable) of the  standard solar models,
including the principal physical variables at the center of the Sun and at the
base  of the  convective zone,  as well  as the  initial composition.  For the
reader's convenience,  we also  present compact tables  of the profile  of the
solar sound speed and the density. Using quadratic interpolation, these tables
can be used to reproduce precise values of the sound speed and density through
the Sun. Finally, we present quantities that are useful in precise analysis of
solar  neutrino   propagation,  including  the  radial   distribution  of  the
production of  the individual  neutrino sources, as  well as the  electron and
neutron number densities as a function of solar radius.

\item[The Depth of the Convective Zone and the Surface Helium Abundance.] 
The measured depth of the solar  convective zone is in good agreement with the
predictions of standard  solar models constructed with the  GS98 heavy element
abundances  (see   eq.~\ref{eq:gs98conservradiuscz}).  However,  solar  models
constructed with the AGS05 recommended abundances and
uncertainties ('optimistic uncertainties') disagree with the measured depth of
the   solar  convective   zone   by  the   equivalent   of  $3.9\sigma$   (see
eq.~\ref{eq:ags05optimisticradiuscz}). The strong disagreement goes away if we
adopt 'conservative  uncertainties' for the  heavy elements together  with the
AGS05 abundances (see eq.~\ref{eq:ags05conservradiuscz}); this results because
the optimistic uncertainties are large enough to reproduce a solar composition
that resembles that of GS98. 

The measured surface helium abundance is  in very good agreement with solar
models   constructed   with   the   GS98   heavy   element   abundances   (see
eq.~\ref{eq:GS98conservYhelio}).  The  agreement is  poor,  however, if  AGS05
abundances are  used: the discrepancy  between measured and  predicted surface
helium  abundance   is  $3.6\sigma$   (effective)  if  the   AGS05  optimistic
uncertainties   are   used  and   $2.8\sigma$   (effective)  if   conservative
uncertainties    are    adopted    (see   eq.~\ref{eq:AGS05optimYhelio}    and
eq.~\ref{eq:AGS05conservYhelio}). We  conclude that the measured  depth of the
solar  convective zone  and the  surface helium  abundance both  indicate that
models constructed  with the GS98  heavy element abundances  are significantly
closer to the actual Sun than are models constructed with the AGS05
recommended  abundances.  Figure~\ref{fig:rczsurfacey} shows,  for
different  assumed abundances  and  uncertainties, the  distribution of  solar
models with different values of the depth of the convective zone and different
values of the surface helium abundance.

\item[Sound Speed and Density Profiles.] 
We present in  \S~\ref{sec:soundsppeddensityprofiles} the distributions of the
sound speed  and density difference  rms.  Table~\ref{tab:rms} gives  the most
probable values  for the rms  distributions and the  respective uncertainties.
The Monte  Carlo simulation  of models constructed  with the  GS98 composition
have distributions  strongly peaked  very close  to zero for  all the  six rms
defined  and used  in  this work.   These  distributions reflect  a very  good
agreement  between solar  models  constructed with  the  GS98 composition  and
results from helioseismology, reinforcing the conclusions drawn from the depth
of the convective zone and  surface helium abundance discussed above.  The set
of models constructed  with the AGS05 composition, on  the contrary, give rise
to rms distributions that make evident  the discrepancy in the sound speed and
density  profiles introduced  by the  adoption of  the AGS05  composition.  In
addition,   Figures~\ref{fig:soundspeed}~and~\ref{fig:density}  allow   us  to
conclude that the  uncertainties in all the other  input parameters entering a
standard solar model  cannot compensate for the degradation  introduced by the
new recommended set of solar abundances.

\item[The Predicted Solar Neutrino Fluxes.] 
We  analyze  in \S~\ref{sec:neutrinofluxes}  the  calculated distributions  of
solar  neutrino fluxes.   Table~\ref{tab:neutrinofluxuncertainties}  gives the
total $1\sigma$ uncertainty for each neutrino source and for all three choices
of heavy element composition and  their uncertainties. The results are in very
good agreement with the uncertainties estimated using power-law dependences of
the fluxes as  a function of input parameters.   The distribution of
calculated  fluxes for  each neutrino  source is  well described  by  either a
normal or a  lognormal distribution, depending on what  is the dominant source
of uncertainty, with the tabulated standard deviation. 

The calculated $^8$B  solar neutrino flux is in good  agreement with the value
measured  by solar neutrino  experiments. The  theoretical uncertainty  in the
prediction of the $^8$B neutrino flux (11\% to 17\%, depending upon the choice
of heavy element composition and uncertainties) is larger than the uncertainty
(5\%) in the experimental determination. For all other solar neutrino sources,
the experimental uncertainties greatly exceed the solar model uncertainties.

The $^7$Be  solar neutrino flux  will be measured  in the next few  years. The
solar model uncertainties  in the prediction of the  $^7$Be neutrino flux vary
from 9.3\% to 10.5\% depending upon the choice of heavy element abundances and
their  uncertainties.  It is  possible  that the  pep  neutrino  flux will  be
measured  in one  of the  same experiments  as the  $^7$Be neutrino  flux. The
predicted anti-correlation between the pep and $^7$Be neutrino fluxes is given
in  eq.~(\ref{eq:deltapepvsdelta7Be})  and  eq.~(\ref{eq:betapepbe7})  and  is
shown in figure~\ref{fig:corr-pepbe7ppbe7}.

The  solar model  predictions for  the  p-p and  the pep  neutrino fluxes  are
strongly  correlated   with  each   other  and  the   p-p  neutrino   flux  is
anti-correlated with  the $^7$Be neutrino  flux.  These relations  between the
predicted  neutrino fluxes  represent  important testable  predictions of  the
solar   models   and   are    discussed   and   analyzed   quantitatively   in
\S~\ref{sec:neutrinofluxes}.

\item[The Correlation Coefficients of the Neutrino Fluxes.]  
The  correlations  between  the   different  neutrino  fluxes  are  succinctly
summarized     by     the      matrix     of     correlation     coefficients.
Table~\ref{tab:GS98correlationcoefficients}                                and
Table~\ref{tab:AGS05correlationcoefficients}     present    the    correlation
coefficients  of the  neutrino fluxes  for, respectively,  GS98  heavy element
abundances and  conservative uncertainties and AGS05  heavy element abundances
and optimistic  uncertainties. These correlation  coefficients can be  used to
make a more constrained and precise analysis of solar neutrino oscillations.

\item[Nuclear Energy Generation Pathways.]
Table~\ref{tab:fusionfractions}  summarizes  the  calculated fraction  of  the
total nuclear energy  generation that is produced by  different nuclear fusion
pathways. For  all choices of the  surface heavy element  abundances and their
uncertainties, the  p-p chain is responsible  for more than 99\%  of the total
energy  generation. The  estimated uncertainty  in the  p-p  energy generation
fraction is less than or of order  0.2\%. The CNO energy fraction is less than
1\%. About  88\% to 90\% of the  p-p energy generation is  from reactions that
are terminated by the $^3$He($^3$He, 2p)$^4$He reaction with an uncertainty of
about 1\%. Although  most of solar neutrino astronomy has  so far been focused
on the high energy $^8$B neutrinos,  the nuclear fusion reactions that lead to
the production  of $^8$B  represent less  than 1\% of  the total  solar energy
generation (best-estimate varies from 0.81\%  to 0.91\% with an uncertainty of
only 0.08\%).

Future solar neutrino experiments that measure different solar neutrino fluxes
can determine empirically the nuclear energy generation fractions and test the
solar model predictions given in Table~\ref{tab:fusionfractions}.

\end{description}

J. N.  B. and A. M.  S. were supported in  part by NSF grants PHY-0070928 and
PHY-0503584 to the Institute for Advanced Study.  A.M.S. was also supported by
the  W.  M.  Keck  Foundation  through a  grant-in-aid  to  the Institute  for
Advanced Study.  S.  B. was  partially supported by NSF grants ATM-0206130 and
ATM-0348837.    This  work   utilizes   data  from   the  Solar   Oscillations
Investigation /  Michelson Doppler Imager (SOI/MDI) on  the Solar Heliospheric
Observatory (SOHO).  The MDI project is  supported by NASA  grant NAG5-8878 to
Stanford University.   SOHO is a project of  international cooperation between
ESA  and  NASA.   We  are  grateful   to  M.  Chen,  P.  Goldreich,  E.  Lisi,
  A.   McDonald,  M.  Pinsonneault,  and  S.   Tremaine,  and
particularly to C.  Pe\~na-Garay for valuable
comments,  discussions, and suggestions  during the  extended period  in which
this project was carried out  and the results analyzed. Computations presented
in this paper were done with the Scheides Beowulf cluster at the Institute for
Advanced Study.

\begin{appendix}
\section{Lognormal distribution} 

We briefly summarize some important properties of lognormal probability
distribution  functions used  in  this  work.  We  also  define the  $1\sigma$
confidence level interval  we have adopted throughout this  paper when a given
quantity is lognormally distributed.

A random variable $x$ is  lognormally distributed when its probability density
function $f(x)$ is given by 
\begin{equation}
f(x)=       \left[      s       \sqrt{2\pi}   ( x-\theta)      \right]^{-1}
\exp{\left[-\frac{(\log{(x-\theta)} - m)^2}{2 s^2} \right]}. 
\end{equation}
Here, $m$ and $s$ are the scale and shape parameters respectively. $\theta$ is
the location parameter which we assume equal to 0 for simplicity. 
The  most  probable  value  (mode)  and  the  mean  values  of  a  lognormally
distributed quantity, respectively, are 
\begin{equation}
Q=\exp{(m-s^2)} \ \ \ ;\ \ \ \ \ \ \mu=\exp{(m+s^2/2)}. 
\end{equation}
In general, we adopt as the $1\sigma$ confidence level interval that given by
the limits of the integral 
\begin{equation}
\int^{x_+}_{x_-} f(x) \,dx = 0.683, \ \ \ \ {\rm with} \ f(x_-)=f(x_+). 
\label{eq:confint}
\end{equation}
Equation~\ref{eq:confint} uniquely defines the $1\sigma$ confidence
level interval $\left[ x_-,x_+\right]$.  An interesting relation between $x_-$
and $x_+$ is 
\begin{equation}
x_-x_+= Q^2.
\end{equation}
The $1\sigma$ confidence level interval is conveniently expressed with respect
to  the mode  $Q$  by  introducing the  quantities  $\sigma_+$ and  $\sigma_-$
defined by 
\begin{equation}
\sigma_+= x_+ - Q \ \ \ \ \ \ \sigma_-= Q - x_-.
\end{equation}
When relative  values for $\sigma_+$ or  $\sigma_-$ are quoted  in this paper,
they are effectively calculated as $\sigma_+/Q$ and $\sigma_-/Q$.

In practice, for a given dataset $\left\{x_i\right\}_{i=1}^N$, $m$ and $s$ are
estimated as 
\begin{equation}
m= N^{-1} \sum_{i=1}^N \log{x_i}
\label{eq:m}
\end{equation}
and
\begin{equation}
s^2= (N-1)^{-1} \sum_{i=1}^N (\log{x_i}-m)^2\,. 
\label{eq:s}
\end{equation}

\end{appendix}

\end{document}